\documentclass[fleqn,usenatbib]{mnras}

\usepackage{newtxtext,newtxmath}

\usepackage[T1]{fontenc}
\DeclareRobustCommand{\VAN}[3]{#2}
\let\VANthebibliography\thebibliography
\def\thebibliography{\DeclareRobustCommand{\VAN}[3]{##3}\VANthebibliography}

\usepackage{graphicx}	
\usepackage{amsmath}	
\usepackage{subcaption}
\usepackage{amsfonts}
\usepackage{orcidlink}
\usepackage{xspace}
\usepackage{hyperref}
\newcommand{\hi} {H\,{\sc I}\xspace}

\def\omegahi{\Omega_{\rm H\small{I}}}
\def\omegastar{\Omega_{\rm star}}
\newcommand{\mhi}{M_{\rm H\sc{I}}}
\newcommand{\htwo}{{\rm H}_2}
\newcommand{\fmag}{\ensuremath{\mathcal{F}_{\rm mag}}\xspace}
\newcommand{\fcolor}{\ensuremath{\mathcal{F}_{\rm color-mag}}\xspace}
\def\mstar{M_{\rm star}}

\def\rhostar{\rho_{\rm star}}

\def\nstar{n_{\rm star}}

\def\msun{M_{\odot}}

\def\vmax{V_{\text{max}}}
\def\veff{V_\text{eff}}
\def\vsurvey{V_{\rm survey}}
\def\ngal{N_\text{gal}}
\def\vhelio{v_{\textrm{helio}}}
\def\zhelio{z_{\textrm{helio}}}
\def\wfifty{W_{50}}
\def\wtwenty{W_{20}}
\def\sflux{S_{21}}
\def\alphah{\alpha.100}
\def\zcmb{z_{\textrm{cmb}}}
\def\mkcorrect{M_{\texttt{kcorrect}}}
\def\mtaylor{M_{\texttt{Taylor}}}
\def\mgswlc{M_{\texttt{GSWLC}}}
\def\sigmamtaylor{\sigma_{M_{\texttt{Taylor}}}}
\def\sigmamgswlc{\sigma_{M_{\texttt{GSWLC}}}}
\def\mkcorrectgswlc{M_{\texttt{kcorrect+GSWLC}}}
\newcommand{\software}[1]{{\large #1}} 
\def\kcorrect{\texttt{KCORRECT}\,\,}
\def\gswlc{\texttt{GSWLC}\,\,}
\def\taylor{\texttt{TAYLOR}\,\,}
\def\kgswlc{\texttt{KCORRECT+GSWLC}\,\,}
\def\hseventy{h_{70}}
\def\reducedchisq{\chi_{\rm red}^2}

\title[The Stellar Mass Function of Gas-Rich Galaxies]{The Stellar Mass Function of Gas-Rich Galaxies and the Underlying $\mhi-\mstar$ Scaling Relation in the Local Universe} 
\author[T. Tripty et al.]{
Tanya Tripty,$^{1,2}$\thanks{E-mail: 
\href{mailto:tanya.tripty@niser.ac.in}{tanya.tripty@niser.ac.in} }
\orcidlink{0009-0005-0224-0096}
Nishikanta Khandai,$^{1,2}$\orcidlink{0000-0003-3081-0189}
Saili Dutta,$^{3}$\orcidlink{0000-0002-8858-5845}
Khushi Sharma$^{1,2}$\orcidlink{0009-0004-2480-6866}
\\
$^{1}$School of Physical Sciences, National Institute of Science Education and Research, Jatni 752050, Odisha, India\\
$^{2}$ Homi Bhabha National Institute, Training School Complex, Anushaktinagar, Mumbai 400094, Maharashtra, India\\
$^{3}$ National Centre for Radio Astrophysics, Tata Institute of Fundamental Research, Pune University Campus, Pune 411007, India }
\date{Accepted XXX. Received YYY; in original form ZZZ}

\pubyear{2026}

\begin{document}
\label{firstpage}
\pagerange{\pageref{firstpage}--\pageref{lastpage}}
\maketitle

\begin{abstract}
We estimate the Galaxy Stellar Mass Function (GSMF) of \hi gas-rich galaxies using the 100\% ALFALFA ($\alphah$) catalog, $\sim 98\%$ of which have optical counterparts
in the Sloan Digital Sky Survey (SDSS) and a subset of them have counterparts in GALEX SDSS WISE Legacy Catalogue-2(GSWLC-2). 
We use the mass estimates from this subset which combines UV, optical and IR bands with individual dust corrections to recalibrate 
optical stellar mass estimates. We use a non-parametric method to estimate the GSMF of these gas-rich galaxies. The resulting, \hi-selected GSMF is consistent with a single Schechter function with best-fit parameters $\left\{\phi_*  (10^{-3}\, h_{70}^{3}\,\mathrm{Mpc}^{-3}\,\mathrm{dex}^{-1}), \log_{10} (M_*/M_{\odot}) + 2\log_{10} h_{70},  
\alpha \right\} = \left\{2.30^{+0.12}_{-0.12},
\,10.83^{+0.01}_{-0.01},\,
-1.14^{+0.02}_{-0.02}\right\} $. 
Additionally, the red and blue populations are each well described by a single Schechter function. After correcting for selection effects, we find that the red population accounts for only $\sim18\%$ of gas-rich galaxies by number, yet contributes $\sim54\%$ of the total stellar mass, with the blue population accounting for the rest. Using an optically selected sample and a joint optical-\hi sample, we find gas-rich galaxies represent $\sim 33\%$ of the total stellar mass density and $\sim 39\%$ of the total galaxy number counts in the local Universe. We use the GSMF and the \hi mass function (HIMF) of the \hi-selected sample to obtain the $\mhi-\mstar$ relation, which is free from selection bias.  
\end{abstract}




\section{Introduction}
\label{Sec_intro}
Understanding the relation between neutral atomic hydrogen (\hi) gas and the stellar content of galaxies is fundamental to galaxy formation and evolution. \hi serves as the primary reservoir for future star formation whereas the stellar mass is a record of the 
integrated star formation history of a galaxy. In the right environment, \hi cools further and transitions to molecular hydrogen, $\htwo$, to eventually form stars in nebulae. Star formation is a complex process \citep{2007ARA&A..45..565M} and active star formation occurs in regions buried deep inside nebulae that are the most $\htwo$-dense and dust-rich environments in the interstellar medium. The star formation rate (SFR units of $\msun/{\rm yr}$) of a galaxy is the aggregate of the local SFR in these cold star forming regions. Indeed the local star formation rate is strongly correlated 
with the local density of $\htwo$ in observations \citep{2008AJ....136.2846B,2008AJ....136.2782L} which in turn depends on the supply 
of \hi on larger, galactic scales. 

Single dish telescopes, like  Parkes and Arecibo, have been used to carry out 
large, blind \hi surveys, like the \hi Parkes All-Sky Survey (HIPASS) \citep{2001MNRAS.322..486B} and  the Arecibo Legacy Fast ALFA (ALFALFA) survey 
\citep{2005AJ....130.2598G}, in the local Universe.  
These surveys have, in turn, been used to constrain the abundance of gas-rich galaxies, the \hi mass function \citep{2003AJ....125.2842Z,2011AJ....142..170H,2018MNRAS.477....2J,2020MNRAS.494.2664D,2022MNRAS.509.3268O}, and the 
cosmic \hi density parameter, $\omegahi$, at $z \simeq 0$. Recently data from the radio  interferometer, MeerKAT\footnote{\href{https://www.sarao.ac.za/science/meerkat/}{https://www.sarao.ac.za/science/meerkat/}}, have been used to constrain the HIMF in the 
MeerKAT International GigaHertz Tiered Extragalactic Exploration (MIGHTEE)-HI Survey
\citep{2023MNRAS.522.5308P}.

Using optical counterparts from the Sloan Digital Sky Survey (SDSS\footnote{\href{https://www.sdss.org/}{https://www.sdss.org/}}) of the ALFALFA 40\% sample ($\alpha.40$) \citep{2011AJ....142..170H}, \cite{2020MNRAS.494.2664D} have shown that red galaxies contribute $\approx17\%$ in $\omegahi$, whereas the blue population contributes $\approx80\%$ (the remaining $\approx 3\%$ is accounted for by optically dark, gas-rich galaxies). Thus \hi-rich galaxies are largely associated with the blue population  while the red population tend to be gas-poor. 
Using SDSS, \cite{2004ApJ...600..681B}) show that the, optically selected galaxy population in stellar mass is bimodal in color. The blue population, which are 
mostly star-forming galaxies, are the dominant population at lower stellar mass, 
whereas the red population, which are mostly quiescent galaxies, are the dominant 
population at the high mass end\citep{2012MNRAS.421..621B}. 

The \hi-selected sample and optically-selected sample therefore probe different 
galaxy populations. The \hi-selected sample are biased towards gas-rich (blue) galaxies, whereas optically selected  sample includes both gas-rich (blue) as well as gas-poor (red) galaxies \citep{2012ApJ...756..113H}.  As a result, the relative contribution of \hi selected galaxies to the overall galaxy population in terms of stellar mass, especially when separated into red and blue populations, remains an important question.
The extent to which \hi selected galaxies trace these populations, and how their contribution varies in terms of number and mass density, is not yet fully quantified. 

In this work, we aim to answer these questions 
and estimate the GSMF of \hi-gas-rich galaxies using the 100\% ALFALFA ($\alphah$) catalog, $\sim 98\%$ of which are cross-matched with  optical counterparts \citep{2020AJ....160..271D} in the Sloan Digital Sky Survey (SDSS) and a subset of them are also matched with ultraviolet (UV) and infrared (IR) counterparts from the Galaxy Evolution Explorer (GALEX\footnote{\href{http://www.galex.caltech.edu/}{http://www.galex.caltech.edu/}}), and the Wide-field Infrared Survey Explorer (WISE\footnote{\href{https://wise2.ipac.caltech.edu/docs/release/allsky/}{https://wise2.ipac.caltech.edu/docs/release/allsky/}}). We use two stellar mass estimates based on optical photometry and the third  is based on GALEX SDSS WISE Legacy Catalogue-2 (GSWLC-2) mass estimates \citep{2016ApJS..227....2S}. The latter utilizes all three surveys in the UV, optical and IR bands and accounts for individual dust corrections to better estimate stellar masses. We also construct an optically selected sample from SDSS which overlaps with $\alpha.100$ and a joint optical-HI sample to look at the detection fraction of $\alpha.100$ with respect to an optically selected sample.

Observed galaxy scaling relations, defined as correlations between different galaxy properties, help us better understand galaxy structure, formation and evolution \citep{2021FrASS...8..157D}. Well-known scaling relations such 
as the ($\Sigma_{\rm HI+H_2}-\Sigma_{\rm SFR}$) Kennicutt–Schmidt law
\citep{1959ApJ...129..243S,1963ApJ...137..758S,1989ApJ...344..685K,1998ApJ...498..541K},
SFR-$\mstar$ (stellar mass) relation of the star forming main sequence \citep{2007ApJ...660L..43N},  
the mass–metallicity relation\citep{2004ApJ...613..898T}, 
and the baryonic scaling laws \citep{2000ApJ...533L..99M,2016ApJ...816L..14L} show that galaxy growth is determined by a complex interplay between gas accretion, star formation,
gas enrichment, feedback processes and gas recycling  
\citep{2013ApJ...772..119L,2015ARA&A..53...51S,2017ARA&A..55..389T}.

In this context, the relation between \hi mass ($\mhi$) and $\mstar$ is of particular significance, as it is a physical indicator of a galaxy's raw fuel reserves relative to its current size, directly dictating its future star-formation capacity and evolutionary stage. Observations from \hi surveys, such as the ALFALFA survey, show that \hi selected galaxies follow a broad but well-defined relation between $\mhi$ and $\mstar$ \citep{2012ApJ...756..113H,2015MNRAS.447.1610M,2020MNRAS.494.2664D}. These studies show that low stellar mass galaxies have higher gas fraction compared to the massive galaxies. However, this relation depends strongly on how galaxies are selected. \hi selected samples preferentially trace gas-rich, star-forming galaxies, so they tend to show higher gas content at a given stellar mass compared to optically selected samples, e.g. the  GALEX Arecibo SDSS Survey \citep[GASS][]{2010MNRAS.403..683C} and extended GALEX Arecibo SDSS Survey \citep[xGASS][]{2018MNRAS.476..875C}.
The \hi-selected $\mhi-\mstar$ scaling relations of 
\cite{2012ApJ...756..113H,2015MNRAS.447.1610M,2020MNRAS.494.2664D} do not account
for the $\alpha.100$ selection function and the observed relation is therefore biased towards
higher $\mhi$ at fixed $\mstar$. As a result scaling relations derived from \hi selected sample may not fully capture the underlying galaxy population of the \hi-selected sample. 

We account for the $\alpha.100$ selection function and derive a $\mhi-\mstar$ 
scaling for the \hi-selected sample and validate this relation by comparing with that derived from a volume-limited sample, which is complete. We also obtain this relation 
for the red and blue population of galaxies in our sample.

The paper is organized as follows. In section~\ref{sec_data}, we describe  the \hi and optical data set used in this paper. In section~\ref{sec_methods}, we present the 
HIMF and the GSMF of gas-rich galaxies and use it to obtain the $\mhi-\mstar$
scaling relation for \hi selected galaxies in local Universe.
In section~\ref{sec_optical-hi-selection} we look at the effect of \hi selection on an optically selected sample. We compare the GSMF of the optically selected sample  with the joint optical-\hi sample. In section~\ref{sec_underlying_rel} we present 
the $\mhi-\mstar$ scaling relation for the \hi-selected and joint optical-\hi selected  samples and compare them with recent results from the MIGHTEE-HI Survey
\citep{2023MNRAS.522.5308P,2023MNRAS.525..256P}. 
We summarize our results and discuss their implications in section~\ref{sec_discussion}.

Unless specifically mentioned we assume  a flat $\Lambda$CDM cosmology with  parameters  $(\Omega_\Lambda, \Omega_{\rm m}, h) = (0.7, 0.3, 0.7)$, where $h$ is the dimensionless Hubble's constant $H_0 = 100h\, \textrm{km.s}^{-1}$ and 
$\Omega_\Lambda$ and $\Omega_{\rm m}$ are the density parameters at $z=0$ associated with the cosmological constant and matter respectively. We use the notation adopted earlier \citep{2022MNRAS.511.2585D}, to define a normalised Hubble's constant, 
$h_{70} = h/0.7 = H_0/(70 \textrm{km.s}^{-1}\textrm{Mpc}^{-1})$. For the value $h=0.7$, used here, $h_{70} = 1$.
The values of mass (\hi or stellar) quoted in this paper, 
will be in units of $M \equiv \log_{10} (M/\msun) + 2\log_{10}h_{70}$. We define three stellar mass estimates: Taylor (\citep{2011MNRAS.418.1587T}), \kcorrect (\citep{2007AJ....133..734B}), and \kcorrect calibrated using \gswlc masses ($\mgswlc$), denoted by $\mtaylor$,$\mkcorrect$, and $\mkcorrectgswlc$, respectively. E.g. the expressions $\{\mkcorrect \geq 8.0\}$ and $\{\log_{10}(\mkcorrect/(\msun/\hseventy^2)) \geq 8.0\}$ are equivalent and will 
be used interchangeably.
\section{Data}
\label{sec_data}
Our gas-rich sample is from the ALFALFA 100\%\footnote{\href{https://egg.astro.cornell.edu/alfalfa/data/index.php}{https://egg.astro.cornell.edu/alfalfa/data/index.php}}
data release ($\alphah$) \citep{2018ApJ...861...49H}.  The $\alphah$ catalog covers
two disjoint regions the Right Ascension(RA)-Declination(Dec) plane
- $22^h < \textrm{RA} < 3^h, 0^\circ < \textrm{Dec} < 36^\circ$ and $7^h30^m < \textrm{RA} < 16^h30^m,0^\circ< \textrm{Dec} <36^\circ$
in the sky with an
area of  $\sim 7000\,\,  \textrm{deg}^2$,
containing $\sim 31,500$ galaxies. 
The catalog contains the following columns:
1. Entry Number in the Arecibo General Catalog,
2. Common name for the corresponding Optical Counterpart (OC) if available,
3. Centroid (J2000) of the \hi line source (in hhmmss.sSddmmss),
4. Centroid of the most probable OC (in hhmmss.sSddmmss),
5.  Heliocentric Velocity $\vhelio = c\zhelio$(in km/s),
6. Velocity width of the \hi line profile, $\wfifty$ (in km/s),
computed as the width at $50\%$ of the peak flux density, and its uncertainty $\left(\sigma_{\wfifty}\right)$,
7. Velocity width of \hi line profile, $\wtwenty$ (in km/s), computed as the width
at  $20\%$ of peak flux density,
8. Integrated \hi line flux density, $\sflux$ (in Jy km/s), and its uncertainty ($\sigma_{\sflux}$),
9. Signal to Noise Ratio (SNR),
10. Noise in spatially integrated spectral profile,
11. Luminosity distance ($D_{\textrm{L}}$) and associated uncertainty ($\sigma_{\textrm{D}}$) (in Mpc).
For objects with $c\zhelio \geq 6000$ km/s, it is   $c\zcmb/H_0$ where $c\zcmb$
is the recessional speed of the galaxy in the Cosmic Microwave Background (CMB) frame of reference.
For lower velocities, a local flow model \citep{2005PhDT.........2M} is used  to calculate the distance.
12. \hi mass ($\mhi$ in units of solar mass) calculated using $\mhi = 2.356 \times 10^5 D_{\textrm{L}}^2\sflux$ and its uncertainty($\sigma_{\mhi}$).
13.The \hi code which can be 1 for galaxies with SNR > 6.5 (25,433 Galaxies) or 2 otherwise (6,068 Galaxies). 

\begin{figure}
    \includegraphics[width=\columnwidth]{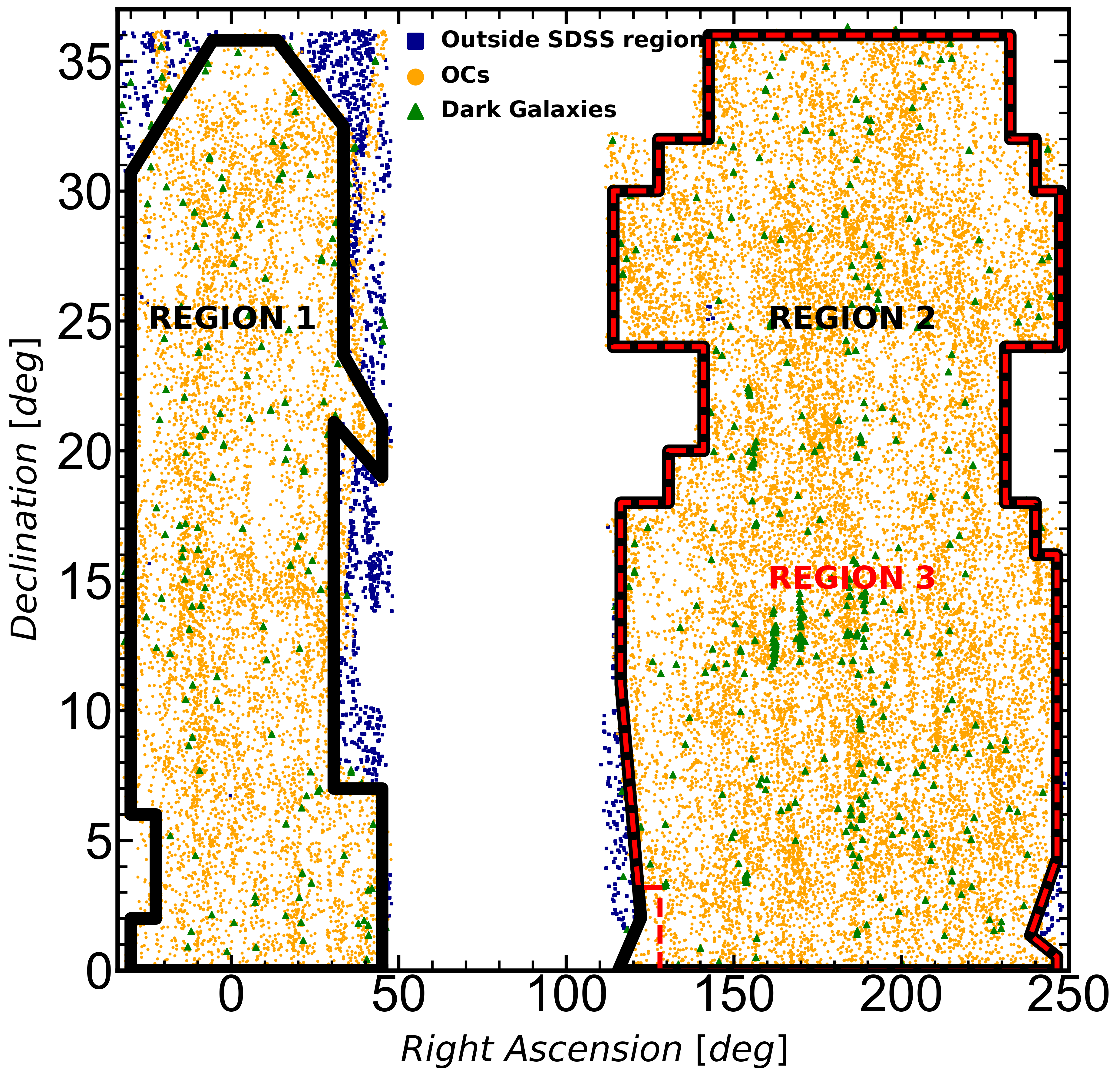} \\
    \caption{The \hi sample used in this work plotted in the RA-Dec plane. Each point is a galaxy with an \hi detection.
      Galaxies inside the solid black boundary (regions 1 and 2) are used to construct \hi selected sample described in Sec. \ref{sec._HI_selc._sample}. 
      The blue points represent galaxies outside the SDSS footprint. The yellow (green) points 
      are galaxies which have (do not have) an OC. All yellow points have photometric detections. 
      The green points represent $\simeq 2\%$ of all points within the black boundary of regions 1 and 2.
      The red dashed boundary, contained within region 2, is region 3. 
      The optically selected and joint optical-\hi sample described 
      in sections \ref{sec_optical_sel._sample} and \ref{sec_joint_HI_optical_sample} respectively, 
      are in region 3.
      The spectroscopic detections are in region 3 (in addition to  
      some patches of region 1, which we do not consider) which is used in constructing the optically selected and the joint optical-\hi sample.       
      }
  \label{fig_ra-dec}
\end{figure}

The ALFALFA-SDSS galaxy catalog \citep{2020AJ....160..271D} is a value-added catalog of optical counterparts of ALFALFA sources.
It has optically determined properties, e.g. stellar mass, rest-frame magnitudes, star formation rates, for 29638 galaxies ($\sim 94\%$)
of the 31501 galaxies from ALFALFA.
We will use two stellar mass estimates from this catalog and one estimated directly from SDSS photometry using \kcorrect \citep{2007AJ....133..734B}.
All the three stellar mass estimates are based on the \cite{2003MNRAS.344.1000B}{(BC03)} stellar population synthesis (SPS) models. 
\begin{itemize}

\item \kcorrect stellar masses ($\mkcorrect$) - This stellar mass is derived using SDSS 
  cmodel magnitudes queried using the SDSS
  Cross-ID tool\footnote{\href{https://casjobs.sdss.org/casjobs/}{https://casjobs.sdss.org/casjobs/}},
  along with their uncertainties determined 
  using the \kcorrect package(5.1.8)\citep{2007AJ....133..734B}. The 
  redshift used here is derived from the distance quoted in the 
  ALFALFA catalog, which is corrected for  peculiar velocities
  and the magnitudes are  extinction corrected \citep{1998ApJ...500..525S} due to our galaxy.

    \item Taylor stellar masses ($\mtaylor$) - 
      This stellar mass is based on an empirical scaling relation
      between stellar mass, $i-$ band luminosity, $L_i$, and the restframe $g-i$ color,
      $\mtaylor/L_i = -0.68 + 0.70(g-i)$ 
      for galaxies with error in $g$ or $i$ band $< 0.05$, 
      based on data from the Galaxy and Mass Assembly (GAMA) survey \citep{2011MNRAS.418.1587T}. 
      A \cite{2000ApJ...533..682C}
      dust obscuration law is used while estimating stellar masses.
      \cite{{2020AJ....160..271D}} correct the SDSS $g-i$ color 
      for internal extinction due to inclination before using the above relation to
      estimate stellar masses along with their associated uncertainties($\sigmamtaylor$).
      
    \item \gswlc-2 stellar masses ($\mgswlc$) - 
      These estimates along with their associated uncertainty ($\sigmamgswlc$) are taken from
      Galaxy Evolution Explorer (GALEX) - SDSS -
      Wide-Field Infrared Survey Explorer (WISE) Legacy catalog-2 (\gswlc)
      using UV/optical/IR SED (Spectral Energy Distribution) fitting \citep{2016ApJS..227....2S}. 
      The added photometry in UV and IR bands allows for separate dust attenuation curves to be used 
      for each galaxy \citep{2018ApJ...859...11S} to better estimate stellar masses.
\end{itemize}

The $\mgswlc$ ($\mkcorrect$) stellar mass estimates are the most (least) accurate due to the
 effect of including (excluding) attenuation due to dust.


\begin{figure*}
  \begin{tabular}{cc}
   \includegraphics[width=3.4in]{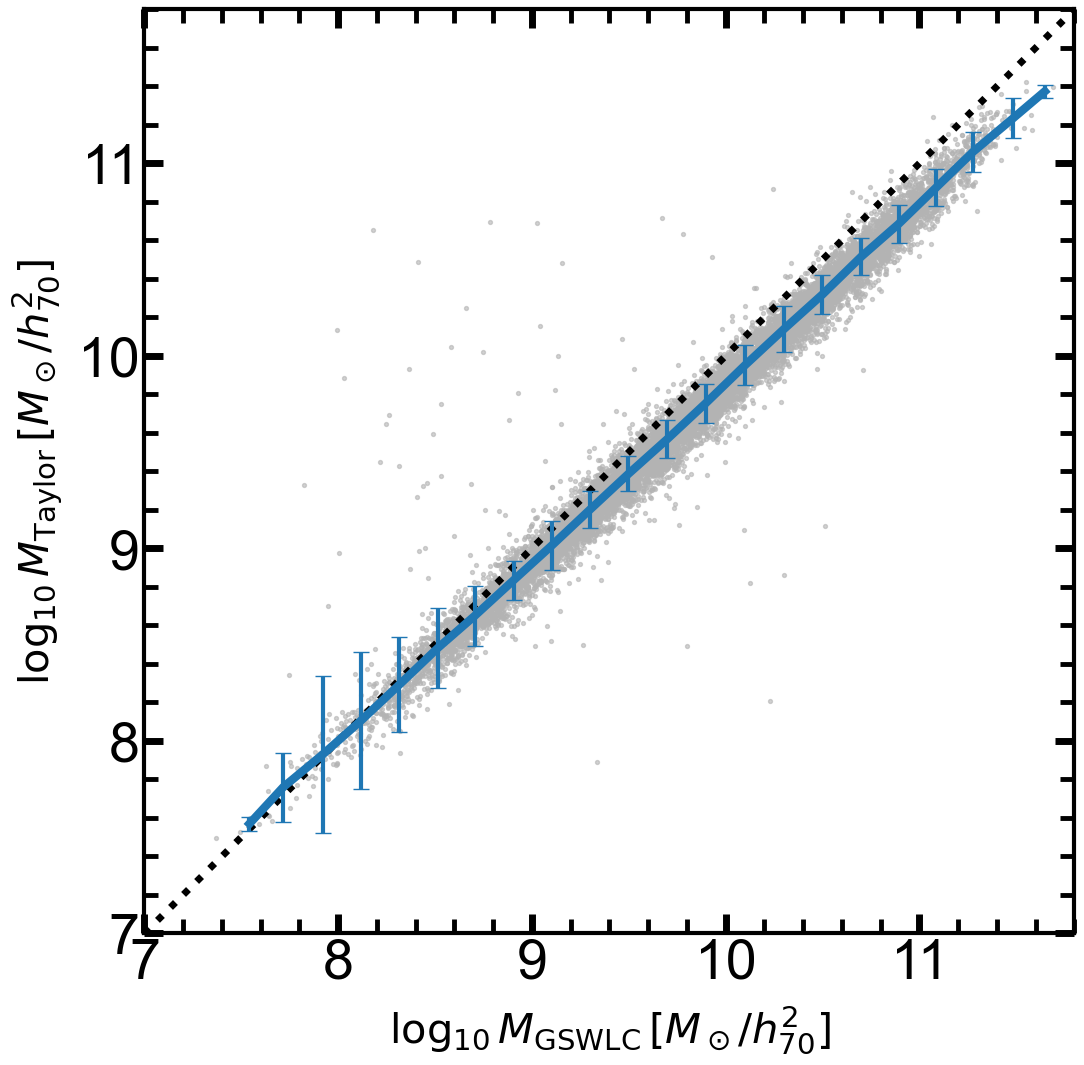} 
   \includegraphics[width=3.4in]{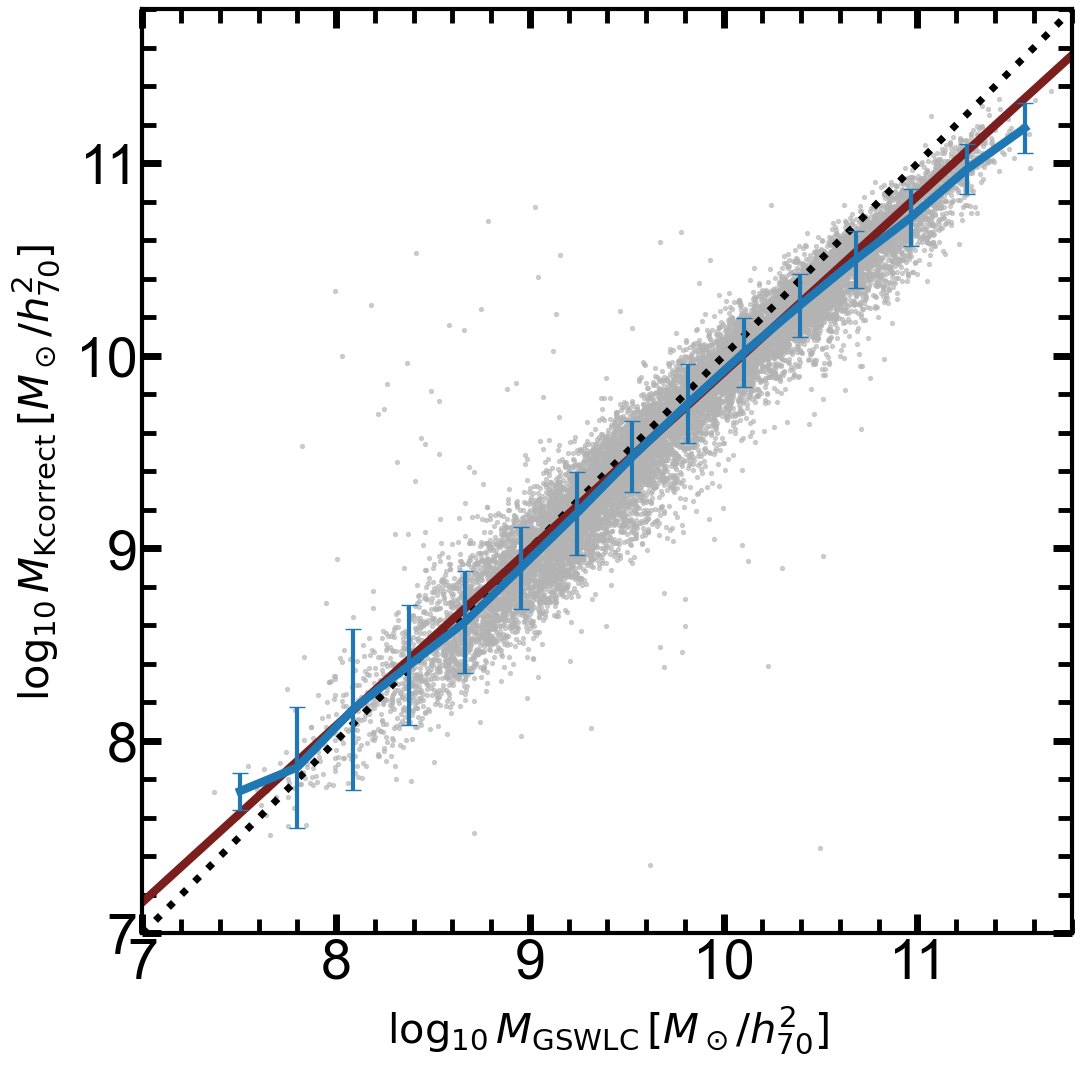}\\
  \end{tabular}
   \caption{\emph{Left}: The scaling relation between \gswlc and Taylor stellar mass estimates 
   for each galaxy (points).  
   \emph{Right}: The scaling relation between \gswlc and \kcorrect stellar mass estimates.
   The blue line with error bars is the mean relation in mass bins and their variance 
   in each bin. The black dotted line is the $y=x$ curve. 
   The solid maroon line in the right panel shows the best fit linear 
   relation $\mkcorrect = 0.917 \mgswlc+0.740$, used to calibrate the \kcorrect mass estimates.}
    \label{fig_mtaylor-mkcorrect-mgswlc}
\end{figure*}

\subsection{\hi selected Sample}\label{sec._HI_selc._sample}
We use the slightly conservative boundary which moves in further in RA $\sim 0.5$hr compared to the targeted $\alphah$ boundary,
as outlined in \cite{2018MNRAS.477....2J}. The ambiguity in the boundary
is due to the variation in the start and stop times at the targeted boundary for the drift-scan mode (moving in RA and not Dec) 
of the observations \cite[See figure 1 of ]{2018MNRAS.477....2J}. The points in figure~\ref{fig_ra-dec} are the ALFALFA detections in this conservative
boundary. In order to maximize the overlap between ALFALFA and SDSS we further shrink this region as outlined in the solid line in figure~\ref{fig_ra-dec}.
The blue points are outside the SDSS footprint and the rest are inside. The yellow points have an OC and the green points do not. We refer to the latter as dark galaxies.
The details of our boundary's vertices are provided in Appendix \ref{sec_boundary}.
Approximately 98.1\% (27,843 of 28,373), inside the solid boundary, have optical counterparts with the remaining galaxies being too faint to be detected in SDSS.
These numbers are similar to our previous sample based on the $\alpha.40$ sample \citep{2020MNRAS.494.2664D}. The dark galaxies
would need deeper followup in order to be detected in optical and their inferred stellar masses are expected to be lower
than the limiting value of $\mstar = 7$ which we adopt here (this is discussed in section \ref{sec_methods} and appendix~\ref{sec_incompl_GSMF}).

We follow \cite{2018MNRAS.477....2J} and exclude galaxies which are affected by RFI and have large uncertainty in their peculiar velocities.
We first consider the redshift range $0 < \zcmb < 0.05$, which leaves us with 27,103 galaxies (OCs = 26,592). Here $\zcmb$ is the redshift in the CMB frame.
At fixed \hi mass and redshift, edge-on galaxies have broader spectra compared to face-on galaxies. Their detection probability
is therefore lower compared to face-on galaxies. The \hi selection function therefore depends on both the flux, $\sflux$, and the observed velocity width, $\wfifty$.  
In ALFALFA the 50\% completeness relation used in determining the selection function can be derived from the data itself \citep{2011AJ....142..170H,2022MNRAS.509.3268O}.
We use the $\alphah$ completeness relation given by \cite{2022MNRAS.509.3268O}.
 \begin{equation}
 \log \sflux =\begin{cases} 
 0.5\log \wfifty -1.170 & :\log \wfifty < 2.5\\
 \log \wfifty -2.420 & :\log \wfifty \ge 2.5
 \end{cases}
 \label{eq_completeness_equation}
 \end{equation}
After applying this completeness cut, we are left with 21412 galaxies (OCs = 20,963).
Following \cite{2022MNRAS.509.3268O} we apply an additional quality cut, SNR > 6.5, 
to ensure that equation \ref{eq_completeness_equation} is appropriately applicable to our sample, which leaves us with 19,838 galaxies (OCs = 19,412).
Our HIMF (discussed in section~\ref{sec_methods}  and figure~\ref{plot_himf})
is based on this sample of 19,838 galaxies, 
with an area of 6070 $\mathrm{deg}^2$ and a corresponding comoving volume, $ 5.84 \times 10^6 \textrm{Mpc}^3$.

Our selection boundary is slightly more conservative than that adopted by \cite{2022MNRAS.509.3268O} and \cite{2018MNRAS.477....2J}; 
our sample is therefore smaller.
The selection criteria used by \cite{2018MNRAS.477....2J} includes SNR > 6.5, $\zcmb < 0.05$, a mass limit of \( \log(\mhi/\msun) > 6 \),
a FWHM limit of \( \log_{10}(\wfifty/\text{km/s}) > 1.2 \),
and a completeness cut of \cite{2011AJ....142..170H} based on the $\alpha.40$ sample.
These criteria resulted in a sample of 22,831 ($\sim 15\%$ larger than our sample) galaxies covering an area of 6501 deg\(^2\) ($7\%$ larger area than this work).
In comparison, \cite{2022MNRAS.509.3268O}
did not impose the mass cut, which led to the inclusion of four additional galaxies in their sample.
The most significant difference in their approach is the consistent adoption of an updated 50\% completeness cut derived 
using the $\alphah$ sample,
which reduced their final sample to 21,827 galaxies ($10\%$ larger than our sample).

Our aim is not to compute the HIMF but to make sure its estimate from our sample is consistent with that
derived earlier by \cite{2022MNRAS.509.3268O}. This is a crucial consistency check since the weights derived for computing
the HIMF,  will be used in computing the \hi selected GSMF and the
\hi mass and stellar mass scaling relation.

To compute the \hi selected GSMF, we will start with the data used to calculate the HIMF.
Redshifts (or distance estimates) corrected for peculiar velocities are more uncertain for local galaxies.
Rest frame magnitudes and \hi masses are very sensitive to these estimates. We therefore restrict our
sample further to  $0.0025 < z < 0.05$, which results in 19,638 galaxies, of which 19,253 have OCs.

We put an additional cut of \( r_{\text{petromag}} < 17.77 \) \citep{2002AJ....124.1810S} on galaxies which have OCs. 
We exclude those galaxies whose spectral class is "STAR",
"QUASARS" and those galaxies which have spectral class "GALAXY" in the SDSS database, but lack spectral data.
We suspect that we cannot trust their cmodelmags values because of these reasons, so we will not include them in our sample.
This reduces the final sample to 16,955 galaxies.
Out of the 16,955 galaxies, 16,716 have Taylor mass estimates.
The remaining 239 galaxies have errors in the g or i band greater than 0.5, hence  their Taylor mass estimates
derived from the empirical formula, are not considered reliable.
We have derived \kcorrect stellar masses for 16,954 out of 16,955 galaxies.
One of these galaxies (AGC 7087) has its cmodel magnitude and corresponding error in the z band recorded as -9999.0
in SDSS, indicating missing or unreliable data. 
We have \gswlc mass estimates of 8,280 out of 16,955 galaxies.
Out of the three stellar mass estimates, we consider the \gswlc  estimate as the most accurate one since
it uses UV, optical, and mid-infrared bands to obtain a much clearer picture of different stellar populations in galaxies, which is crucial
for determining accurate mass to light ratios.

In our prepared sample, only 49\% of ALFALFA galaxies have \gswlc mass estimates, which is insufficient
for any accurate analysis. We explored various additional selection criteria that could potentially increase this
fraction when applied to the sample, but did not identify any suitable properties.
Thus, we will only use \gswlc stellar masses to calibrate \kcorrect mass estimates since the latter have no correction
due to dust extinction, which biases stellar mass estimates. 
We do not apply corrections to the Taylor mass estimates, 
as they have already been corrected for dust extinction with a single \cite{2000ApJ...533..682C} curve
and further corrected for reddening due to inclination \citep{2020AJ....160..271D}.

In figure~\ref{fig_mtaylor-mkcorrect-mgswlc}  
we compare the Taylor ($\mtaylor$) and \kcorrect ($\mkcorrect$) mass estimates with the 
\gswlc-2 ($\mgswlc$) mass estimates in the left and right panels respectively.
The points are the data. The binned relation (mean, with error bars) is shown as the solid line. The 
dotted line is the $y=x$ curve.  We use the full sample here from \cite{2020AJ....160..271D} 
which contains $\mgswlc$ estimates in addition to $\mkcorrect$ or $\mtaylor$ mass estimates for these plots.
The plots can be used to look at the consistency, and any potential biases, between pairs of mass estimates. 
Overall, we see a tight relation between the mass estimates throughout the mass range. 
The scatter in the $\mtaylor - \mgswlc$ relation is smaller than that in $\mkcorrect-\mgswlc$ relation.
There are two reasons for this: i. $\mtaylor$ estimates are only calculated for galaxies with errors 
in $g$ or $i$ cmodel magnitudes less than 0.05. ii. Since both the $\mtaylor$ and $\mgswlc$ mass estimates correct for dust, although 
with different approaches, they should be more strongly correlated compared to $\mkcorrect$ and $\mgswlc$ (or $\mtaylor$).
However, in both cases, scatter appears to be roughly symmetric and relatively constant throughout the mass range, 
for which we conclude that whatever factors affects the scatter in these mass estimates are not mass-dependent.  

There is, however, a systematic offset between the two mass estimates, which becomes more pronounced as we go to higher masses. 
This offset is more pronounced for $\mtaylor$ as compared to $\mkcorrect$. 
The mean $\mtaylor - \mgswlc$ relation is systematically biased for $\mgswlc \gtrsim 8.5$
and the mean $\mkcorrect - \mgswlc$ relation is systematically biased for $\mgswlc \gtrsim 10.0$. 
The $\mgswlc$ mass estimate is larger than the other two above these corresponding thresholds.
At masses below these thresholds, both $\mtaylor$ and $\mkcorrect$ are consistent with $\mgswlc$.
Although the mean $\mkcorrect$ estimates show closer agreement with $\mgswlc$ for a considerably larger mass range compared to $\mtaylor$, 
the larger scatter between the two will suppress the \hi selected GSMF with $\mkcorrect$  as compared to the $\mtaylor$
as will be seen in section~\ref{sec_gsmf_estimates}. 

Both the $\mtaylor$  and $\mkcorrect$ mass estimation methods utilize the same population synthesis model \citep{2003MNRAS.344.1000B} and corrections due to peculiar velocities to obtain rest frame magnitudes.
For ALFALFA the local flow model of \cite{2005PhDT.........2M} is used, whereas the \cite{2011MNRAS.418.1587T}
 flow model is used for estimating the $\mtaylor$ masses. Secondly the $\mtaylor$ estimates are based on the GAMA
sample which may or may not be consistent with the SDSS sample used here. 

We use the $\mgswlc$ 
to recalibrate $\mkcorrect$, rather than $\mtaylor$, for dust attenuation since the $\mkcorrect$ estimates
do not correct for dust.
In order to prepare our final sample with stellar mass estimates, we retain galaxies with available $\mgswlc$ mass estimates and use calibrated estimates for galaxies that lack $\mgswlc$ mass estimates but have $\kcorrect$ mass estimates, substituting them for stellar mass using a linear relation, as shown in figure 
\ref{fig_mtaylor-mkcorrect-mgswlc}.
We use the $\mgswlc-\sigmamgswlc$ relation to assign errors in $\mkcorrectgswlc$. 
This process results in a total of 16,957 galaxies. This number includes three additional galaxies compared to the $\kcorrect$ mass estimates (AGC No. 231122, 7087, and 728368). We calibrate the rest 8,677 galaxies which do not have \gswlc mass estimates using the \gswlc sample and include the \gswlc mass estimates of 8,280 galaxies into our final sample to maximize the number of galaxies with stellar mass estimates. We call this sample \gswlc calibrated $\kcorrect$ mass estimates. 
\begin{itemize}
    \item Number of Taylor Mass Estimates = 16,716
    \item Number of \kcorrect Mass Estimates = 16,954
    \item Number of \gswlc calibrated \kcorrect Mass Estimates = 16,957
\end{itemize}

\subsection{Optically Selected Sample}\label{sec_optical_sel._sample}
To construct the optical sample (sample selected on the basis of $r$-band magnitude), we first queried SDSS galaxies in the region: \(9.5^\text{h} < \text{RA} < 15.4^\text{h}\), \(0^\circ < \text{Dec} < 36^\circ\),  which contains the ALFALFA survey region. The solid, black boundary of 
figure~\ref{fig_ra-dec} is then used to further remove galaxies outside the \hi selected ALFALFA region used in this work.
For the \hi selected sample, we used \hi redshifts to compute rest frame magnitudes with $\kcorrect$. This allowed us to 
include galaxies which do not have spectroscopic redshifts. For the optically selected sample we 
only include galaxies with $z_{\rm spec}$ (since $z_{\rm photo}$ is not 
accurate at very low redshifts considered 
in this work \citep{2012ApJS..201...32S}) with $r_{\rm petromag} < 17.77$(\citep{2002AJ....124.1810S}).
Spectroscopic detections are in region 3 (red dashed boundary) contained within region 2 of figure~\ref{fig_ra-dec}.
There are small patches in region 1 with spectroscopic detections, which we ignore. Our optical sample
is therefore limited to region 3 of figure~\ref{fig_ra-dec} with an 
area of 3947 $\mathrm{deg}^2$ and a corresponding comoving volume, $ 3.81 \times 10^6 \textrm{Mpc}^3$. Region 3 represents 
$\sim 65\%$ of the area (and volume) of the \hi selected sample (regions 1 and 2).
The peculiar velocity corrected distances are derived using Cosmicflows-4 \citep{2020AJ....159...67K}. 
The distances are converted into the corresponding redshift in the heliocentric frame, which are then converted into the CMB frame.   
We restrict our sample to galaxies within the redshift range $0.0025<z_{\rm cmb}^{\rm spec}<0.05$, 
as done in earlier studies \citep{2017ApJ...846...61G}, resulting in a total of 45,609 galaxies.
The $1/\vmax$ method \citep{1968ApJ...151..393S,2008MNRAS.388..945B} is 
used to estimate the GSMF for this flux limited optically selected sample.  
It should be noted that for some galaxies the value of $\vmax$ was much larger than expected, resulting in an unphysical mass function. Upon closer inspection, we found that the errors associated with their model magnitudes are quite large. Consequently, the corresponding absolute magnitudes computed by \kcorrect for those galaxies are also unreliable. We therefore removed these galaxies from our analysis. Finally, we are left with 45,609 galaxies, which 
constitutes our optical sample.

\subsection{Joint Optical-\hi sample}\label{sec_joint_HI_optical_sample}
We impose the same cuts ($0.0025<z_{\rm cmb}^{\rm spec}<0.05$, $r_{\rm petromag} < 17.77$) and 
restrict galaxies in region 3 (the same as the optically selected sample) on the \hi selected sample to make 
a fair comparison of the GSMF between the two in section \ref{sec_HI-sel.}.
We call this the joint optical-\hi sample, which contains 8,870 galaxies. It is the intersection of the optically and \hi selected samples.

\vspace{3mm}
We use the \gswlc calibrated \kcorrect stellar masses ($\mkcorrectgswlc$), 
for both the optical selected and the joint optical-\hi sample, in addition 
to the \hi-selected sample.

\subsection{Optical colors and the red-blue dichotomy}
We divide our samples into two disjoint sub-samples based on their location relative to the curve given by equation~\ref{eq_tanh_cut} in the color-magnitude plane (\citep{2004ApJ...600..681B}).  
\begin{eqnarray} 
    C^\prime_{\rm ur}(M_{\rm r}) &=& 2.06 - 0.244 \tanh \left[ \frac{M_{\rm r}+20.07}{1.09} \right]  
    \label{eq_tanh_cut}  
\end{eqnarray}  
Galaxies above this curve (equation~\ref{eq_tanh_cut}) are classified as red galaxies and those below it as blue galaxies. The number of galaxies in the \hi selected red-blue population is 3,823 and 12,892 (AGC 7087 has been removed as discussed earlier due to large errors in cmodel magnitudes) for Taylor mass estimates, 3,892-13,062, 
for \kcorrect and \gswlc calibrated \kcorrect mass estimates. 
AGC 7087, 231122 and 728368 are neither classified as red nor blue since we are unable to compute its absolute 
magnitude due to the reason mentioned earlier, although we have included them in the total population. 
Since these galaxies are typical gas-rich galaxies, with masses close to the knee of the HIMF, 
their removal does not affect our GSMF.
In the case of optically (joint optical-\hi) selected sample, the observed number of red and blue galaxies 
which are complete are 18,023 (1,488) and 27,586 (7,382) respectively.
\section{Methods}
\label{sec_methods}

\subsection{The $1/\veff$ method}\label{sec_veff_calc}
For a blind survey like ALFALFA, the limiting flux depends on velocity width $\wfifty$, which can be 
calibrated from the data itself, equation~\ref{eq_completeness_equation}, if the sample is reasonably 
large. 
We use an implementation of the non-parametric two-dimensional step wise maximum likelihood (2DSWML) method
\citep{2000MNRAS.312..557L}, adapted to estimate 
the bivariate \hi mass - \hi velocity width, $\phi(\mhi,\wfifty)$, function 
\citep{2003AJ....125.2842Z,2010ApJ...723.1359M,2011AJ....142..170H,2018MNRAS.477....2J,2020MNRAS.494.2664D,
2022MNRAS.509.3268O}. We briefly discuss the 2DSWML method and point the reader 
to the appendix of \citep{2020MNRAS.494.2664D} for details. 

In the 2DSWML, the binned bivariate  mass-velocity width function $\phi_{jk} \equiv \phi(\mhi^{j},\wfifty^{k})$ is obtained iteratively using the completeness relation of equation~\ref{eq_completeness_equation} and the survey volume. Here the range in \hi mass  and \hi velocity width is divided 
logarithmically into $n_{\rm M}$ and $n_{\rm W}$ bins, with ${j} \in [0, n_{\rm M}[$ and 
${k} \in [0,n_{\rm W}[$. The probability of detecting the $i^{\rm th}$ galaxy with properties $(\mhi^i, \wfifty^i)$ at a distance $D^i$ is
\begin{eqnarray}
P_i &=& \frac{\displaystyle\sum\limits_{j=0}^{n_{\rm M}-1} \sum\limits_{k=0}^{n_{\rm W}-1} V_{ijk} \phi_{jk}}
{\displaystyle\sum\limits_{j=0}^{n_{\rm M}-1} \sum\limits_{k=0}^{n_{\rm W}-1} H_{ijk} \phi_{jk} \Delta M \Delta W}
\label{eq_probability_i} 
\end{eqnarray}
where $\Delta M$ and $\Delta W$ are the bins in  
mass $M=\log_{10}[\mhi/\msun]$ and velocity width $W=\log_{10}[\wfifty/(\text{km.s}^{-1})]$.
$V_{ijk}$ is an occupation number and takes a value of 1 (0 otherwise) 
if  the galaxy \emph{i} is binned in the '$jk$' bin, 
$\Sigma_i V_{ijk} = N_{jk}$,
where $N_{jk}$ is the observed number of galaxies in '$jk$' bin.
$H_{ijk}$ represents the fractional area (values 0 to 1) in the $\mhi-\wfifty$ plane that is accessible to the 
\emph{i}$^{\rm th}$ galaxy and takes care of the completeness limit of the sample 
\citep[See][]{2020MNRAS.494.2664D}.
Maximizing the joint likelihood, $\mathcal{L} = \prod_{i=1}^{N_{\text{gal}}} P_i$ 
with respect to $\phi_{jk}$,
we obtain: 
\begin{eqnarray} \label{eq_phijk}
\phi_{jk} &=& N_{jk}\left[\sum\limits_{i=1}^{N_{\rm gal}} \frac{H_{ijk}}
{\sum\limits_{m=0}^{n_{\rm M}-1} \sum\limits_{n=0}^{n_{\rm W}-1} H_{imn} \phi_{mn}}\right]^{-1}
\end{eqnarray}
The above equation is solved iteratively and for the $l^{\rm th}$ iteration we have
\begin{eqnarray} 
\phi_{jk}^{(l)} &=& N_{jk}\left[\sum\limits_{i=1}^{N_{\rm gal}} \frac{H_{ijk}}
{\sum\limits_{m=0}^{n_{\rm M}-1} \sum\limits_{n=0}^{n_{\rm W}{-1}} H_{imn} \phi_{mn}^{(l-1)}}\right]^{-1}
\end{eqnarray}
The iteration is started 
by setting $\phi_{jk}^{(0)} = N_{jk}/V_{\rm survey}$, which is the observed bivariate mass-velocity width function. 
The iteration is stopped when a minimum 1\%  accuracy is achieved for all $\phi_{jk}$.  
The normalization of $\phi$ has to be fixed since the 2DSWML fixes 
only the shape of $\phi(\mhi,\wfifty)$ (the normalization factors 
cancel out in equation~\ref{eq_completeness_equation}). We fix the normalization by matching 
the observed abundances at the high mass end which is complete and does not suffer from 
the selection function, as in \cite{2020MNRAS.494.2664D}.
Integrating $\phi_{jk}$ over mass (velocity width), we obtain $\phi_k \equiv \phi (\wfifty)$ 
($\phi_j \equiv \phi(\mhi) $).

In the 2DSWML method, the corrections (or weights)
which take you from $(\phi_{jk}^{(0)} = N_{jk}/\vsurvey) \rightarrow \phi_{jk}$ are determined 
by the full galaxy catalog. In the limit of a very large sample (or survey volume, i.e. the 
homogeneous limit) these weights converge. 
The weights in the 2DSWML method are associated with the $jk$ bin rather than the individual contributions of galaxies (indexed as $i$) in the $jk$ bin (as in the $1/V_{\rm max}$) \citep{1968ApJ...151..393S}. 
When working with smaller mutually disjoint populations of galaxies 
(e.g. populations which are chosen based on their color, rest-frame magnitudes or any other intrinsic property) whose union is the parent catalog, it becomes difficult to obtain the underlying mass functions 
(or $\phi(\mhi,\wfifty)$ in general) of these populations which should add up to 
the mass function of the full sample. The reason is that the weights of each galaxy contributing 
to any $jk$ bin in the $\mhi-\wfifty$ space are now determined 
from the subsample and not the parent sample, which leads to inconsistent results.  
In \cite{2020MNRAS.494.2664D} these consistency conditions were imposed by hand but did not have an elegant 
interpretation as the $1/V_{\rm max}$ method. 

In this work, we associate an effective max volume $\veff^i$ (interpreted as $V_{\rm max}^i$) 
to every galaxy $i$ after determining the normalized $\phi(\mhi,\wfifty)$ of the full sample
using the 2DSWML. 
$\veff^i$ is the maximum volume that a galaxy can occupy and be detected 
by ALFALFA. Once $\veff^i$ is determined, it can be used freely to obtain \hi selected 
abundances of other intrinsic properties, e.g. $\mhi$, $\wfifty$, $M_r$, $\mstar \,\,...$ to name a few.
This is done by binning the appropriate property of the galaxy and boosting the count in the bin
by the weight $\vsurvey/\veff^i$. We refer the reader to the appendix of \cite{2013PhDT.......391P}
for a more detailed discussion of the $1/\veff$ method. A galaxy i is defined as complete
if $\veff^i \geq \vsurvey$, in which case $\veff^i = \vsurvey$.
\begin{eqnarray}
   \veff^i &=& \sum\limits_{l=1}^{\ngal}\frac{1}
    {\sum\limits_{m=0}^{n_{\rm M}-1} \sum\limits_{n=0}^{n_{\rm W}-1} H_{lmn} \phi_{mn}^{\rm norm}} \\
 \nonumber   \phi_{jk}^{\rm norm} &=& \frac{1}{\vsurvey}\sum\limits_{i}\frac{\vsurvey}{\veff^i} \\ 
   &=& \sum\limits_{i}\frac{1}{\veff^i}, \quad \text{for all galaxies in the $jk$ bin}  
\end{eqnarray}
Here $\phi_{jk}^{\rm norm} \equiv \phi(\mhi,\wfifty)$ is the normalized bivariate mass function for the full 
sample. The $1/\veff$ method is identical to the 2DSWML method when using the full catalog of galaxies 
to compute  $\phi(\mhi,\wfifty)$.
\begin{figure*}
    \begin{tabular}{cc}
    \includegraphics[width=3.4in]{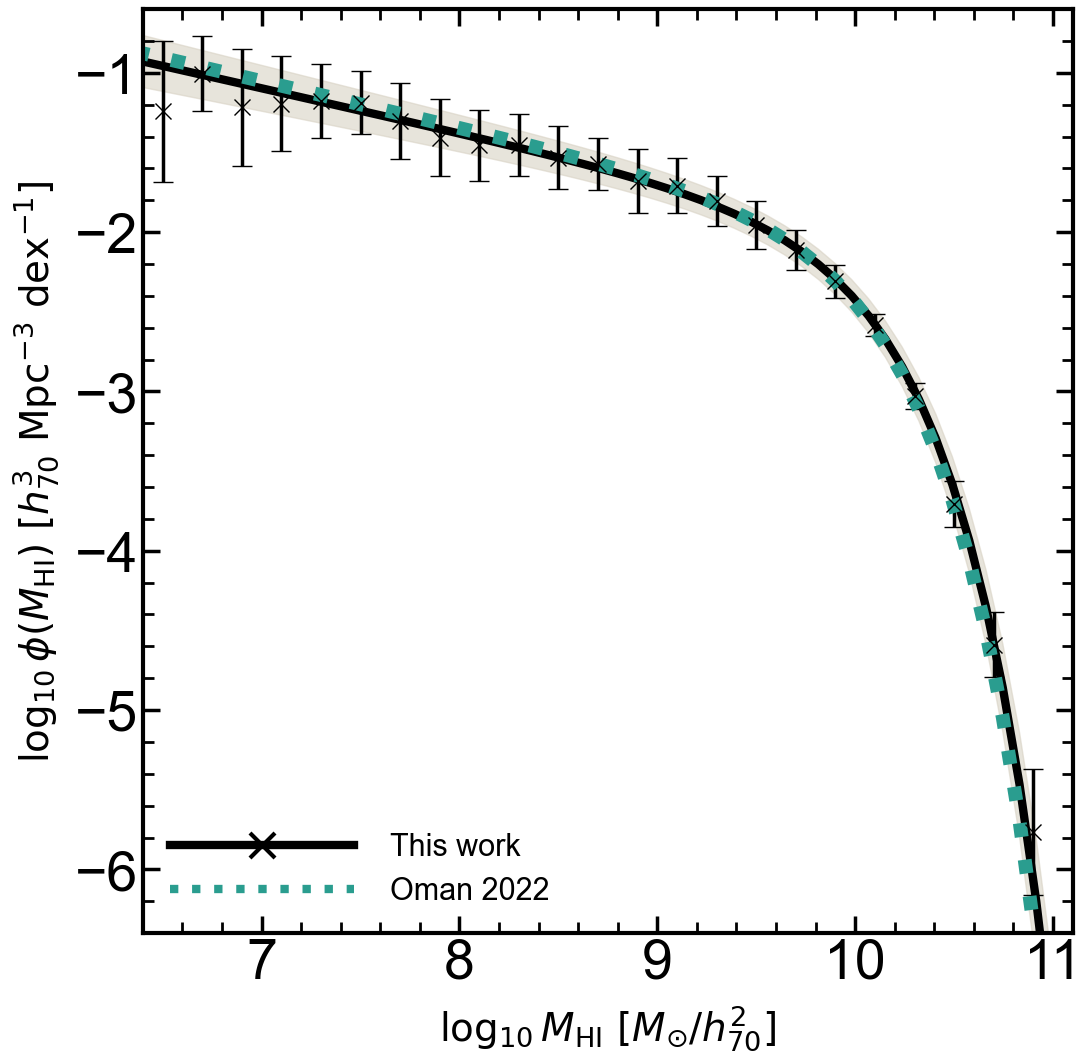} 
    \includegraphics[width=3.4in]{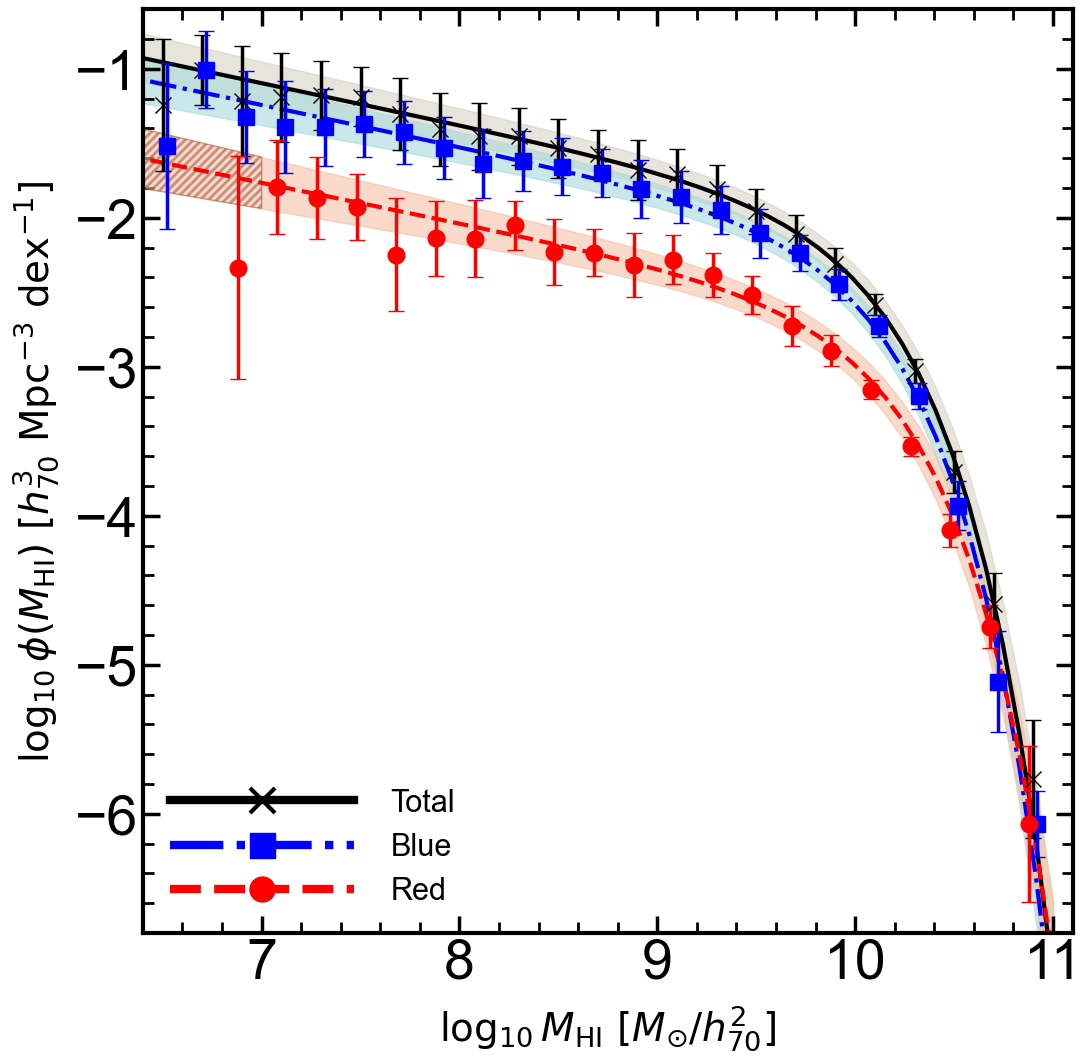} \\
    \end{tabular}
    \caption{\emph{Left}: The $\alphah$ HIMF for our sample (data points with errors) is estimated based on the
      the $1/\veff$ method. The solid line and the shaded region are the best-fit Schechter function
      and its uncertainty respectively. Comparison is made with the HIMF derived from the $\alphah$ sample,
      based on similar quality cuts, survey  volume,  and completeness relation ~\ref{eq_completeness_equation} of \citet{2022MNRAS.509.3268O} (dotted line)
      \emph{Right}: The HIMF for the total (black), red (red) and blue (blue) population (data points with error bars) and their corresponding
      Schechter function fits and uncertainties, solid lines and shaded regions. For the red sample, the hatched region represents the
      extrapolation of the HIMF to lower masses.
      The data points corresponding to blue (red) galaxies are shifted to the right(left) by 0.02 dex for a clearer comparison.}
    \label{plot_himf}
\end{figure*}

\subsection{Error Analysis}
\label{sec_erroranalysis}
We discuss various errors which translate to uncertainties in the HIMF ($\phi(\mhi)$) 
and the GSMF ($\phi(\mstar)$) 
estimated from an optically selected sample, an \hi selected sample and a
joint optical-\hi selected sample. 
\subsubsection{Error Analysis for HIMF}\label{sec_err_HIMF}
\begin{itemize}
    \item {\bf Poisson Errors}: We include Poisson errors $\propto \dfrac{1}{\sqrt{N}}$ due to finite 
    counts in bins. These errors are large, especially at low and high masses.
    \item {\bf Mass Errors}: As the \hi mass is determined from the galaxy's integrated flux and distance $\mhi \propto \sflux D_{\rm L}^2$, the uncertainties in  $\sflux$ and $D_{\rm L}$ translate to uncertainties in $\mhi$.
    To quantify errors arising due to mass, we generate 1000 Gaussian realizations
    of $\sflux$ and $D_{\rm L}$ for each galaxy based on their measured values and uncertainties. 
    The 1000 catalogs are then used to estimate the uncertainties in $\veff$, which translate to 
    uncertainties in $\phi(\mhi)$.  
    For lower redshift galaxies ($c\zcmb < 6000 {\rm km/sec}$), distances are based on the local flow model of \cite{2005PhDT.........2M}, which has a velocity dispersion $\sigma_v = 160 {\rm km/sec}$ \citep{2018MNRAS.477....2J}.  For these galaxies, we take distance errors to be the maximum of $\sigma_v$ and 10\% of the distance. For higher redshift galaxies ($c\zcmb > 6000 {\rm km/sec}$), which are less affected due to peculiar velocities, distance errors are assumed to be 10\% \citep{2018MNRAS.477....2J}.
    We use errors in flux from the ALFALFA catalog to generate realizations.  
    \item {\bf Jackknife Errors}: We estimate statistical errors using jackknife samples.  We split our full coverage area into 46  domains of nearly equal area. A jackknife sample is constructed by dropping one such domain. We have, therefore, 46 jackknife samples.  The jackknife uncertainty in the $j^{\rm th}$ mass bin $\phi^j(\mhi)$ is computed as $\sigma_{\phi^j} = \frac{N-1}{N}\sum_{i=1}^{N=46} (\bar{\phi}^j - \phi^j_i)^2$ where $\bar{\phi}^j$ is the mean of jackknife samples and  $\phi^j_i$ is the value obtained from the i$^{\text{th}}$ jackknife sample.
    
\end{itemize}
\subsubsection{Error Analysis for \hi selected GSMF}
\label{sec_err_HI_GSMF}
\begin{itemize}
    \item Mass Errors: 
    \begin{itemize}
        \item {\bf \taylor stellar masses}: To account for mass errors in the case of \taylor mass estimates we generate 1000 realizations, taking the stellar mass value as the mean and the associated uncertainty as the standard deviation.
       
        \item {\bf \kcorrect stellar masses}: In case of \kcorrect mass estimates, we do not have mass errors, so we generate realizations based on their cmodel magnitude in r band and its associated error. 
    
        \item {\bf \kgswlc  stellar masses}: For those objects which have \gswlc mass estimates we use 
        $\sigmamgswlc$ as errors. For the rest which do not have \gswlc mass estimates and are calibrated 
        to $\mkcorrectgswlc$ we use the relation between \gswlc masses and their errors
        ($\sigmamgswlc=-0.021\mgswlc+0.250$) to estimate their errors. Both the masses and errors are on a logarithmic scale.
  
        \end{itemize}  
    
        \item Additionally we consider Poisson and  jackknife errors. Since this is an \hi selected sample
        we also consider errors in $\veff$.

\end{itemize}

\subsubsection{Error Analysis for optically selected GSMF}\label{sec_err_opt_GSMF}
\begin{itemize}
    \item {\bf Errors in $\vmax$}:  We compute the uncertainty in $\vmax$ by assuming a 15\% relative uncertainty (E. Kourkchi 2025, private communication)in the distance estimates derived from Cosmicflows-4 \citep{2020AJ....159...67K} and errors in $r_{\rm cmodel}$.

\item   As in section~\ref{sec_err_HI_GSMF} we consider Poisson, jacknife and mass errors.
\subsubsection{Error Analysis for joint optical-\hi GSMF}\label{sec_err_opt_HI_GSMF}
\item We follow the same procedure for calculating errors for the joint optical \hi-GSMF as described in Sec. \ref{sec_err_HI_GSMF}. 
\end{itemize}

\begin{figure*}
   \begin{tabular}{cc}
    \includegraphics[width=\columnwidth]{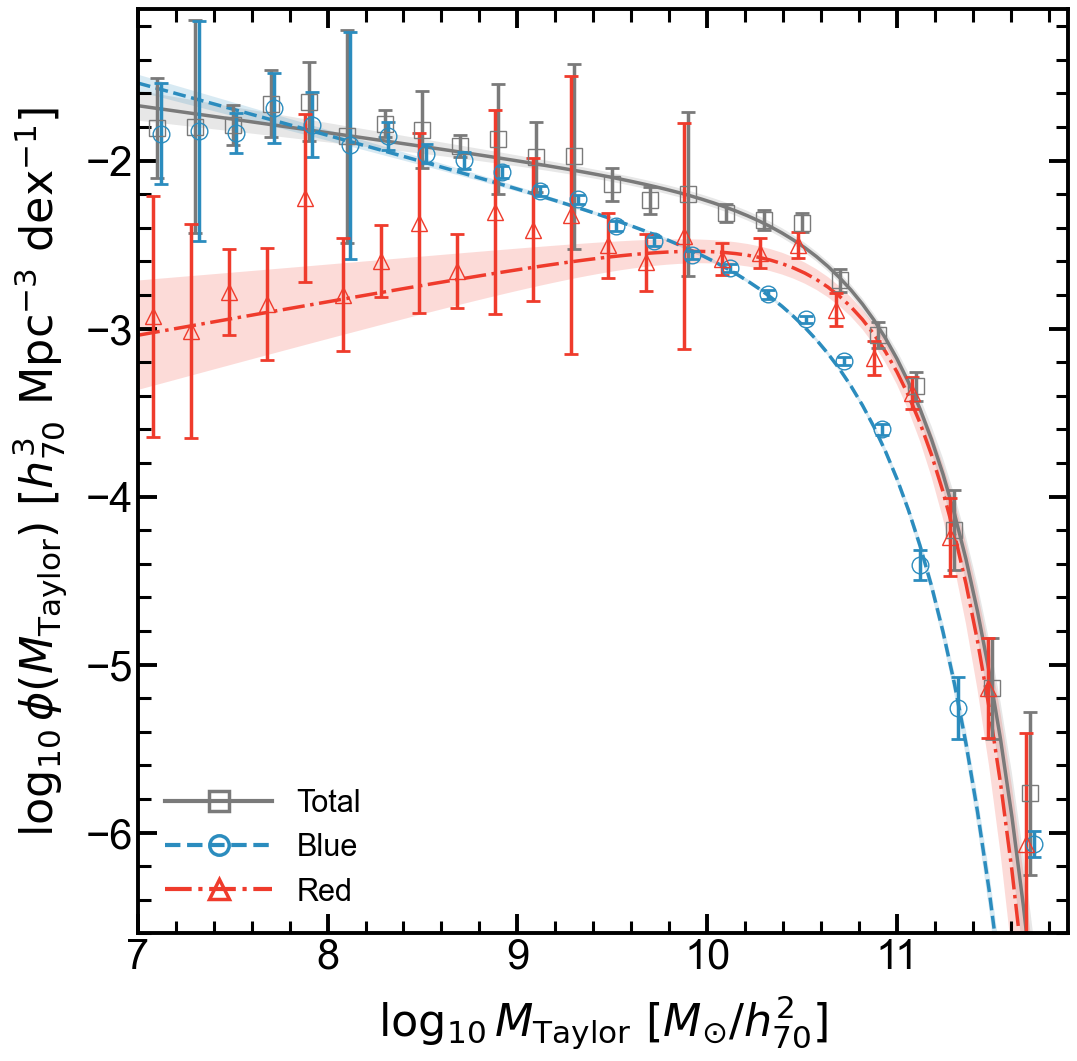}& 
    \includegraphics[width=\columnwidth]{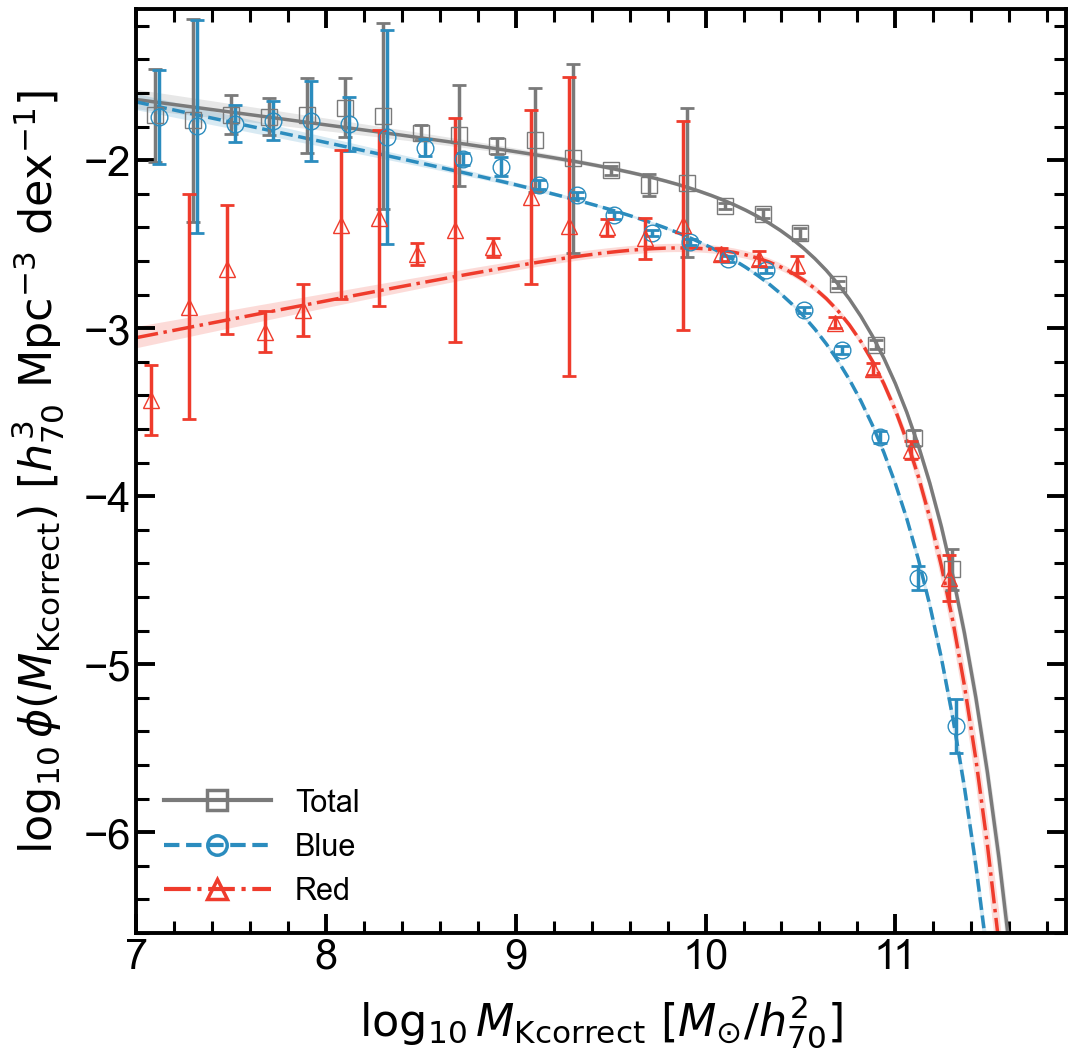} \\
    \includegraphics[width=\columnwidth]{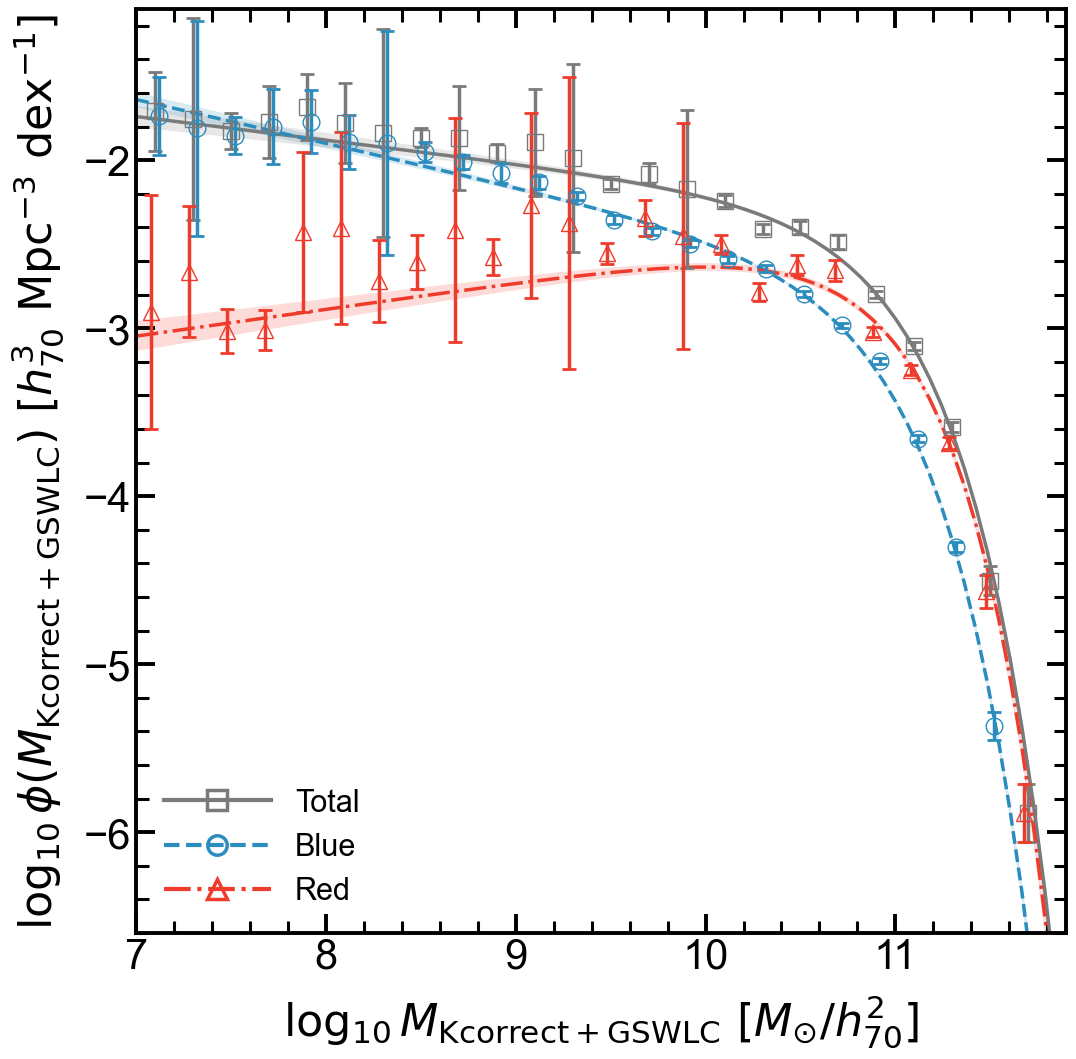}&
    \includegraphics[width=\columnwidth]{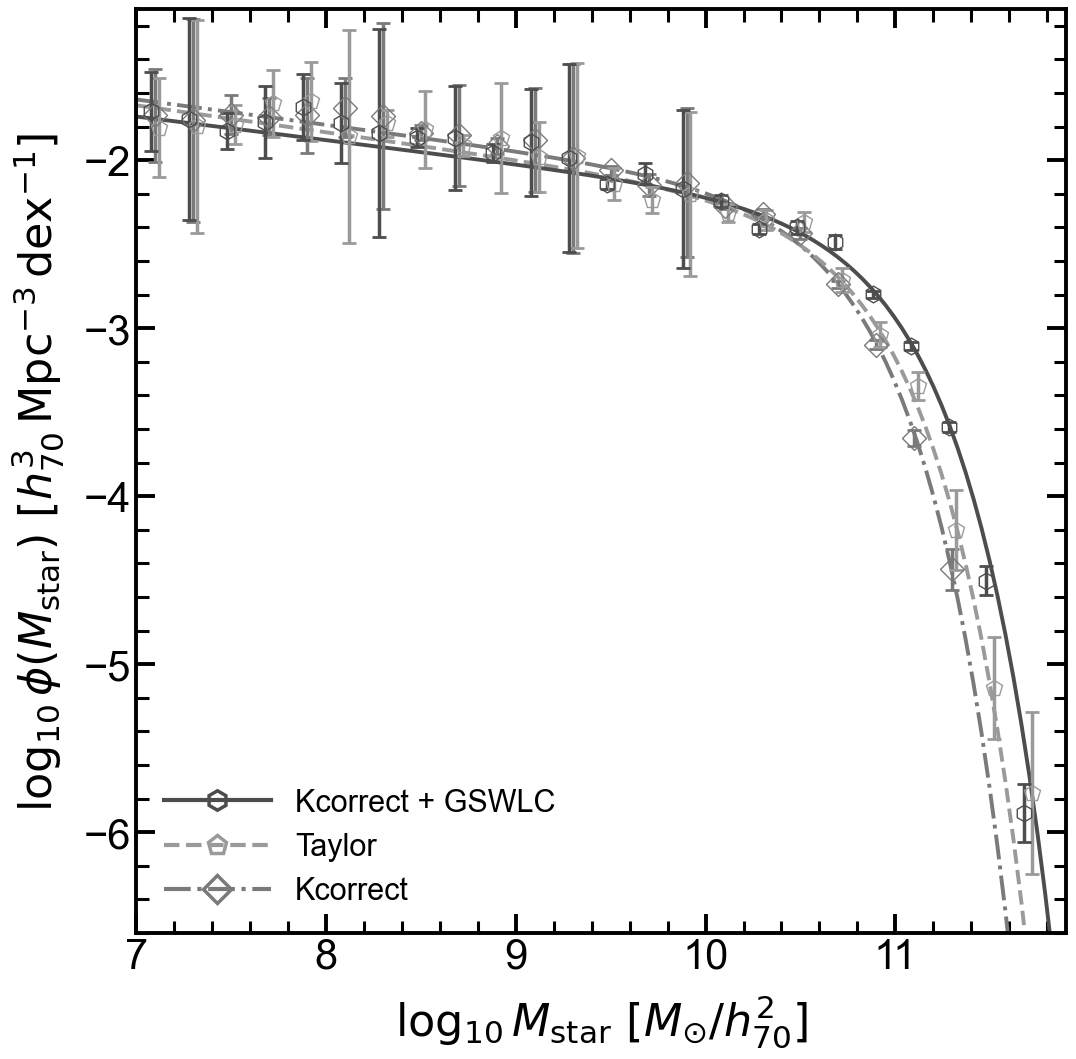} 
    \end{tabular}
    \caption{Comparison of \hi selected GSMFs derived using different stellar mass estimates.\emph{Top left}: GSMF estimated using $\mtaylor$. 
                     \emph{Top right}: GSMF using $\mkcorrect$. 
                     \emph{Bottom left}: GSMF using $\mkcorrectgswlc$. 
                     In each panel, the total (grey squares), blue (blue circles), and red (red triangles) populations are shown with their error bars, along with their corresponding best-fit Schechter functions (solid lines) and associated uncertainties (shaded regions).
 \emph{Bottom right}: Comparison of the \hi selected GSMFs for the total sample, for the three stellar mass estimates. The best fit Schecter function parameters are listed in 
 table~\ref{tab:combined_schechter_params}.}
    \label{fig_GSMF_hi-selected}
\end{figure*}
\begin{table*}
\begin{center}
\begin{tabular}{|l|c|c|c|c|}
\hline
 & \multicolumn{4}{c|}{\textbf{Taylor Mass Estimates}} \\
\hline
Population 
& $\log (M_*/M_{\odot}) + 2\log h_{70}$ 
& $\alpha$ 
& $\phi_* \ (10^{-3}\, h_{70}^{3}\,\mathrm{Mpc}^{-3}\,\mathrm{dex}^{-1})$ 
& $\reducedchisq$ \\
\hline

Red 
& $10.63^{+0.07}_{-0.05}$ 
& $-0.80^{+0.06}_{-0.12}$ 
& $2.11^{+0.31}_{-0.44}$ 
& 0.28 \\

Blue 
& $10.59^{+0.02}_{-0.02}$ 
& $-1.31^{+0.02}_{-0.02}$ 
& $0.97^{+0.06}_{-0.06}$ 
& 0.95 \\

Total 
& $10.70^{+0.04}_{-0.04}$ 
& $-1.16^{+0.03}_{-0.04}$ 
& $2.36^{+0.26}_{-0.33}$ 
& 0.40 \\

\hline
 & \multicolumn{4}{c|}{\textbf{\kcorrect Mass Estimates}} \\
\hline

Red   
& $10.52^{+0.02}_{-0.02}$ 
& $-0.78^{+0.02}_{-0.03}$ 
& $2.27^{+0.17}_{-0.17}$ 
& 1.51 \\

Blue  
& $10.52^{+0.01}_{-0.01}$ 
& $-1.24^{+0.02}_{-0.02}$ 
& $1.38^{+0.07}_{-0.07}$ 
& 2.88 \\

Total 
& $10.60^{+0.02}_{-0.02}$ 
& $-1.15^{+0.02}_{-0.02}$ 
& $2.88^{+0.18}_{-0.18}$ 
& 0.49 \\

\hline
 & \multicolumn{4}{c|}{\textbf{\kgswlc Mass Estimates}} \\
\hline

Red 
& $10.80^{+0.02}_{-0.02}$ 
& $-0.84^{+0.02}_{-0.03}$ 
& $1.58^{+0.09}_{-0.11}$ 
& 1.36 \\

Blue 
& $10.76^{+0.01}_{-0.01}$ 
& $-1.26^{+0.02}_{-0.02}$ 
& $1.05^{+0.04}_{-0.04}$ 
& 1.98 \\

Total 
& $10.83^{+0.01}_{-0.01}$ 
& $-1.14^{+0.02}_{-0.02}$ 
& $2.30^{+0.12}_{-0.12}$ 
& 1.13 \\

\hline
\end{tabular}
\end{center}
\caption{\hi selected GSMF: Best-fitting Schechter parameters and their $1\sigma$ uncertainties for $\phi(M_*)$ for the total, red, and blue galaxy populations using \taylor, \kcorrect, and \gswlc mass estimates. These  mass functions are plotted in 
figure~\ref{fig_GSMF_hi-selected}.}
\label{tab:combined_schechter_params}
\end{table*}
\subsection{Estimating the Mass Functions}
\label{sec_gsmf_estimates}
We compute the HIMF using the $1/\veff$ method 
applied to our sample of total, red, and blue populations. 
In the left panel of figure~\ref{plot_himf} we compare our HIMF for our total sample
with that of \citet{2022MNRAS.509.3268O}.  Our survey area is slightly smaller than that of 
 \cite{2022MNRAS.509.3268O} which in turn is smaller than the $\alphah$ area \citep{2018MNRAS.477....2J}.
Our results compare well with \cite{2022MNRAS.509.3268O} with differences showing up at the high mass end.
The HIMF is well fit by a Schechter function \citep{1976ApJ...203..297S}
\begin{eqnarray}
    \phi (M) &=& \phi_* \left( \frac{M}{M_*}\right)^{\alpha} \exp \left(- \frac{M}{M_*}\right)
    \label{eq_schechterfn}
\end{eqnarray}
Here $\phi_*$, $M_*$ and $\alpha$ represent the amplitude, characteristic mass and the slope at the low-mass end.
The abundance of galaxies drop exponentially above the characteristic mass $M_*$. 
Various sources of errors translate to errors in the estimated mass function as described in section~\ref{sec_erroranalysis}.
The fitting is done using \texttt{EMCEE} \citep{2013PASP..125..306F} for the 3-parameter Schechter function after removing
the initial burn-in phase of the Markov Chain Monte Carlo. The shaded region represents the 1$\sigma$ uncertainty  
around the best fit mass function.

Our best fit values of the Shechter function for the HIMF are
$\left\{\phi_* \ (10^{-3}\, h_{70}^{3}\,\mathrm{Mpc}^{-3}\,\mathrm{dex}^{-1}), 
\log_{10} (M_*/M_{\odot}) + 2\log_{10} h_{70},  
\alpha\right\} = \left\{5.19^{+0.73}_{-0.95},
\,9.95^{+0.03}_{-0.04},\,
-1.28^{+0.04}_{-0.04}\right\}$. 

These values are consistent with \cite{2022MNRAS.509.3268O} within uncertainties, 
who report 
$\phi_* \ (10^{-3}\, h_{70}^{3}\,\mathrm{Mpc}^{-3}\,\mathrm{dex}^{-1}), 
\log_{10} (M_*/M_{\odot}) + 2\log_{10} h_{70}, 
\alpha\} = 
\{5.50^{+0.25}_{-0.25},\,9.92^{+0.01}_{-0.01},\,-1.29^{+0.02}_{-0.02}\}$.

In the right panel of figure~\ref{plot_himf} we look at the HIMF for red and blue galaxies and compare 
them with the full sample.  
The best fit Schechter parameters, $\{\phi_*, M_*, \alpha\}$ 
(in the same units as quoted earlier) for the red and blue populations are 
$\{1.14^{+0.16}_{-0.22}, 10.02^{+0.03}_{-0.04}, -1.27^{+0.05}_{-0.05}\}$ and
$\{3.73^{+0.48}_{-0.62}, 9.94^{+0.03}_{-0.03}, -1.28^{+0.04}_{-0.04}\}$ respectively.
The data points of total sample are by construction the sum of red and blue populations. 
However in the fitting procedure we have not imposed this constraint since 
it does not significantly affect the results. Our results are broadly consistent 
with previous HIMF estimates based on color in the $\alpha.40$ sample 
\citep{2020MNRAS.494.2664D,2022MNRAS.511.2585D}.  

For the \hi selected sample, the best fit values change and becomes
$\phi_* \ (10^{-3}\, h_{70}^{3}\,\mathrm{Mpc}^{-3}\,\mathrm{dex}^{-1}), 
\log_{10} (M_*/M_{\odot}) + 2\log_{10} h_{70},  
\alpha\} = \{5.03^{+0.50}_{-0.58},
\,9.94^{+0.03}_{-0.03},\,
-1.16^{+0.04}_{-0.04}\}$.
The best fit Schechter parameters, $\{\phi_*, M_*, \alpha\}$ 
(in the same units as quoted earlier) for the red and blue populations are 
$\{1.19^{+0.14}_{-0.17}, 9.99^{+0.03}_{-0.04}, -1.16^{+0.05}_{-0.05}\}$ and
$\{3.89^{+0.39  }_{-0.46}, 9.92^{+0.03}_{-0.03}, -1.17^{+0.04}_{-0.04}\}$ respectively.

In our HIMF calculations, we assign a weight to each galaxy ($\vsurvey/\veff \geq 1$) while binning 
in the corresponding \hi mass bin. 
For calculating the  GSMF for our \hi selected sample, 
we bin them according to their stellar masses, giving them the same weight 
($\vsurvey/\veff \geq 1$) since this is an \hi selected sample. 
Using the errors on the mass function or mass estimates, as described in section~\ref{sec_err_HI_GSMF}, 
we can then fit this binned data with a Schechter function to get the corresponding Galaxy Stellar Mass Function.

We fit the GSMFs from $\log_{10}(\mstar/\msun) > 7.0$, because of the completeness of the mass above this threshold. 
The observed $\mhi-\mstar$ relation (figure~\ref{figure_vol_lim}) has a scatter and our sample has an intrinsic 
low-mass threshold of $\log_{10}(\mhi/\msun) = 6.0$, which makes the stellar mass incomplete at lower masses.
To quantify this effect, we adopt two approaches. First, we apply a range of $\mhi$ cuts to the GSMF and identify the stellar mass scale where the results converge. In the second approach, 
we attempt to correct for the applied $\mhi$  cuts. Based on these two methods, we find that above $\log_{10}(\mstar/\msun)=7.0$, the intrinsic  threshold $\log_{10}(\mhi/\msun)=6.0$ no longer affects the GSMFs. These methods have 
been described in detail in  Appendix~\ref{sec_incompl_GSMF}.
As was done for the HIMF we also obtain the GSMF for red and blue galaxies.


In figure \ref{fig_GSMF_hi-selected} we plot the \hi selected GSMF for the three stellar mass estimates,  
$\mtaylor$ (top left), $\mkcorrect$ (top right), $\mkcorrectgswlc$ (bottom left) for the total (open grey squares), 
red (open red triangles) and blue (open blue circles). In the bottom right panel we compare the 
three mass functions for the total sample. 
For the sake of clarity, the red (blue) data points have been shifted by 0.02 dex to the right (left) with respect to the total sample. 
The shaded regions depict the uncertainties associated with the fitted function.
The lines in all the plots are the best fit Schechter functions.  
We note that the abundances of the blue and red samples add up to that of the total sample, but 
in the fitting procedure we do not impose this condition. However 
the fitted functions are consistent with this condition, i.e. the union of red and blue samples 
is the total sample, as will be discussed in reference to table~\ref{tab:combined_mass_estimates_schechter}. 

We can see that in all the estimates of the \hi selected GSMFs 
the mass function corresponding to the red galaxies dominates at the high mass end, whereas 
the blue galaxies are the dominant population at the low-mass end,
with the crossover between the two occurring at $\log_{10}(\mstar/\msun) \simeq 10.0$, which is below the characteristic mass ($\log_{10}(M_*/\msun) \simeq 10.7 )$ (see table~\ref{tab:combined_schechter_params}). 
The best fit parameters of the fitted Schechter functions are summarised 
in table~\ref{tab:combined_schechter_params}.

The \hi selected GSMF for the three mass estimates show qualitatively similar behaviour. The slope of the mass function, 
determined at the low-mass end, is negative ($\alpha + 1 < 0$) for the total and blue populations and positive ($\alpha +1 > 0$) 
for the red population. Of the three parameters of the Schechter function, $\alpha$ is the only one which shows a value 
consistent among the three mass estimates. $\alpha \simeq -0.81, -1.27, -1.15$ for the red blue and total samples respectively (table~\ref{tab:combined_schechter_params}).
There are 4 bins between $\mstar < 10\msun$ which have quite large error bars. It is due to the large scatter in the $\mhi-\mstar$ relation for   $\mstar < 10\msun$  
as shown in the right panel of figure \ref{figure_vol_lim}. These come from 
a long tail in the \hi distribution at these stellar masses, 
which results in a large variation in $1/\veff$
and corresponding errors.

The knee of the mass function, $M_*$, is the smallest (largest) for the $\mkcorrect$ ($\mkcorrectgswlc$) estimates. 
For each sample (red, blue, total), the increase is $\sim 0.1\, \rm dex$ when we go from $\mkcorrect$ to $\mtaylor$ to $\mkcorrectgswlc$.
At fixed $\alpha$, an increase in $M_*$ results in a decrease in amplitude $\phi_*$ for a fixed sample. This is the trend (reverse) we 
see in $\phi_*$. $\phi_*$ systematically decreases as we go from $\mkcorrectgswlc$
to $\mtaylor$ to $\mkcorrect$. 

By comparing the reduced chi-square ($\reducedchisq$) (last column of table~\ref{tab:combined_schechter_params}) 
we find that the quality of the fit for the $\mkcorrectgswlc$ mass estimate is better than the other two. 
A single Schechter function describes the data reasonably well. The $\reducedchisq \simeq 1$ for the red and total samples 
and $\reducedchisq \simeq 2$ for the blue sample. For $\mkcorrect$ and $\mtaylor$ we also find that the $\reducedchisq $ for the blue sample is larger by a factor of at least 2 than the red and total samples. This suggests that we may be underestimating the error bars 
for the blue sample. For the $\mtaylor$ mass estimate, the $\reducedchisq < 1$. Here, we may be overestimating the error bars. 
This is seen in the figure, especially for the red sample whose error bars and the associated uncertainties in the mass function 
at the low-mass end are larger than the other two. 
\begin{table*}
\begin{center}
\begin{tabular}{|l|c|c|c|c|c|}
\hline
Population & $\rhostar$ & $\omegastar$ & $\nstar$ 
& $\rhostar/\rhostar^{\rm total}$ 
& $\nstar/\nstar^{\rm total}$ \\
& $[10^{8}h^{-1}_{70}\,M_{\odot}\,\mathrm{Mpc}^{-3}]$ 
& $[10^{-4}h^{-1}_{70}]$ 
& $[10^{-3}\,\mathrm{Mpc}^{-3}]$ 
& & \\
\hline

\multicolumn{6}{|c|}{\textbf{\taylor Stellar Mass Function (Schechter)} ($\log \mstar \ge 7$)} \\
\hline
Red 
& $0.826 \pm 0.081$
& $6.081 \pm 0.596$
& $7.704 \pm 1.722$
& $0.623 \pm 0.071$
& $0.187 \pm 0.047$\\

Blue 
& $0.494 \pm 0.010$
& $3.636 \pm 0.071$
& $36.472 \pm 2.595$
& $0.373 \pm 0.023$
& $0.888 \pm 0.118$\\

Total 
& $1.326 \pm 0.076$
& $9.759 \pm 0.557$
& $41.099 \pm 4.602$
& $0.995 \pm 0.084$
& $1.075 \pm 0.142$ \\

\hline
\multicolumn{6}{|c|}{\textbf{\kcorrect Stellar Mass Function (Schechter)} ($\log \mstar \ge 7$)} \\
\hline
Red 
& $0.686 \pm 0.026$
& $5.050 \pm 0.189$
& $7.687 \pm 0.438$
& $0.538 \pm 0.023$
& $0.170 \pm 0.015$\\

Blue 
& $0.553 \pm 0.011$
& $4.067 \pm 0.081$
& $33.255 \pm 2.140$
& $0.434 \pm 0.013$
& $0.735 \pm 0.070$  \\

Total 
& $1.274 \pm 0.027$
& $9.377 \pm 0.202$
& $45.217 \pm 3.195$
& $0.973 \pm 0.030$
& $0.905 \pm 0.080$ \\

\hline
\multicolumn{6}{|c|}{\textbf{\kgswlc Stellar Mass Function (Schechter)} ($\log \mstar \ge 7$)} \\
\hline
Red 
& $0.927 \pm 0.036$
& $6.820 \pm 0.264$
& $6.747 \pm 0.539$
& $0.541 \pm 0.025$
& $0.176 \pm 0.020$\\

Blue 
& $0.747 \pm 0.018$
& $5.498 \pm 0.135$
& $33.353 \pm 2.309$
& $0.436 \pm 0.015$
& $0.870 \pm 0.093$\\

Total 
& $1.714 \pm 0.042$
& $12.615 \pm 0.311$
& $38.346 \pm 3.137$
& $0.977 \pm 0.034$
& $1.046 \pm 0.106$ \\

\hline
\end{tabular}
\end{center}
\caption{Derived parameters from the \hi selected GSMF: The top, middle and lower rows are the results derived from $\mtaylor, \mkcorrect$ and $\mkcorrectgswlc$
mass estimates for the red, blue and total samples. 
Columns 2-6 Stellar mass density $\rhostar$, dimensionless stellar mass density ($\omegastar$), 
number density of galaxies ($\nstar$), along with the fractional contributions 
of $\rhostar$ and $\nstar$ with respect to the total sample. The fractional contribution for the total sample
is the ratio of (red + blue)\/total. Uncertainties represent $1\sigma$ confidence intervals and have been propagated from uncertainties of the mass function parameters of table \ref{tab:combined_schechter_params}.}
\label{tab:combined_mass_estimates_schechter}
\end{table*}
\begin{figure}
  \includegraphics[width=\columnwidth]{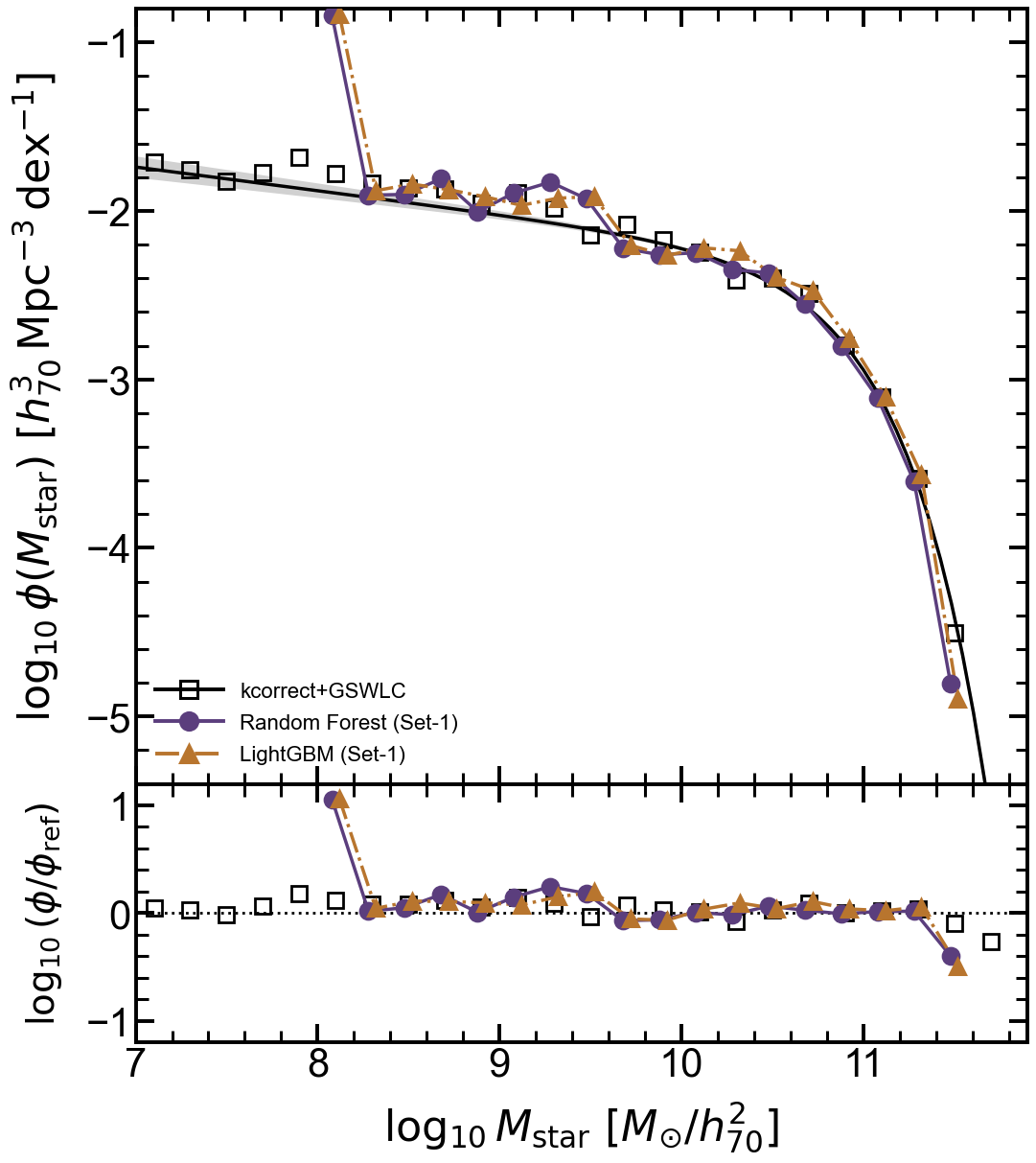} \\
  \caption{The stellar mass function of the \hi-selected sample derived using $\mkcorrectgswlc$ mass estimates (open squares, with the best-fit Schechter function and its $1\sigma$ uncertainty shown as the black curve and grey shaded band) compared with Machine Learning models: Random Forest \citep{2001MachL..45....5B} (filled circle) and LightGBM \citep{Ke2017LightGBMAH} (filled triangle) predicted mass estimates trained on $\mgswlc$ mass estimates. The data points corresponding to Random Forest (LightGBM) predicted mass estimates are shifted to the left (right) by 0.02 dex with respect to the $\mkcorrectgswlc$ mass estimates for a better comparison. The lower panel shows residuals with respect to the GSMF derived from $\mkcorrectgswlc$ mass estimates. We refer the reader to appendix~\ref{sec_ML} for details of the Machine Learning models.}
\label{plot_ml_estim}
\end{figure}

In the bottom right panel of figure~\ref{fig_GSMF_hi-selected} we compare the \hi selected GSMF for the total sample 
with the three mass estimates. A comparison of the red and blue samples can be gleaned from table~\ref{tab:combined_mass_estimates_schechter}.
The solid, dashed and dot-dashed lines represent the best fit Schechter functions compiled from the individual 
mass estimates in other panels.
As discussed earlier, the differences between the three estimates of the mass functions show up at the large mass end.
Corrections to the stellar mass due to dust have the largest effect at larger masses, consistent with figure~\ref{fig_mtaylor-mkcorrect-mgswlc}.
The knee of the mass function increases by $\sim 0.1 - 0.2 \,\rm{dex}$ when compared to the case with no dust correction ($\kcorrect$) consistent 
with \citep{2016ApJS..227....2S}.

 In table~\ref{tab:combined_mass_estimates_schechter} we 
 look at integrated quantities of the \hi selected GSMF, the number density ($\nstar$) and cosmic stellar mass density $\rhostar$, based on the three 
 mass estimates, for the red, blue and total samples. The integral of the Schechter function ($0^{\rm th}$ moment) is $\nstar$ 
 and the first moment is $\rhostar$. 
 The dimensionless stellar density parameter, $\omegastar$ is: 
\begin{equation}
    \Omega_{\text{star}} = \dfrac{\rho_{\text{star}}}{\rho_c} = 
    \frac{M_* \phi_*}{\rho_c}\Gamma(\alpha+2)
    \label{eq_omegah1}
\end{equation}
Here $\rho_c$ is the critical density. 
The second part of the above equation assumes that the integration limits are taken from zero to infinity. 

Table \ref{tab:combined_mass_estimates_schechter} presents $\rhostar$, $\omegastar$ and $\nstar$ (columns 2 - 4), along with the fractional contributions 
of $\rhostar$ and $\nstar$ with respect to the total sample (columns 5 - 6). 
This is done for the three mass estimates. Uncertainties on these quantities are estimated using the three-dimensional parameter space, within $1\sigma$,
sampled by \texttt{EMCEE}. 
The numbers have been obtained by integrating the Schechter functions from $\mstar = 7.0$ to infinity. We find that the numbers do 
not change much if we change the upper integration limit to the maximum mass in the data(differences $ \le 10^{-7}$). 
Similarly, when integrating the tabulated data and comparing it with the quoted values, the fractional differences are within $10^{-2}$.
In the fitting procedure we do not impose the condition that the mass functions of the red and blue samples add up to the mass function 
of the total sample. 
Therefore, the mass and number densities in the total sample are not the sum of the red and blue samples by construction;
it is however consistent with the sum of the red and blue samples. This is more evident by looking at the ratios of densities with respect to the total density for every third row (total) in columns 5 and 6. The numbers are consistent 
with 1 with deviations of a few percent with the exception of the $\kcorrect$ mass estimate, where the deviations are greater than 10\% and more than $1\sigma$.

For the total, red and blue samples $\rhostar$ (and therefore $\omegastar$) is the largest with the $\mkcorrectgswlc$ mass estimate, followed 
by the $\mtaylor$ and $\mkcorrect$ estimates. The differences between the $\rhostar$ and $\omegastar$ values for the three mass estimates 
are greater than $1\sigma$, the number densities $\nstar$ on the other hand are within $1\sigma$. The relative fractions tell us that 
in terms of mass, $\sim 54\%-62\%$ of stellar mass density in the \hi selected sample is associated with red galaxies and the rest with blue galaxies.
On the other hand the \hi selected sample predominantly picks galaxies from the blue cloud, with $\sim 74\% - 89\%$ of the detections 
(corrected for the $\alphah$ survey selection) associated with blue galaxies. 

We end this section by checking the consistency of our HI-selected GSMF using  $\mkcorrectgswlc$ mass estimates with those from model-independent method using machine learning (ML) algorithms. These algorithms are trained and tested on 
\gswlc masses to predict stellar masses for the sample which does not have \gswlc stellar mass estimates. We then combine these two samples to obtain the HI-selected GSMF and compare with the estimate made using $\mkcorrectgswlc$ masses. We do not finetune 
any of the hyperparameters of the algorithms nor do we feed an optimal set of features to them. We rather examine if these algorithms give results which are consistent with each other with this initial setup. We refer the reader to appendix~\ref{sec_ML} for further 
details.

In figure~\ref{plot_ml_estim} we compare the HI-selected GSMF
based on mass predictions of two algorithms,  
Random Forest \citep{2001MachL..45....5B} (filled circle) and LightGBM \citep{Ke2017LightGBMAH} (filled triangle).
Both are fed feature set 1 (see appendix~\ref{sec_ML}), which consists of five 
optical rest frame magnitudes from SDSS and three \hi properties, mass, velocity width
and redshift to predict $\mgswlc$ masses. The catalog is then used to compute  the
HI-selected GSMF using the $1/\veff$ weights. We find excellent agreement between 
the two ML models and the HI-selected GSMF with $\mkcorrectgswlc$ mass estimates
in the range $8.15 \leq \mstar \leq 11.25$. At lower and higher masses the predictions
deviate significantly since at these two ends the observed data is sparse, making 
it difficult to train the ML algorithm. Additionally correlations between galaxy properties exhibit an intrinsic scatter, due to various environmental effects, therefore it is only natural that predictions will be accurate up to this uncertainty. 
At lower and higher masses, small fluctuations in predicted masses result in large Poisson fluctuations where the median count does not average out over neighboring bins.  
This is a limitation of the statistics at these two mass ends, rather than the algorithm.
Indeed we find that by training it with more data from $\mkcorrect$ mass estimates at the low mass end, we are able to better recover the mass function with these algorithms. 
This is an inconsistent way to address the problem, but works since the mass functions of the three mass estimates are consistent with each other as can be seen from 
the bottom right panel of figure~\ref{fig_GSMF_hi-selected}.

\begin{figure}
    \includegraphics[width=\columnwidth]{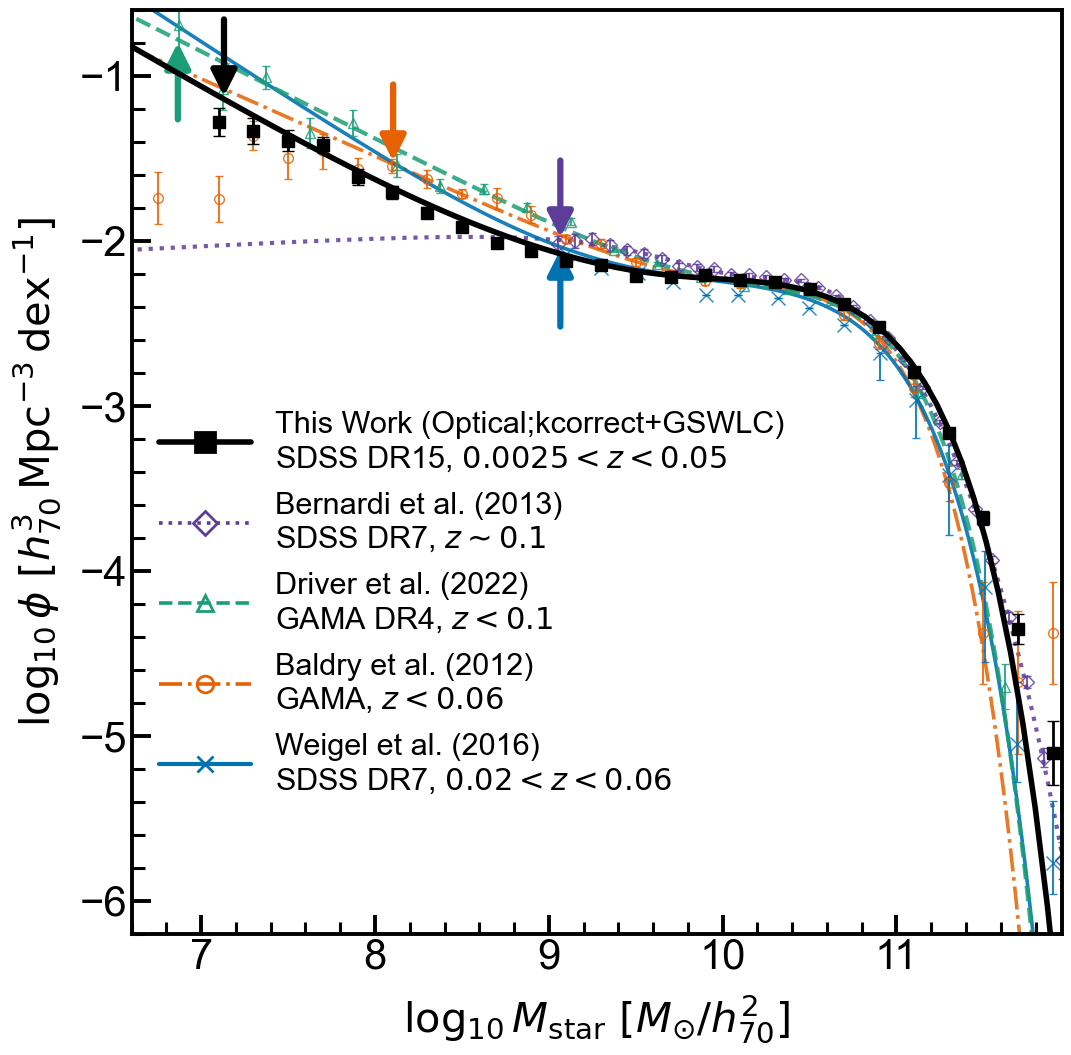} 
    \caption{Optically selected GSMF: comparison between this work (filled squares) using SDSS DR15 in the redshift range $0.0025 < z < 0.05$ with the
    $\mkcorrectgswlc$ stellar mass estimate and previous measurements in the local Universe. These measurements 
    are \citet{2012MNRAS.421..621B} (open circles) and \citet{2022MNRAS.513..439D} (open triangles) with the GAMA survey and   
    \citet{2013MNRAS.436..697B} (open tilted square) and \citet{2016MNRAS.459.2150W} (cross) with the SDSS survey.
    The curves represent Schechter function fits to the data that is reliable, indicated by the solid arrow. The fits have however been extrapolated below this mass scale. This work uses the six-parameter (equation~\ref{eqn_double_schechter}) double Schechter function, 
    whereas  \citet{2012MNRAS.421..621B,2016MNRAS.459.2150W,2022MNRAS.513..439D} use a restricted five parameter double Schechter function
    (with $M_{\rm b}^* = M_{\rm f}^* = M^*$). \citet{2013MNRAS.436..697B} use a modified double Schechter function with seven parameters 
    (see their equation 1).
    }
    \label{fig_gsmf_optical_compare}
\end{figure}

\section{Effect of \hi selection on Galaxy Stellar Mass Function}\label{sec_HI-sel.}
\label{sec_optical-hi-selection}
\begin{table*}
    \begin{center}
\begin{tabular}{|l|c|c|c|c|c|c|c|}
\hline
\multicolumn{8}{|c|}{\textbf{Optically Selected Sample}} \\
\hline
 & $\log_{10} (M^*_{\rm f}/M_{\odot})$ 
 & $\alpha_{\rm f}$ 
 & $\phi^*_{\rm f}$ 
 & $\log_{10} (M^*_{\rm b}/M_{\odot})$ 
 & $\alpha_{\rm b}$ 
 & $\phi^*_{\rm b}$ 
 & $\reducedchisq$ \\
 &  $+ 2\log_{10} h_{70}$
 &  
 & $(10^{-3} h_{70}^{3}\,\mathrm{Mpc}^{-3}\,\mathrm{dex}^{-1})$ 
 & $+ 2\log_{10} h_{70}$
 &  
 & $(10^{-3} h_{70}^{3}\,\mathrm{Mpc}^{-3}\,\mathrm{dex}^{-1})$ 
 &  \\
\hline
 Red
 & $9.44^{+0.10}_{-0.56}$ 
 & $-1.68^{0.11}_{0.11}$ 
 & $0.35^{+0.29}_{-0.34}$ 
 & $10.88^{+0.02}_{-0.01}$ 
 & $-0.62^{+0.02}_{-0.03}$ 
 & $2.59^{+0.08}_{-0.09}$ 
 & 2.62 \\
 Blue 
 & $10.10^{+0.18}_{-0.27}$ 
 & $-1.50^{+0.06}_{-0.05}$ 
 & $0.54^{+0.14}_{-0.12}$ 
 & $10.66^{+0.03}_{-0.02}$ 
 & $-0.82^{+0.08}_{-0.12}$ 
 & $1.90^{+0.14}_{-0.10}$ 
 & 0.96 \\
 Total 
 & $9.54^{+0.10}_{-0.41}$ 
 & $-1.62^{0.05}_{0.05}$ 
 & $0.96^{+0.48}_{-0.75}$ 
 & $10.90^{+0.024}_{-0.013}$ 
 & $-0.90^{+0.01}_{-0.05}$ 
 & $3.52^{+0.12}_{-0.20}$ 
 & 1.71 \\
\hline
\end{tabular}
\end{center}
 \caption{Best-fitting double Schechter parameters (equation~\ref{eqn_double_schechter}) for the optically selected sample. The double Schechter functions are plotted in the left panel of figure~\ref{plot_HI_optical}.}
\label{tab:schechter_params_optical_sel}
\end{table*}

In this section we look at the effect of \hi selection on an optically selected sample.
We compare the GSMF of the optically selected sample  with the joint optical-\hi sample. 

In figure \ref{fig_gsmf_optical_compare}, we compare our estimate of the optically selected GSMF in the overlapping 
patch of SDSS and $\alphah$ with other estimates based on SDSS \citep{2013MNRAS.436..697B,2016MNRAS.459.2150W} and GAMA \citep{2012MNRAS.421..621B,2022MNRAS.513..439D}. 
Our optically selected sample is restricted to $0.0025 \leq z_{\rm cmb} \leq 0.05$, which is shallower than the other surveys. 
The survey area of our  optically selected sample, $A_{\rm surv}\sim 3947 \,\,\rm{deg}^2$ 
is smaller than the SDSS DR7 spectroscopic coverage, $A_{\rm surv}\sim 9380 \,\,\rm{deg}^2$ \citep{2016MNRAS.459.2150W},
due to the condition that it should overlap with $\alphah$.
It is however, larger than GAMA DR4 \citep{2022MNRAS.513..439D} with $A_{\rm surv} \sim 230 \,\,\rm{deg}^2$ and GAMA DR1 \citep{2012MNRAS.421..621B} with $A_{\rm surv} \sim 143 \,\,\rm{deg}^2$. 
GAMA  detects fainter galaxies 
compared to $r_{\rm petromag} (\equiv r_{\rm sdss}) \leq 17.77$ of SDSS, a difference of roughly two magnitudes. 
A combination of limiting magnitude, survey area and redshift depth makes our sample smaller by a factor of roughly $\sim 4.3 \rm{x}$ than that of \cite{2022MNRAS.513..439D}, $\sim 2.4 \rm{x}$ to that of \cite{2016MNRAS.459.2150W} and $\sim 5.7{\rm x}$ 
to that of  \cite{2013MNRAS.436..697B}. Our sample is a factor $\sim 8.7{\rm x}$ larger than the GAMA DR1 sample 
considered in \cite{2012MNRAS.421..621B}. Apart from differences in the sample, we emphasize that 
our stellar mass estimates are calibrated from the \gswlc mass estimates \citep{2016ApJS..227....2S} 
which account for individual dust reddening based on additional UV and IR bands of GALEX and WISE. 
\begin{table*}
    \begin{center}
\begin{tabular}{|l|c|c|c|c|}
    \hline
     & \multicolumn{4}{c|}{\textbf{Joint Optical-\hi selected Sample}} \\
    \hline
     & $\log (M_*/M_{\odot})+ 2\log h_{70}$ 
     & $\alpha$ 
     & $\phi_* (10^{-3} h_{70}^{3} \mathrm{Mpc}^{-3}\,\mathrm{dex}^{-1})$ 
     & $\reducedchisq$ \\
    \hline
     Red 
     & $10.89_{-0.07}^{+0.05}$ 
     & $-0.95_{0.06}^{0.06}$ 
     & $0.53_{-0.14}^{+0.10}$ 
     & 0.95 \\
     Blue 
     & $10.68_{-0.04}^{+0.03}$ 
     & $-1.21^{0.03}_{0.03}$ 
     & $1.02_{-0.18}^{+0.14}$ 
     & 0.35 \\
     Total 
     & $10.86^{0.02}_{0.02}$ 
     & $-1.23^{0.02}_{0.02}$ 
     & $1.02^{0.07}_{0.07}$ 
     & 2.20 \\
    \hline
\end{tabular}
\end{center}
 \caption{Best-fitting single Schechter (equation~\ref{eq_schechterfn}) parameters for the joint optical-\hi sample. The single Schecher functions are plotted in the right 
 panel of figure~\ref{plot_HI_optical}.}
 \label{tab:schechter_params_HI_sel}
\end{table*}
The vertical arrows in figure~\ref{fig_gsmf_optical_compare} represent the stellar mass completeness limits. The Schechter function 
fits (lines) are based on data (points with error bars) down to these limiting masses, but have been extended below them. 
Our estimate of the optically selected GSMF in the $\alphah$ survey area qualitatively agrees well with the other estimates.
All the mass functions agree well around the knee of $9.5 \lesssim \mstar \lesssim 10.4$, with differences occurring at 
larger and smaller masses. At smaller masses our abundances are lower than the estimates from the GAMA survey 
\citep{2012MNRAS.421..621B,2022MNRAS.513..439D}. We cannot make a fair comparison with the SDSS estimates from  \cite{2013MNRAS.436..697B,2016MNRAS.459.2150W} at masses below their limiting mass of $\mstar \simeq 9.1$. 
At larger masses, $\mstar \gtrsim 10.4$, our abundances match with \cite{2013MNRAS.436..697B} but are larger
than the estimates from GAMA \cite{2012MNRAS.421..621B,2022MNRAS.513..439D} and the slightly more local measurements of 
\cite{2016MNRAS.459.2150W} from SDSS.

The prominent bump around the knee of the optically selected GSMF can be described as a sum of two different populations, 
each following a single Schechter function. 
Following \cite{2009ApJ...707.1595D}, 
we fit our optically selected GSMF with a double Schechter function given below:
\begin{equation}\label{eqn_double_schechter}
\begin{aligned}
\phi(M)\, dM 
&= \phi_{\rm f}(M)\, dM + \phi_{\rm b}(M)\, dM \\
&= \phi_{\rm f}^{*} \left( \frac{M}{M_{\rm f}^{*}} \right)^{\alpha_{\rm f}}
\exp\!\left( -\frac{M}{M_{\rm f}^{*}} \right) dM  \\
& +  \phi_{\rm b}^{*} \left( \frac{M}{M_{\rm b}^{*}} \right)^{\alpha_{\rm b}}
\exp\!\left( -\frac{M}{M_{\rm b}^{*}} \right) dM .
\end{aligned}
\end{equation}

\begin{figure*}
    \begin{tabular}{cc}
    \includegraphics[width=3.4in]{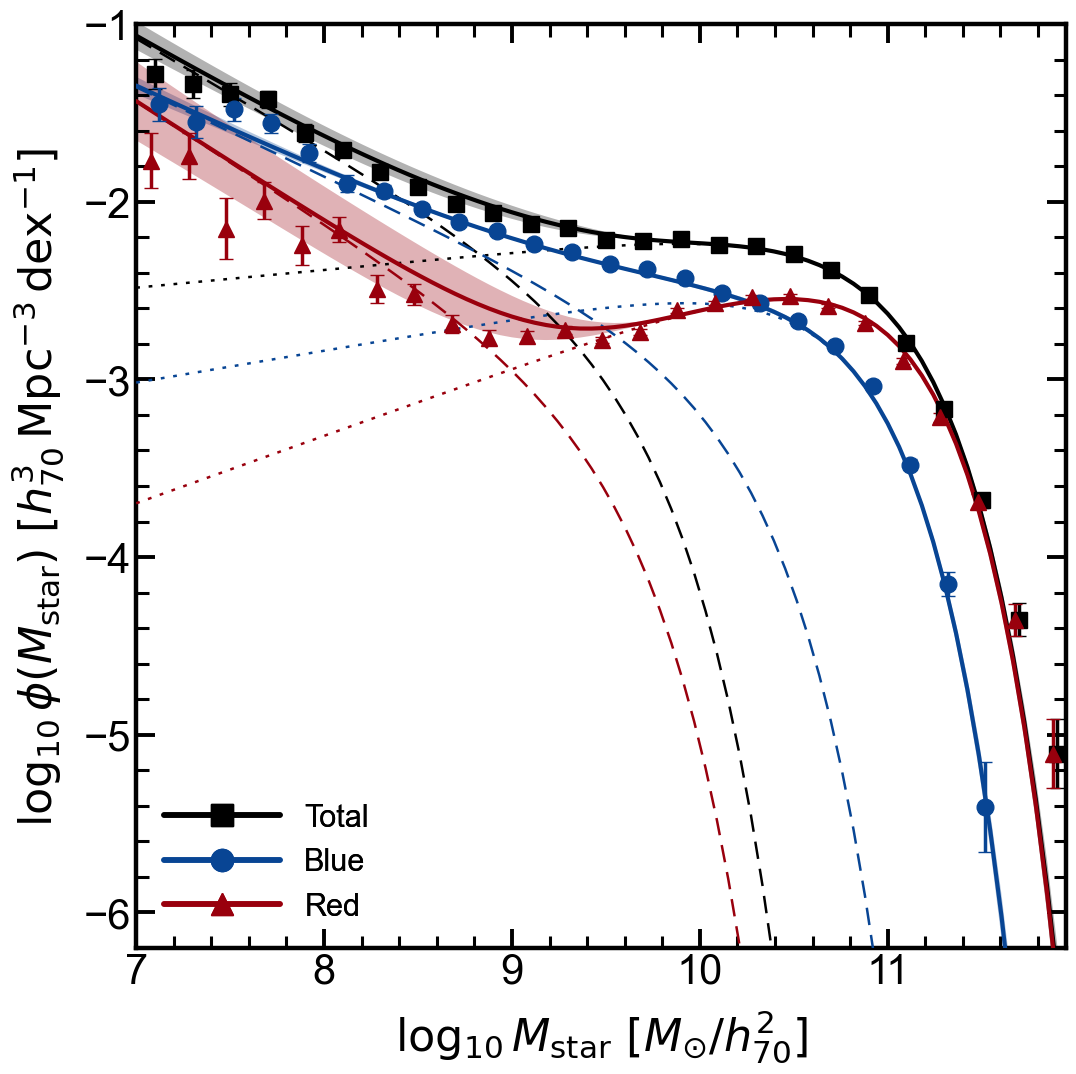} 
    \includegraphics[width=3.4in]{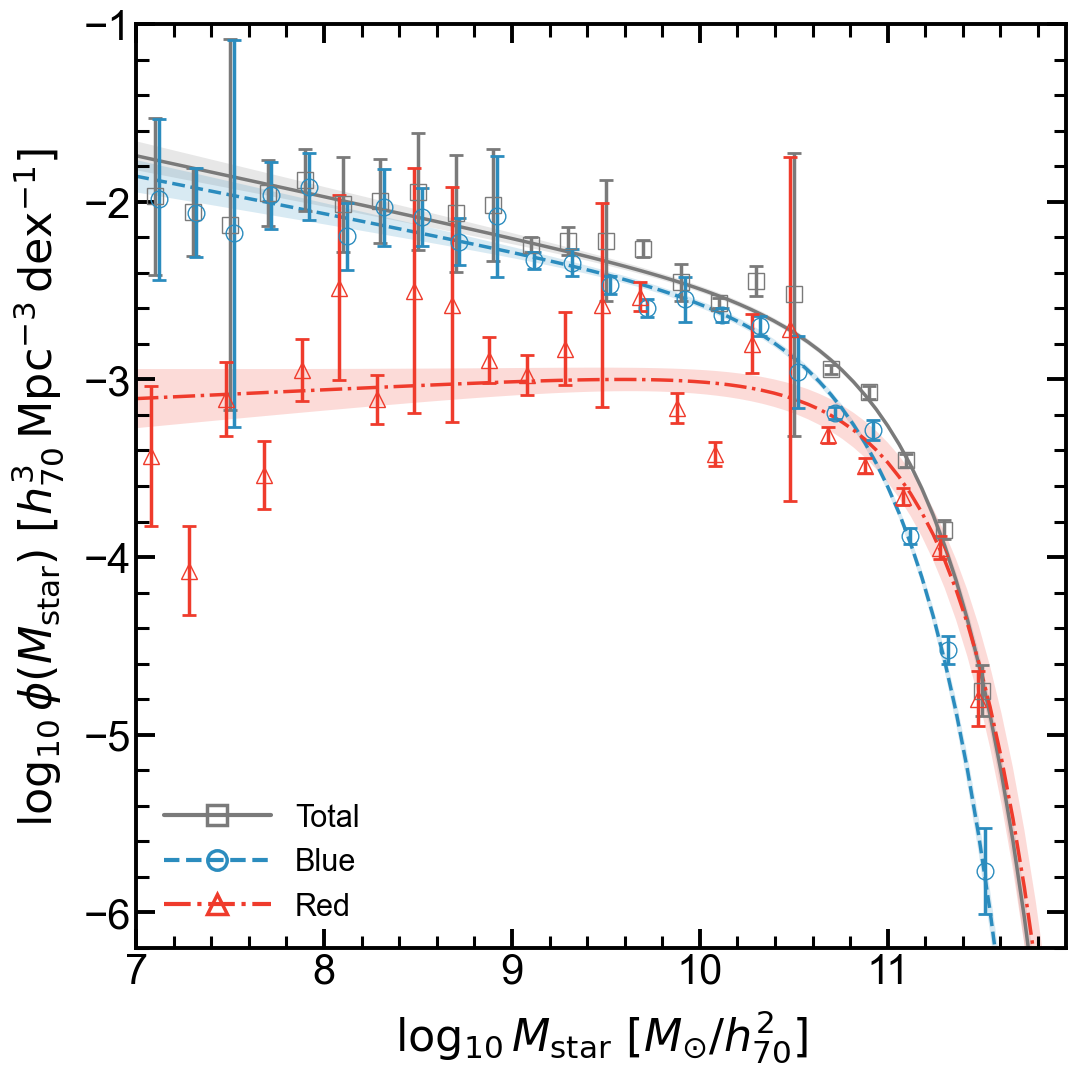} \\
    \end{tabular}
    \caption{\emph{Left}: optically selected GSMF for the total(black), red(red) and blue(blue) populations (data points with error bars) and their corresponding double Schecter function (equation~\ref{eqn_double_schechter} fits and uncertainties, solid lines and shaded regions. 
    The data points corresponding to blue (red) galaxies are shifted to the right (left) by 0.02 dex with respect to the total population for a better comparison. The dashed (dotted) curves represent the Schechter functions for the bright (faint) galaxy populations, shown separately for the total, red, and blue samples.
    The best fit parameters of the double Schechter function and their uncertainties
    are given in table~\ref{tab:schechter_params_optical_sel}.
    \emph{Right}: The joint optical-\hi selected GSMF for the total (grey open squares),
     red (red open triangle) and blue (blue open circle) populations and their corresponding single Schechter function fits and uncertainties (lines and shaded regions). 
     The data points corresponding to blue (red) galaxies are shifted to the right (left) by 0.02 dex with respect to the total population for a better comparison. 
    The best fit parameters of the single Schechter function and their uncertainties are
    given in table~\ref{tab:schechter_params_HI_sel}.
     }
    \label{plot_HI_optical}
\end{figure*}

We do this separately for the total, red and blue samples. 
The two terms in equation~\ref{eqn_double_schechter} are single Schechter functions corresponding, 
loosely, to the faint and bright galaxy populations (under the assumption of a monotonic mass-to-light ratio). 
The fit is shown in the left panel of figure \ref{plot_HI_optical} and the corresponding fitted Schechter parameters are shown in 
table \ref{tab:schechter_params_optical_sel}.
As before, the sum of the red and blue 
abundances is the total abundance in the data. As done in section~\ref{sec_gsmf_estimates}, 
while fitting, we do not impose this constraint and fit the three samples independently. As can be seen 
from the last two columns of table~\ref{tab:schechter_fractions}, we obtain $\rhostar$ (total sample) 
within an accuracy of $1\%$ for both the optically selected and the joint optical-\hi samples.  
$\nstar$ (total sample) is accurate within $3\%$ for the optically selected sample and $8\%$ for the joint \hi-optical sample.  
Equation~\ref{eqn_double_schechter} considered in this work has six parameters, 
whereas  \citet{2012MNRAS.421..621B,2016MNRAS.459.2150W,2022MNRAS.513..439D} 
use a restricted five-parameter double Schechter function
(with $M_{\rm b}^* = M_{\rm f}^* = M^*$). 
\citet{2013MNRAS.436..697B} use a modified double Schechter function with seven parameters (see their equation 1).

Unlike the optically selected sample, we do not see a prominent bump around the knee of the mass function for 
the joint optical-\hi sample. We fit a single Schechter function to the joint optical-\hi sample. Our results, for the red, blue and total populations, are shown in the right panel of figure~\ref{plot_HI_optical} and the best fit parameters are listed in table~\ref{tab:schechter_params_HI_sel}.

For the optically selected sample 
the differences between the Schechter parameters of the faint and bright populations are largest for the red sample
as compared to the blue sample and total samples. 
E.g. $\phi_{\rm b}^*/\phi_{\rm f}^* = \{7.4,3.6,3.6\}$,  
$\Delta \alpha = \alpha_{\rm b} - \alpha_{\rm f} = \{1.06, 0.68, 0.72\}$ and 
$\Delta M = \log_{10} (M_{\rm b}^*/M_{\rm f}^*) = \{1.44,0.56,1.36\}$ for the red, blue and total samples.
This translates to a distinct bimodal (faint + bright) shape in the GSMF for the red sample compared to 
the blue sample in figure~\ref{plot_HI_optical}. The red sample dominates the abundances at the high mass (bright) end
and the blue sample is the dominant population at intermediate and low  masses. The Schechter parameters  of 
the bright and faint populations for the red and blue samples reflect this. 
The quality of the fits is reasonable. The red and total samples have outliers which increase their $\reducedchisq$.

For the joint optical-\hi sample, we also see a similar trend to the optically selected sample. 
Abundances at the high mass end are determined by the red sample and the blue sample dominates the 
abundances at intermediate and low masses. But the bimodal (red and blue) shape for the total sample seen for the optically 
selected sample is less obvious here. Indeed a single Schechter function describes the GSMF for the joint optical-\hi sample. 
This is due to a combination of the amplitude and the knee for the red and blue samples. The transition scale, $M_{\rm tr}$, determined by $\phi_{\rm blue}(M_{\rm tr}) = \phi_{\rm red}(M_{\rm tr})$ shifts from $M_{\rm tr} = 10.3$ for the optically
selected sample to $M_{\rm tr} = 10.8$ for the joint optical-\hi sample. This is
because the \hi selection on the optical sample primarily selects
galaxies in the blue cloud as will be discussed next.

In table~\ref{tab:schechter_fractions} we list $\rhostar$, $\omegastar$, $n_{\rm star}$
as well as the relative 
fractions of $\rhostar$ and $n_{\rm star}$ with respect to those of the total sample. 
This is done for both the optically selected sample and the joint optical-\hi sample.
As was done for the \hi selected sample (in section~\ref{sec_gsmf_estimates}), 
we do not impose the condition that the total abundance is the sum 
of the red and blue abundances in the fitting procedure, although the data
explicitly satisfies this condition. The fractional mass densities are consistent 
with this condition within $1\%$, 
although we find larger deviations, albeit within uncertainties, 
for the fractional number densities. 
Number and mass densities for the joint optical-\hi sample are smaller than that of the 
optically selected sample as expected. As mentioned earlier 
our uncertainties in $\rhostar$ and $n_{\rm star}$ are estimated  
by sampling directly the mass function parameter space from \texttt{EMCEE}. 
This ensures that we sample the parameter space within ($1\sigma$) errors that are correlated. 
If we assume that the error contours for the parameters are uncorrelated
and generate uncorrelated gaussian realizations of the parameters, 
we find that for the optically selected total sample 
$\rhostar/(10^8 \msun\,\,{\rm Mpc}^{-3}) = 2.713 \pm 0.479$. The uncertainty in $\rhostar$ is $9\times$ greater than what is quoted in 
table~\ref{tab:schechter_fractions}.
Assuming uncorrelated errors on their best fit parameters and integrating 
with a lower mass threshold of $10^7 \msun$ for the fitted Schechter functions, 
we estimate 
$\rhostar/(10^8 \msun\,\,{\rm Mpc}^{-3}) = 
2.226 \pm 0.362, 2.169 \pm 1.169, 2.951 \pm 0.202$ for 
the optically selected mass functions of   
\cite{2012MNRAS.421..621B,2016MNRAS.459.2150W, 2022MNRAS.513..439D} respectively.
For \cite{2013MNRAS.436..697B} we  obtain 
$\rhostar/(10^8 \msun\,\,{\rm Mpc}^{-3}) = 2.736$ without any estimate of uncertainties
since the best fit mass function parameters are quoted without errors. Our estimate
of $\rhostar$ is consistent with all these estimates. We, however, note that 
this comparison is fair only with the results of \cite{2022MNRAS.513..439D}
since the GSMF estimates are complete to $\log_{10}(M/\msun) = 7.0$.
The comparison with 
\cite{2012MNRAS.421..621B,2013MNRAS.436..697B,2016MNRAS.459.2150W} has been made
by considering the fitted Schechter function below their completeness 
limit of $\log_{10}(M/\msun) = 7.0$. 
\begin{table*}
\begin{center}
\begin{tabular}{|l|c|c|c|c|c|}
\hline
Population 
& $\rhostar$ 
& $\omegastar$ 
& $\nstar$ 
& $\rhostar/\rhostar^{\rm total}$ 
& $\nstar/\nstar^{\rm total}$ \\
& $[10^{8}\,M_{\odot}\,\mathrm{Mpc}^{-3}]$ 
& $[10^{-4}]$ 
& $[10^{-3}\,\mathrm{Mpc}^{-3}]$ 
& & \\
\hline
\multicolumn{6}{|c|}{\textbf{Optically Selected  sample ($\log \mstar \ge 7$)}} \\
\hline
Red 
& $1.774 \pm 0.036$ 
& $13.057 \pm 0.266$ 
& $28.075 \pm 7.573$ 
& $0.654 \pm 0.018$ 
& $0.380 \pm 0.110$ \\
Blue 
& $0.919 \pm 0.021$ 
& $6.761 \pm 0.153$ 
& $43.871 \pm 2.272$ 
& $0.339 \pm 0.010$ 
& $0.594 \pm 0.068$ \\
Total 
& $2.713 \pm 0.053$ 
& $19.964 \pm 0.392$ 
& $73.860 \pm 7.560$ 
& $0.993 \pm 0.025$ 
& $0.974 \pm 0.146$ \\
\hline
\multicolumn{6}{|c|}{\textbf{Joint Optical-\hi selected sample ($\log \mstar \ge 7$)}} \\
\hline
Red 
& $0.400 \pm 0.110$ 
& $2.947 \pm 0.808$ 
& $3.546 \pm 2.112$ 
& $0.452 \pm 0.126$ 
& $0.123 \pm 0.074 $ \\
Blue 
& $0.573 \pm 0.064$ 
& $4.218 \pm 0.469$ 
& $23.077 \pm 2.902$ 
& $0.647 \pm 0.077$ 
& $0.798 \pm 0.136$ \\
Total 
& $0.885 \pm 0.038$ 
& $6.516 \pm 0.283$ 
& $28.932 \pm 3.348$ 
& $1.099 \pm 0.151$ 
& $0.920 \pm 0.163$ \\
\hline
\end{tabular}
\end{center}
\caption{
Stellar mass density, stellar mass density parameter, number density, and their fractional contributions derived from the fitted Schechter function integration for the optically selected sample and the joint optical-\hi sample. 
}
\label{tab:schechter_fractions}
\end{table*}
For the optically selected sample 
the red galaxies account for $\sim 65\%$ of the stellar mass budget with the remaining
accounted by blue galaxies, although in terms of galaxy counts the red-blue divide  
is roughly $40-60$. For the joint optical-\hi sample, these numbers are reversed. $\alphah$ samples the optically selected population primarily in the blue cloud. The stellar mass 
density and number density of blue galaxies of the joint optical-\hi sample is $\sim 65\%$
and $\sim 80\%$ respectively of the total, with the rest being associated with red galaxies.
\begin{figure*}
    \centering
    \begin{tabular}{ccc}
        \includegraphics[width=2.3in]{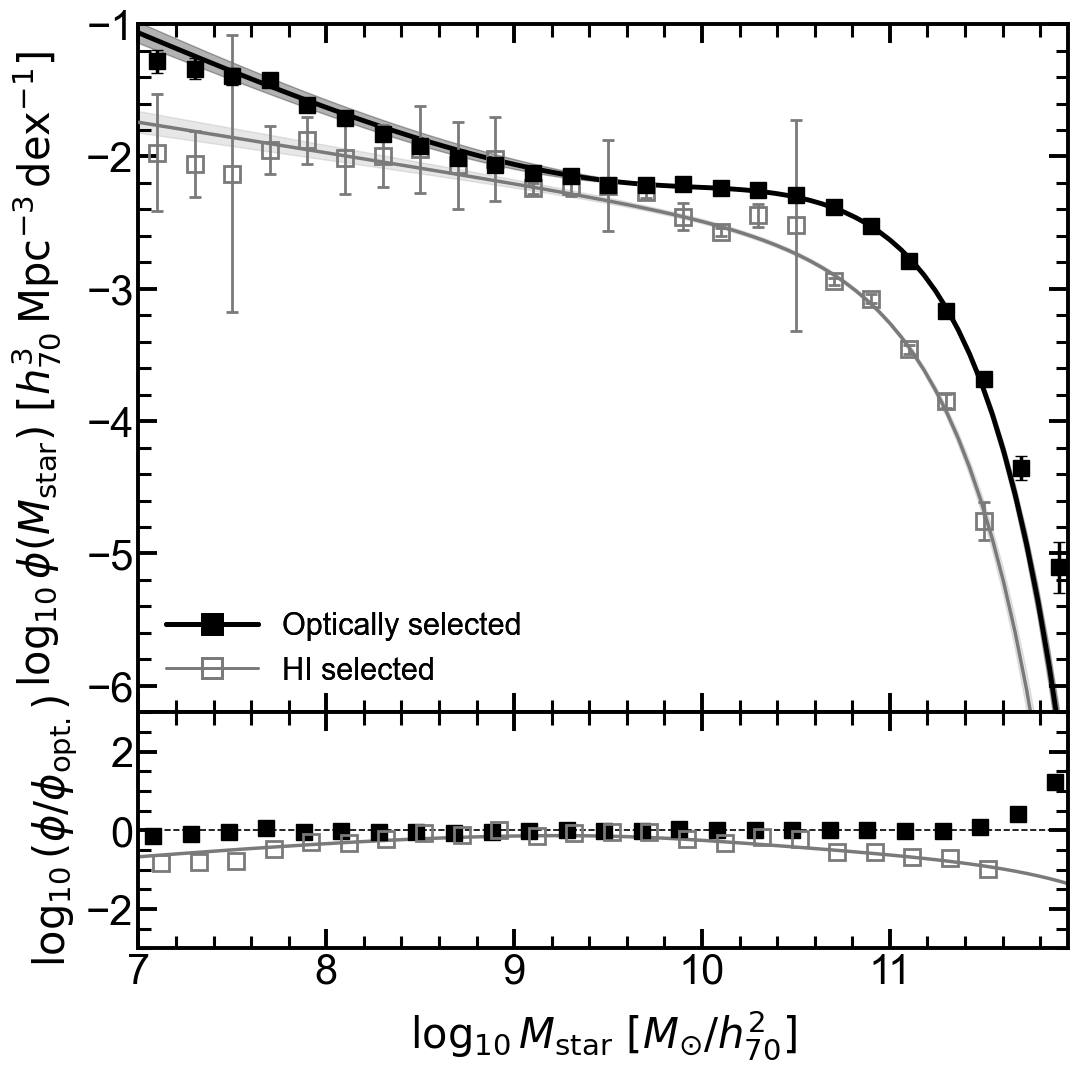} &
        \includegraphics[width=2.3in]{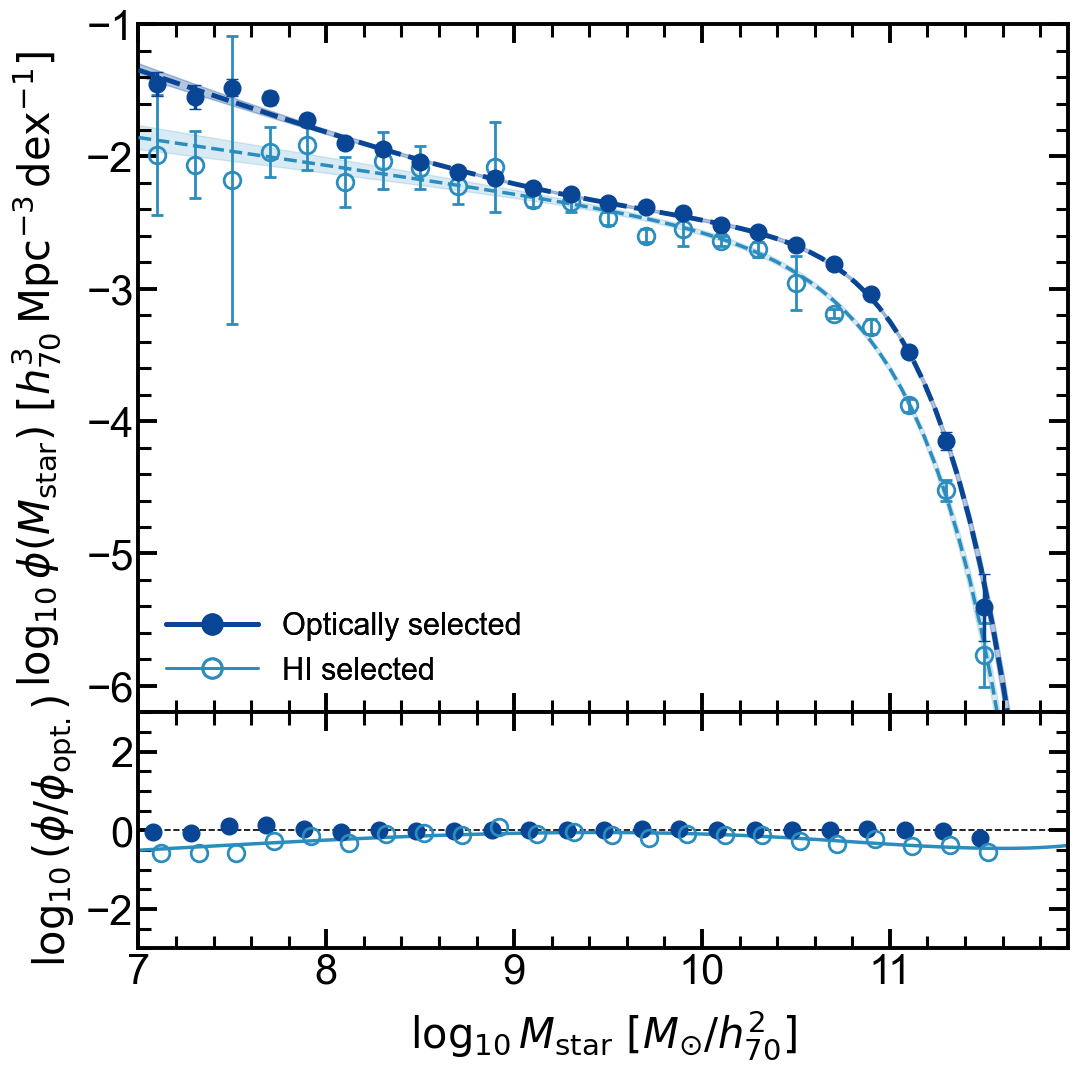} &
        \includegraphics[width=2.3in]{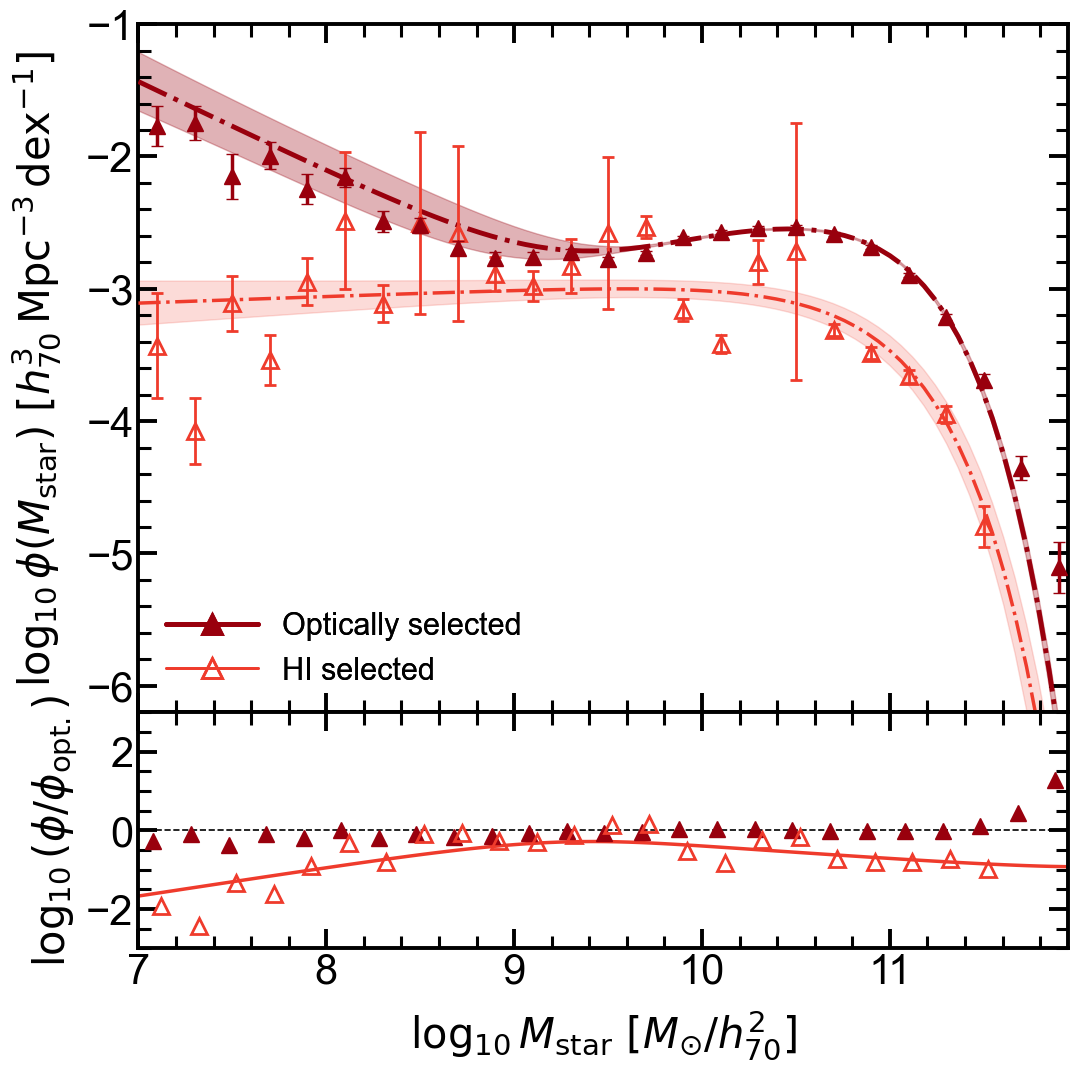} \\
    \end{tabular}
    \caption{Optically and joint optical-\hi selected GSMF for the total(\emph{Left}), blue(\emph{Middle}), and red(\emph{Right}) galaxy populations. In the upper panel, solid and dashed curves show the best-fit optical and \hi Schecter function, respectively, with shaded bands denoting $1\sigma$ uncertainties. Filled and open markers indicate optical and \hi selected measurements. The lower panel shows residuals with respect to the optical-selected fitted function, with error bars obtained by adding in quadrature the data and model uncertainties. The solid curve in the residual panel represents the ratio of the \hi and optical models.}
    \label{plot_comparison_HI_opt}
\end{figure*}

In figure~\ref{plot_comparison_HI_opt} we compare the GSMF of the optically selected and the joint optical-\hi samples for the total (left), blue (middle) and red (right) galaxy 
populations. The bottom row represents the ratio of the mass functions  
with respect to the best fit optically selected Schechter function.
This figure highlights the mass-dependent sampling of $\alphah$
on an optically selected sample, or the mass-dependent detection fraction. 
The ratio of abundances, 
$\phi_{\rm opt-\hi}/\phi_{\rm opt}$, peaks below 
the knee of the mass function. For the total, blue and red populations 
the peak ratio and the mass scale of the peak ratio are 
$(\phi_{\rm opt-\hi}(M_{\rm peak})/\phi_{\rm opt}(M_{\rm peak}), M_{\rm peak})$ =  $(0.73,9.3)$, $(0.87,9.3)$, $(0.51,9.5)$ respectively. 
At the lowest mass scale, $\mstar = 7.1$, and the largest mass scale, $\mstar = 11.9$,
the ratio of these abundances (total, blue, red) decreases
from the  peak values to $(0.23, 0.33, 0.03)$ and $(0.05,0.39,0.12)$ respectively.
In table~\ref{tab:HI_optical_ratio_schechter} we 
list these ratios of integrated mass and number densities.
The $\alphah$ survey is able to account for $33\% \pm 2\%$ of  
stellar mass and $39\% \pm 6\%$ of galaxies of an optically selected sample.
However it detects a larger fraction of blue galaxies compared to red galaxies.
$53\% \pm 8\%$ ($13\% \pm 8\%$) of detections are in the blue (red) cloud. 
$\alphah$ detects $62\% \pm 7\%$ ($23\% \pm 6\%$) of stellar mass  
in the blue (red) cloud. We note that these numbers are higher 
for the \hi selected sample since they include galaxies which
have photometric-only detections as well. However a fair comparison 
is only possible between the optically selected sample and the joint
optical-\hi sample since they have a common volume and a common 
limiting spectroscopic magnitude. As we will show next, the $\mstar-\mhi$ scaling 
relation is not very different between the \hi selected sample and the joint optical-\hi sample.



%
\section{The Underlying Stellar Mass - \hi Mass Scaling Relation Using Abundance Matching}
\label{sec_underlying_rel}

\begin{figure*}
    \centering
    \begin{tabular}{ccc}
        \includegraphics[width=2.3in]{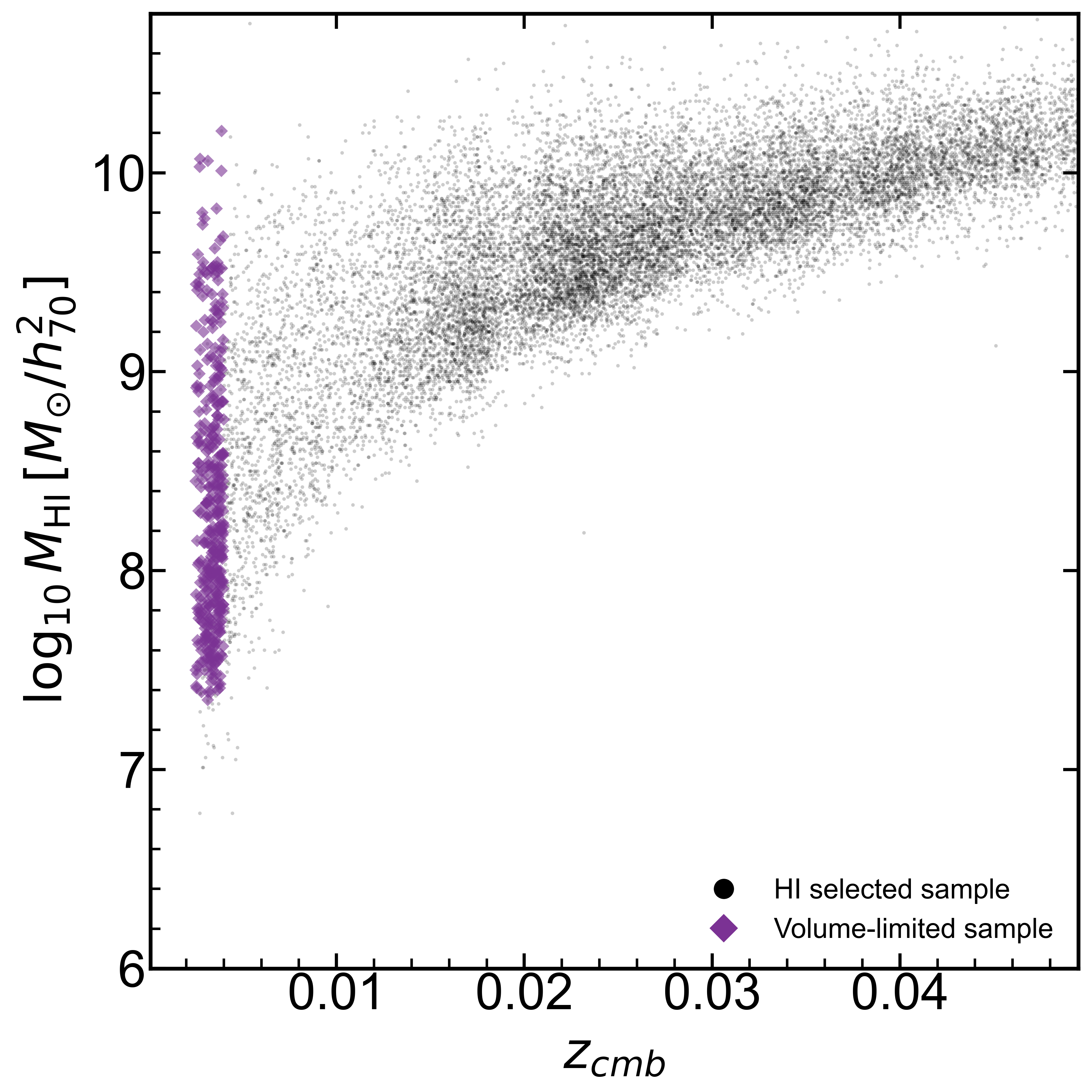} &
        \includegraphics[width=2.3in]{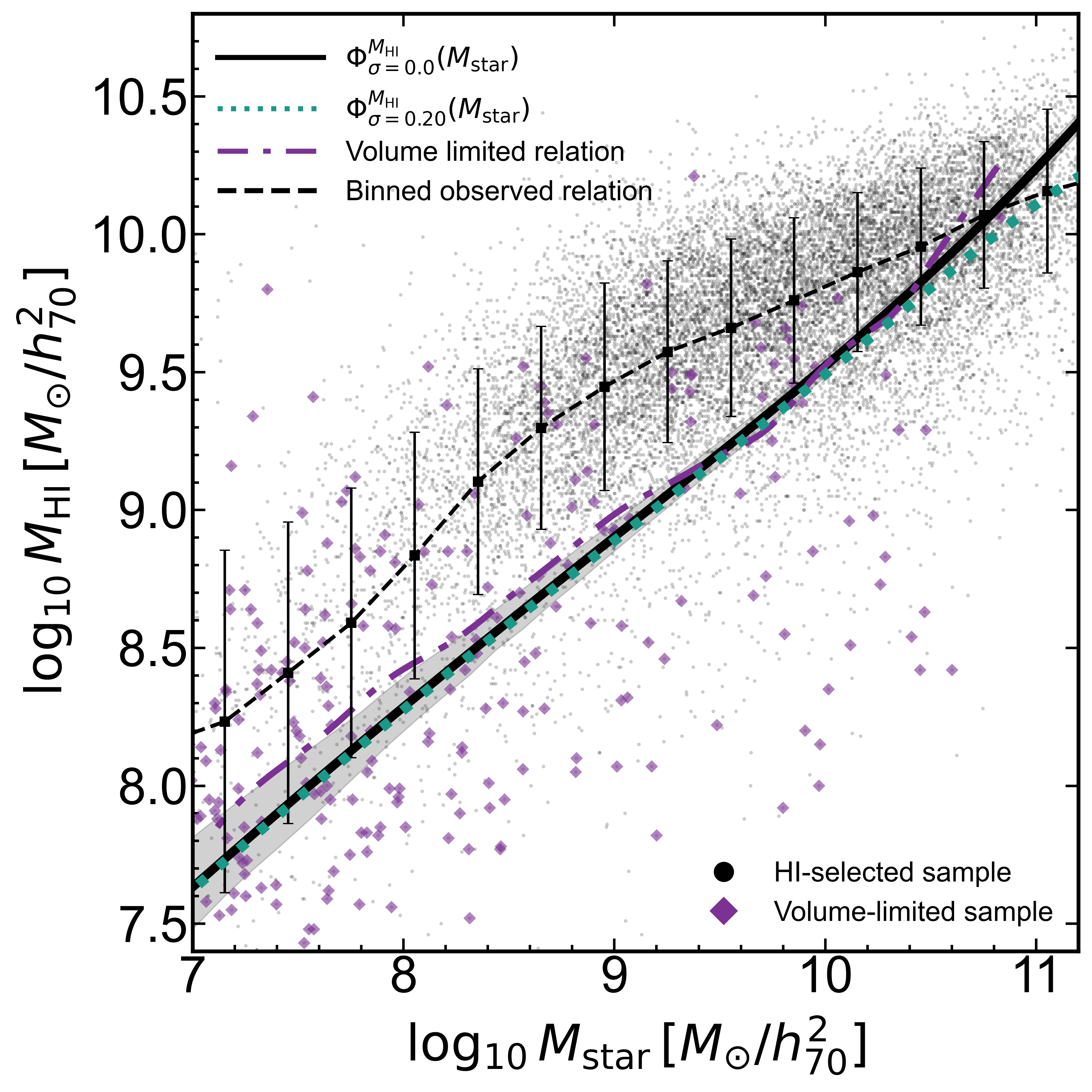} &
        \includegraphics[width=2.3in]{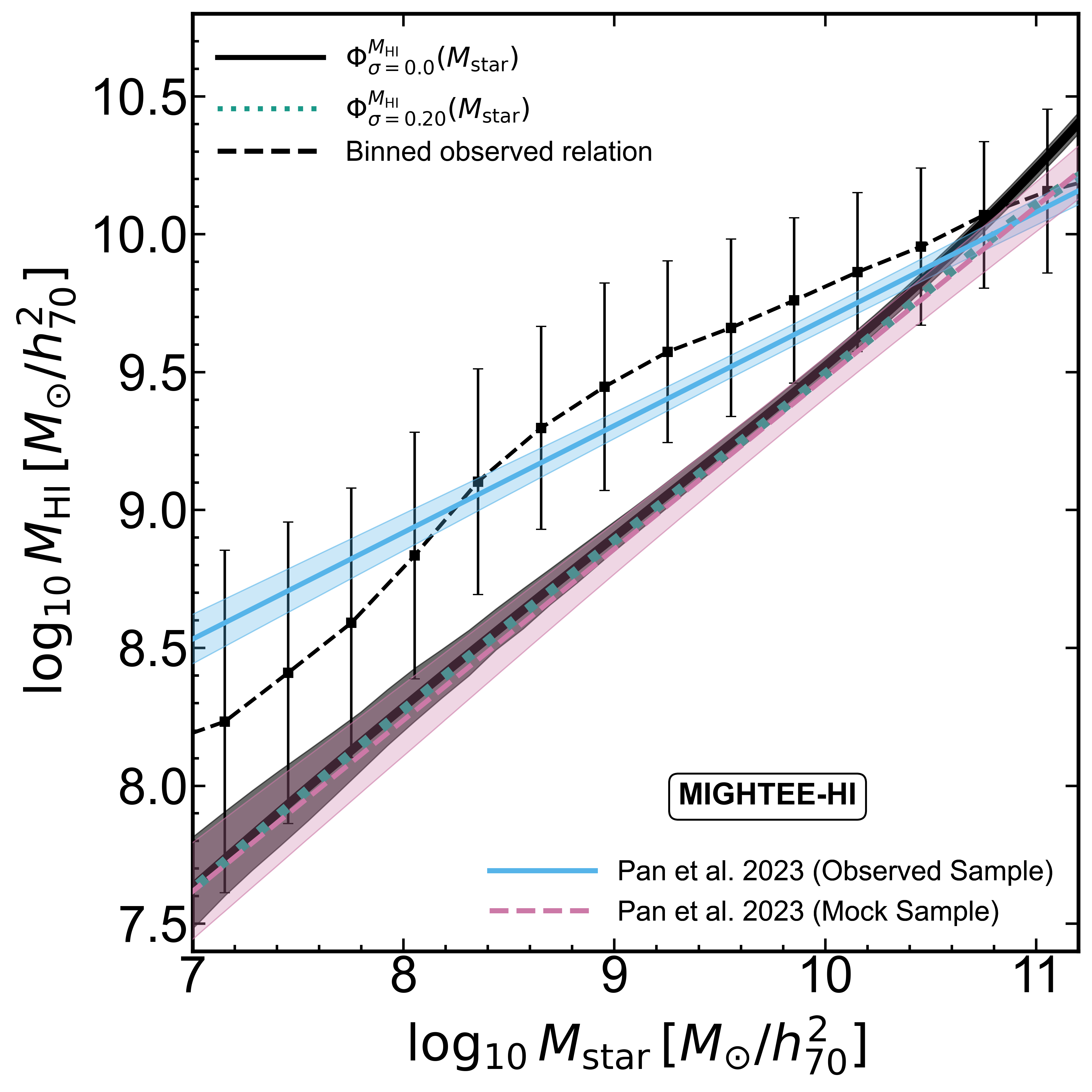} \\
    \end{tabular}
    \caption{\emph{Left}: $\log_{10}(\mhi/\msun)$ as a function of redshift for ALFALFA galaxies (black points). Purple diamonds denote galaxies from the volume-limited sample in the redshift range $0.0025 \leq z \leq 0.004$.
   \emph{Middle}: Comparison of stellar mass and \hi mass scaling relations for ALFALFA galaxies. The solid black line($\Phi^{\mhi}_{\sigma=0.0}(M_\mathrm{star})$)is the \hi selected $\mstar-\mhi$ relation 
   obtained by abundance matching the \hi selected $\phi(\mhi)$ and $\phi(\mstar)$ with  zero scatter. The teal dotted line($\Phi^{\mhi}_{\sigma=0.20}(M_\mathrm{star})$ shows the same abundance-matched relation with scatter of $\sigma=0.20$ dex.
   The shaded region is the uncertainty of this scaling relation obtained from the
   uncertainties in the two mass functions.
   The dashed black line with error bars is the binned, observed relation. Grey points represent the full ALFALFA sample. Purple diamonds denote a local volume-limited sample (see left panel), 
   with the corresponding scaling relation obtained by abundance matching, shown   
   as the purple dot-dash line with zero scatter.
   \emph{Right}: Comparison  of \hi selected relations from $\alpha.100$ (same as middle panel) with the results from MIGHTEE-HI survey \citep{2023MNRAS.525..256P}. The blue solid line (and uncertainty) show the relation for a power law fit to the observed data and the pink dashed line is the relation after accounting for galaxies 
   below the detection threshold (i.e. the selection function) by creating a mock sample.}
    \label{figure_vol_lim}
\end{figure*}

We abundance match the HIMF to the GSMF for both the 
\hi selected and the joint optical-\hi sample to obtain 
the $\mstar-\mhi$ scaling relation. This is an \hi selected scaling relation
since the samples are \hi selected. 
The normalization of both mass functions, HIMF and GSMF, 
is fixed at high $\mhi$, which is complete in \hi mass but incomplete in stellar mass, 
due to the large scatter in $\mstar-\mhi$ (black points),
as can be seen in the middle panel of figure~\ref{figure_vol_lim}. 
The black data points in the middle panel of figure~\ref{figure_vol_lim}
are from the \hi selected sample. The dashed line, joining the points, 
is the observed mean relation in bins of $\mstar$ with the error bars denoting
the observed rms fluctuation in the corresponding bins. We note these 
data are affected by the $\alphah$ selection function and represent 
a biased relation of gas-rich objects (larger $\mhi$) at fixed stellar mass.
We expect that accounting for the selection function would 
bring the scaling relation below this observed relation.
The purple diamonds represent an unbiased data from a volume-limited sample. 
The volume-limited sample is a local sample with $\zcmb \in [0.0025, 0.004]$.
The upper limit of this redshift range determines the mass threshold above which every galaxy is detectable, irrespective of its individual redshift. Thus the constructed 
volume-limited sample is complete within this redshift range and is free from selection
effects. The choice of the volume-limited sample is shown in the left panel where we plot $\mhi$ as a function of $\zcmb$ 
for all galaxies (black, same as the middle panel of figure~\ref{figure_vol_lim})
and galaxies from the volume-limited sample (purple diamonds).
The range of (small) redshifts is chosen so as to maximize 
the range in $\mhi$.  

The abundance-matched $\mstar-\mhi$ scaling relation, 
which we call the intrinsic, unbiased,  scaling relation, for the \hi selected sample is shown as the solid black line in the middle panel of figure~\ref{figure_vol_lim}. The shaded region is the uncertainty in this scaling relation and has been computed from the uncertainties in the HIMF and the GSMF. As argued earlier, this, selection-corrected, intrinsic scaling relation is below the observed scaling relation. However the data from the volume-limited sample are spread evenly across the intrinsic scaling relation, as expected. The dot-dashed line is the abundance-matched scaling relation from the volume-limited sample and is  consistent with the abundance-matched scaling relation of the full 
\hi selected sample. For the volume-limited sample we do not need to use the 
$1/\veff$ binning since it is a complete sample, free from selection effects. We obtain 
the mass function directly by binning in mass.
The small differences in the scaling relation between the \hi selected 
sample and the volume-limited sample, are i/due to the limited 
number of galaxies, in our volume-limited sample and ii/the local sample
is more susceptible to peculiar velocity effects which may not have been completely
accounted for in the flow model.
\begin{table}
\begin{center}
\begin{tabular}{|l|c|c|}
\hline
Population 
& $\rhostar^{\rm \hi} / \rhostar^{\rm Optical}$ 
& $\nstar^{\rm \hi} / \nstar^{\rm Optical}$ \\
\hline

&$\log \mstar \ge 7$
&\\
\hline
Red   & $0.23 \pm 0.06$ & $0.13 \pm 0.08$ \\
Blue  & $0.62 \pm 0.07$ & $0.53 \pm 0.08$ \\
Total & $0.33 \pm 0.02$ & $0.39 \pm 0.06$ \\
\hline
\end{tabular}
\end{center}
\caption{
Ratio of \hi selected to optical selected stellar mass density and number density for red, blue, and total galaxy populations, computed using corresponding Schechter 
function integrations for each sample.}
\label{tab:HI_optical_ratio_schechter}
\end{table}

The black line represents an abundance-matched relation with zero scatter 
in $\mstar$ and $\mhi$.
This is why, even though the sample is nearly complete at the high-mass end, it still does not align (exactly) with the observed relation.
Scatter preferentially affects the high mass end through an Eddington-like bias \citep{2010ApJ...717..379B}, where the steeply declining number density causes more low
mass galaxies to scatter upward than high mass galaxies to scatter downward, effectively flattening the relation. We therefore expect that, 
once scatter is incorporated into our abundance-matched relation, the high-mass end will be brought into better agreement with observations. The abundance-matched scaling 
relation with a scatter of $\sigma = 0.2 {\rm dex}$ is shown as the teal dotted line. Scatter 
suppresses the high-mass relation and aligns it better with the observed relation, whereas 
the low-mass end is unaffected. We explore the effect of scatter on both the scaling relations and clustering measures  in a forthcoming paper.

In the right panel of figure~\ref{figure_vol_lim} we compare our scaling relations 
(same as the middle panel) with  results from the blind, MIGHTEE-HI survey 
based on a HI-selected sample of 249 galaxies \citep{2023MNRAS.522.5308P,2023MNRAS.525..256P}. 
The blue solid line and shaded region denote a linear fit and $1\sigma$ uncertainty 
to the observed data and is consistent with the observed scaling relation in $\alpha.100$ (dashed line). However the observed data points do not account for undetected galaxies.
In this work we use the HI selection of $\alpha.100$ to correct for this, whereas 
due to the limited sample size of the MIGHTEE-HI survey, \cite{2023MNRAS.525..256P} create a mock sample using the GAMA GSMF of \cite{2022MNRAS.513..439D} to account
for the full population of galaxies (detected and undetected in MIGHTEE-HI).
The pink dashed line and shaded region are a power-law fit and $1\sigma$ uncertainty to the mock catalog and matches perfectly with our relation with a scatter of $\sigma = 0.2$ dex.
The scaling relations of \citep{2023MNRAS.525..256P} are model-dependent but nevertheless agree with those from our analysis which is model-independent. 
It is highly encouraging to find that completely different datasets from single-dish telescopes and interferometers, such as ALFALFA and MIGHTEE-HI, yield consistent results.

\begin{figure}
  \includegraphics[width=\columnwidth]{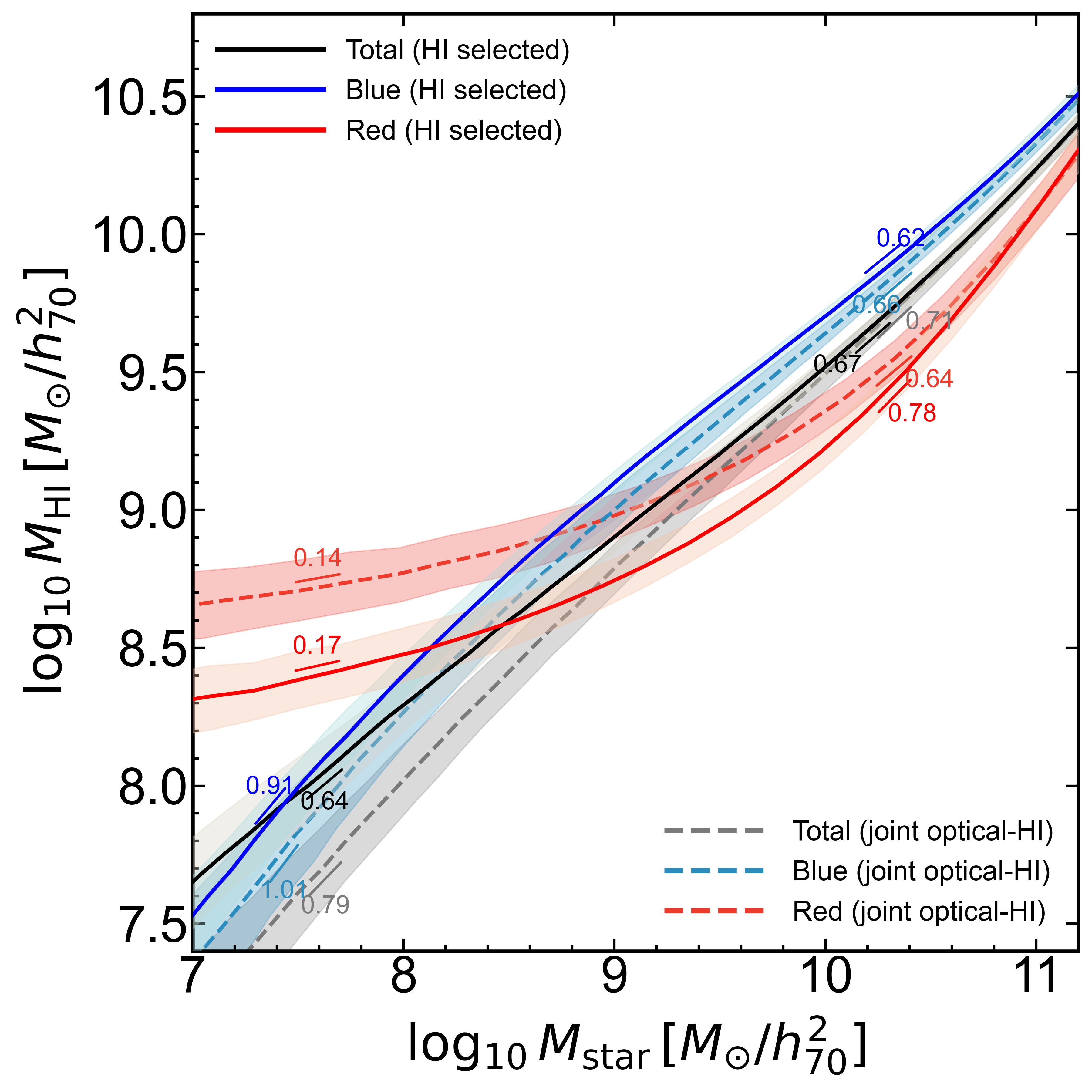} \\
   \caption{The \hi selected $\mstar-\mhi$ scaling relations  and uncertainties (shaded regions) for the total(black), red(red) and blue(blue) populations. 
   The solid (dashed) line corresponds to the \hi (joint optical-\hi) selected sample. The solid line segments with numbers denote the local slope of the 
   scaling relations.}
   \label{figure_scaling_reln}
\end{figure}

In figure~\ref{figure_scaling_reln} we plot the intrinsic scaling relation 
for the \hi selected (solid line) 
and the joint optical-\hi sample (dashed line) for the total, red and blue populations, without scatter. This is done by abundance matching each of the populations separately.
The agreement between both samples is good with some differences for the red population 
at intermediate and lower masses. 
For $\mstar \geq 9.5$ blue 
galaxies are, on average, more gas rich compared to red galaxies. 
Since $\alphah$ primarily samples the blue cloud, the scaling relation 
of blue galaxies is closer to the scaling relation for all galaxies.
The scaling relations for the total and blue populations 
are consistent between the joint optical-\hi sample and the \hi selected sample.
For the red population, the scaling relations differ for $\mstar \lesssim 9.6$
with differences between both samples increasing with decreasing mass.
The slope of the $\mstar-\mhi$ scaling relation, $\beta$, 
increases from $\beta \simeq  0.6$ to 
$\beta \simeq 1$ with decreasing mass for the blue and total populations.
However for the red population the trend in $\beta$ is reversed. It decreases from $\beta \sim 0.8$ to $\beta \sim 0.15$ with decreasing mass, with the 
rapid change occurring at $\mstar \sim 10.0$. This is because the logarithmic slope of the stellar mass function, $\alpha + 1$, for the red population is positive , whereas it is negative for the total and blue populations and it is negative for the three populations for the \hi mass function. 

\section{Summary and Discussion}\label{sec_discussion}
In this work, we have used a sample 
of gas-rich galaxies from the ALFALFA survey ($100\%$ catalog) 
and their optical  counterparts from SDSS to estimate an \hi selected GSMF for three different stellar mass estimates (\taylor, \kcorrect and \kgswlc defined in  section~\ref{sec._HI_selc._sample}). 
The difference in \hi selected GSMFs derived from \taylor, \kcorrect, and \kgswlc stellar mass
estimates highlights the role of dust attenuation in shaping stellar mass estimates. While the three mass functions show consistent values at low stellar masses, significant deviations emerge at the high-mass end.

This is because of dust attenuation which correlates strongly with stellar mass \citep{2010MNRAS.409..421G}, especially for a gas-rich star-forming sequence like the 
$\alphah$ sample.  In this gas-rich sample, massive galaxies are typically more metal and dust-rich, therefore their estimated stellar masses are sensitive to the adopted 
dust-correction model. In contrast, low-mass galaxies are relatively dust-poor, 
thus their stellar mass estimates remain largely unaffected 
(figure~\ref{fig_GSMF_hi-selected}). 
The different dust attenuation models adopted in the stellar mass estimates therefore lead to systematic shifts in the high-mass end of the GSMF
(figure~\ref{fig_GSMF_hi-selected}). The \hi selected stellar mass functions are consistent with a three-parameter single Schechter function for the total, red and blue populations. 
The \kcorrect mass estimates, which are not dust-corrected, produce the lowest characteristic mass of the mass function, while the \gswlc-calibrated \kcorrect mass estimates, 
which  have individual dust-attenuation for all galaxies, 
produce the highest characteristic mass. 
The characteristic mass associated with 
the \taylor mass estimates, which account for dust through inclination-based corrections, 
lies between the two. A similar trend is seen 
(figure~\ref{fig_GSMF_hi-selected} \& table~\ref{tab:combined_schechter_params})
for the red and blue populations. This indicates that the \hi selected stellar mass functions are sensitive to the adopted stellar mass estimation methods.
Our HI-selected stellar mass function using $\mkcorrectgswlc$ mass estimates 
is consistent with ML methods trained on $\mgswlc$ mass estimates (figure~\ref{plot_ml_estim}). This demonstrates
that our linear calibration of $\mkcorrect$ masses to $\mgswlc$ masses does not 
systematically alter the HI-selected GSMF.

These differences between stellar mass estimates are reflected in derived quantities such as $\omegastar$ (or $\rhostar$) and $\nstar$. The variation of $\omegastar$ values for the three mass estimates are greater than $1\sigma$, the number densities $\nstar$, on the other hand, are within $1\sigma$. Although the $\alphah$ sample predominantly picks galaxies from the blue cloud, with $\sim 74\% - 89\%$ of the detections 
(corrected for the $\alphah$ survey selection) associated with blue galaxies, 
$\sim 54\%-62\%$ of stellar mass density, $\omegastar$, in the \hi selected sample is associated with red galaxies and the rest with blue galaxies (table~\ref{tab:combined_mass_estimates_schechter}).
For the remaining part of the paper we only use the $\mkcorrectgswlc$ stellar 
mass estimate.

In order to understand the population of gas-rich galaxies with respect to 
the overall galaxy population, we use a sub-volume (65\%) of the \hi selected sample, with complete spectroscopic coverage 
in SDSS to define an optically selected sample. Galaxies in the optically selected sample
have spectroscopic redshifts, corrected for peculiar velocities, using CosmicFlows-4 
\citep{2020AJ....159...67K}. We also define a joint optical-\hi sample in the same volume
as the optical sample, it is the intersection of the \hi selected sample and the optically
selected sample. The optically selected sample, with 45,609 galaxies, is $\sim 5\times$ 
larger than the joint optical-\hi sample. This allows us to make a fair comparison 
of the GSMF between the optically selected and the joint optical-\hi sample.

We estimate the GSMF of the optically selected sample for the total, red and blue 
galaxies. The GSMF is described by a double Schechter function with six parameters.
Our results are consistent with previous estimates of the GSMF 
based on SDSS \citep{2013MNRAS.436..697B,2016MNRAS.459.2150W} and GAMA \citep{2012MNRAS.421..621B,2022MNRAS.513..439D} (figures~\ref{fig_gsmf_optical_compare}, \ref{plot_HI_optical} and table~\ref{tab:schechter_params_optical_sel}). 
The GSMF for the joint optical-\hi sample, on the other hand, is similar in shape to the \hi selected sample, but with an amplitude that is $\sim 2.3\times$ smaller than the \hi selected sample. It is well described by a single Schechter function (figure~\ref{fig_gsmf_optical_compare} and table~\ref{tab:schechter_params_HI_sel}).
For the optically selected total sample 
$\rhostar/(10^8 \msun\,\,{\rm Mpc}^{-3}) = 2.713 \pm 0.479$, which is consistent 
with $\rhostar/(10^8 \msun\,\,{\rm Mpc}^{-3}) = 2.951 \pm 0.202$ from 
the results of the GAMA survey \citep{2022MNRAS.513..439D}, where a fair comparison 
can be made. Our estimates of $\rhostar$ are also consistent with results from 
SDSS \citep{2013MNRAS.436..697B,2016MNRAS.459.2150W} and earlier results from GAMA
\citep{2012MNRAS.421..621B} when we extrapolate their results to lower 
limiting masses of $\mstar \geq 7.0$.

The red galaxies account for $\sim 65\%$ of the stellar mass content with the remaining 
accounted by blue galaxies in the optically selected sample. In terms of galaxy counts
the red-blue divide is $40-60$. 
The trend is reversed for the joint optical-\hi sample.
$\alphah$ samples the optically selected population primarily in the blue cloud.
The stellar mass density and number density of blue galaxies of the joint optical-\hi sample is $\sim 65\%$ and $\sim 80\%$ respectively of the total, with the rest being associated with red galaxies (table~\ref{tab:schechter_fractions}).

We quantify the detection fraction of the joint optical-\hi sample with respect 
to the optically selected sample. 
The $\alphah$ survey is able to detect  $\sim 40\%$ of galaxies 
accounting for $\sim 33\%$ of stellar mass of an optically selected sample.
However the detection fraction is larger for blue population, as compared to the red.
$\sim 53\%$ $(\sim 13\%)$ of detections are in the blue (red) cloud which accounts 
for $\sim 62\%$ $(\sim 23\%)$ of stellar mass of the blue (red) cloud of an optically selected sample (table~\ref{tab:HI_optical_ratio_schechter}). 

The detection fraction is mass and color-dependent. The mass dependent detection fraction, 
the ratio of the GSMF of the joint optical-\hi and the optically selected samples, 
$\phi_{\rm opt-\hi}(\mstar)/\phi_{\rm opt}(\mstar)$ peaks at $\mstar = M_{\rm peak} = 9.3-9.5$,  below the knee of the mass function $M_* \simeq 10.8$. 
$\phi_{\rm opt-\hi}(M_{\rm peak})/\phi_{\rm opt}(M_{\rm peak}) = (0.87, 0.51)$ for the 
blue and red populations respectively.  
The ratio of these abundances (blue, red) decrease
from the  peak values to $(0.33, 0.03)$ and $(0.35,0.14)$ at the lowest mass scale 
$\mstar = 7.1$ and the largest mass scale, $\mstar = 11.5$, respectively.

We obtain the $\mstar-\mhi$ scaling relation by abundance matching the GSMF and the HIMF for the \hi-selected sample and the joint optical-\hi sample. This is a (\hi) selection-corrected scaling relation different from the scaling relation obtained from the observed \hi selected data. $\alphah$ data is biased in that it samples 
gas-rich objects (larger $\mhi$) at fixed stellar mass. Correcting for the selection bias, we obtain 
an unbiased $\mhi-\mstar$ scaling relation for the \hi-selected sample  that is consistent 
with a volume limited sample, that does not suffer from selection bias, 
in figure~\ref{figure_vol_lim}. Our scaling relations are also consistent 
with those from the MIGHTEE-HI survey using the MeerKAT interferometer, with a sample 
of 249 galaxies \citep{2023MNRAS.522.5308P,2023MNRAS.525..256P}.
We have repeated the analysis for 
the joint optical-\hi sample and present scaling relations for the red, blue and total populations.
These are consistent with the scaling relations of these populations for the \hi-selected sample.
The scaling relations are monotonically increasing functions of mass, but with a slope that is 
decreasing with increasing mass for the blue and total populations and a slope that is increasing 
with increasing mass for the red populations (figure~\ref{figure_scaling_reln}).

The work presented here is a multi-wavelength, from UV to radio, analysis of galaxy properties. 
It allows us to constrain and correlate the cold, atomic gas content of galaxies to its stellar mass, in an unbiased way. Although we have only looked at the $\mhi-\mstar$ relation in this work, we can also use
this catalog to constrain e.g. the r-band luminosity function, $\phi(M_{\rm r})$ which in turn would 
lead to an $\mhi-M_{\rm r}$ relation for \hi-selected galaxies. This can be repeated for any other bands in SDSS, leading to tight constraints and correlations between the atomic gas and stellar population of galaxies  that cosmological simulations need to reproduce. By using conditional \hi abundances in the color-magnitude plane 
of galaxies \citep{2021MNRAS.500L..37D}, our results can also be used to explore 
optimal stacking strategies to not only detect \hi \citep{2018MNRAS.473.1879R}, but to also constrain \hi abundances at high redshifts, which are free from selection bias. 
Stellar masses, on the other hand, have a tight monotonic relation with halo masses \citep{2009ApJ...696..620C,2010ApJ...717..379B,2013MNRAS.428.3121M,2019MNRAS.488.3143B}, which 
can be exploited to connect $\mhi$ to $M_{\rm halo}$ through the galaxy directly, rather 
than indirectly from stacked \hi observations in optically selected galaxies \citep{2022MNRAS.511.2585D}.
This will allow us to model the clustering of gas-rich galaxies and compare them directly to 
observations \citep{2013ApJ...776...43P,2017ApJ...846...61G}. We will report these results in forthcoming papers.

\section*{Acknowledgements}

This work is based on data from the ALFALFA survey and Sloan Digital Sky Survey Data Release 15 (SDSS DR15). We acknowledge the efforts of the ALFALFA collaboration in observing, flagging, and extracting the galaxy properties used in this study. Funding for SDSS and SDSS-II was provided by the Alfred P. Sloan Foundation and a consortium of participating institutions, and the survey is managed by the Astrophysical Research Consortium.

TT and NK acknowledge useful conversations with Jasjeet Bagla, Raghunathan Srianand and Aseem Paranjape.
NK acknowledges support from the IUCAA Associateship Programme.
We thank Adriana Durbala for making the value added catalog of $\alphah$, used in this work, public. TT thanks Adriana Durbala for clarifications regarding the SDSS cross-match corrections in her catalog. We also thank Ehsan Kourkchi for helpful correspondence regarding uncertainty estimation in the Cosmicflows distance-velocity calculations. 
All analyses presented in this work were carried out on the NISER HPC facilities.

\noindent
\textit{Python Packages} :
\software{NumPy} \citep{2011CSE....13b..22V}, 
\software{SciPy} \citep{2020NatMe..17..261V},
\software{astropy} \citep{2013A&A...558A..33A},
\software{Matplotlib} \citep{2007CSE.....9...90H},
\software{emcee}\citep{2013PASP..125..306F}
\software{kcorrect}\citep{2007AJ....133..734B}
\software{jupyter} \citep{2016ppap.book...87K},
\software{mpi4py} \citep{mpi4py2021}.
\software{Scikit-Learn}\citep{2011JMLR...12.2825P},
\software{pandas}\citep{2010scpy.soft.....M},
\software{XGBoost}\citep{2016arXiv160302754C},
\software{LightGBM}\citep{Ke2017LightGBMAH}.

\section*{Data Availability}\label{sec_data availability}
The 100\% ALFALFA catalog and its optical counterpart identified by  
\citet{2020AJ....160..271D} are available on \textit{egg.astro.cornell.edu/alfalfa/data}. 
Data from Sloan Digital Sky Survey Data Release 15 (SDSS DR15) are publicly available via \textit{skyserver.sdss.org}. Distances corresponding to SDSS galaxies are obtained from the Extragalactic Distance Database (EDD) at \textit{edd.ifa.hawaii.edu}.


\bibliographystyle{mnras}
\bibliography{references}

\appendix

\section{Boundary Coordinates of Defined Regions}\label{sec_boundary}
The vertices of Region I are as follows in a clockwise direction: \\ 
$(45.00^\circ, 0.0^\circ), (330.00^\circ, 0.0^\circ), (330.00^\circ, 2.0^\circ),(337.50^\circ, 2.0^\circ),\\
(337.50^\circ, 6.0^\circ), (330.00^\circ, 6.0^\circ),  (330.00^\circ, 30.7^\circ), (353.45^\circ, 35.8^\circ),\\ (13.65^\circ, 35.8^\circ), (33.45^\circ, 32.5^\circ),  
(33.45^\circ, 23.7^\circ),  (45.00^\circ, 21.1^\circ),\\ (45.00^\circ, 19.0^\circ), (30.60^\circ, 21.1^\circ), (30.60^\circ, 7.0^\circ),(45.00^\circ, 7.0^\circ),\\ 
(45.00^\circ, 2.0^\circ)$.\\
The vertices of Region II are as follows in a clockwise direction:  
$(246.45^\circ, 0.0^\circ), (115.50^\circ, 0.0^\circ), (122.25^\circ, 2.02^\circ),(116.25^\circ, 11.0^\circ),\\(116.25^\circ, 16.0^\circ),(116.25^\circ, 18.0^\circ),  (130.50^\circ, 18.0^\circ), (130.50^\circ, 20.0^\circ), \\(141.00^\circ, 20.0^\circ), (141.00^\circ, 24.0^\circ),  
(114.00^\circ, 24.0^\circ), (114.00^\circ, 30.0^\circ), \\(127.50^\circ, 30.0^\circ), (127.50^\circ, 32.0^\circ), (142.50^\circ, 32.0^\circ),  (142.50^\circ, 36.0^\circ), \\(232.50^\circ, 36.0^\circ), (232.50^\circ, 32.0^\circ), (240.00^\circ, 32.0^\circ), (240.00^\circ, 30.0^\circ), \\  
(247.50^\circ, 30.0^\circ),  (247.50^\circ, 24.0^\circ), (231.00^\circ, 24.0^\circ), (231.00^\circ, 20.0^\circ), \\(231.00^\circ, 18.0^\circ), (240.00^\circ, 18.0^\circ),  (240.00^\circ, 16.0^\circ), (246.45^\circ, 16.0^\circ), \\(246.45^\circ, 4.3^\circ), (238.50^\circ, 1.35^\circ), 
(246.45^\circ, 0.53^\circ).$\\
The vertices of Region III are as follows in a clockwise direction:
$(127.95^\circ, 0.0^\circ), (127.95^\circ, 3.2^\circ), (121.50^\circ, 3.2^\circ), (116.25^\circ, 11.0^\circ),\\(116.25^\circ, 16.0^\circ), (116.25^\circ, 18.0^\circ), (130.50^\circ, 18.0^\circ), (130.50^\circ, 20.0^\circ),\\ (141.00^\circ, 20.0^\circ), (141.00^\circ, 24.0^\circ), (114.00^\circ, 24.0^\circ), (114.00^\circ, 30.0^\circ), \\(127.50^\circ, 30.0^\circ), (127.50^\circ, 32.0^\circ), (142.50^\circ, 32.0^\circ), (142.50^\circ, 36.0^\circ),\\ (232.50^\circ, 36.0^\circ), (232.50^\circ, 32.0^\circ), (240.00^\circ, 32.0^\circ), (240.00^\circ, 30.0^\circ), \\(247.50^\circ, 30.0^\circ), (247.50^\circ, 24.0^\circ), (231.00^\circ, 24.0^\circ), (231.00^\circ, 20.0^\circ),\\ (231.00^\circ, 18.0^\circ), (240.00^\circ, 18.0^\circ), (240.00^\circ, 16.0^\circ), (246.45^\circ, 16.0^\circ), \\(246.45^\circ, 4.3^\circ), (238.50^\circ, 1.35^\circ), (246.45^\circ, 0.53^\circ), (246.45^\circ, 0.0^\circ).$

\section{Determining the incompleteness of the GSMF at the Low-Mass End}
\label{sec_incompl_GSMF}
\begin{figure*}
    \begin{tabular}{cc}
    \includegraphics[width=3.4in]{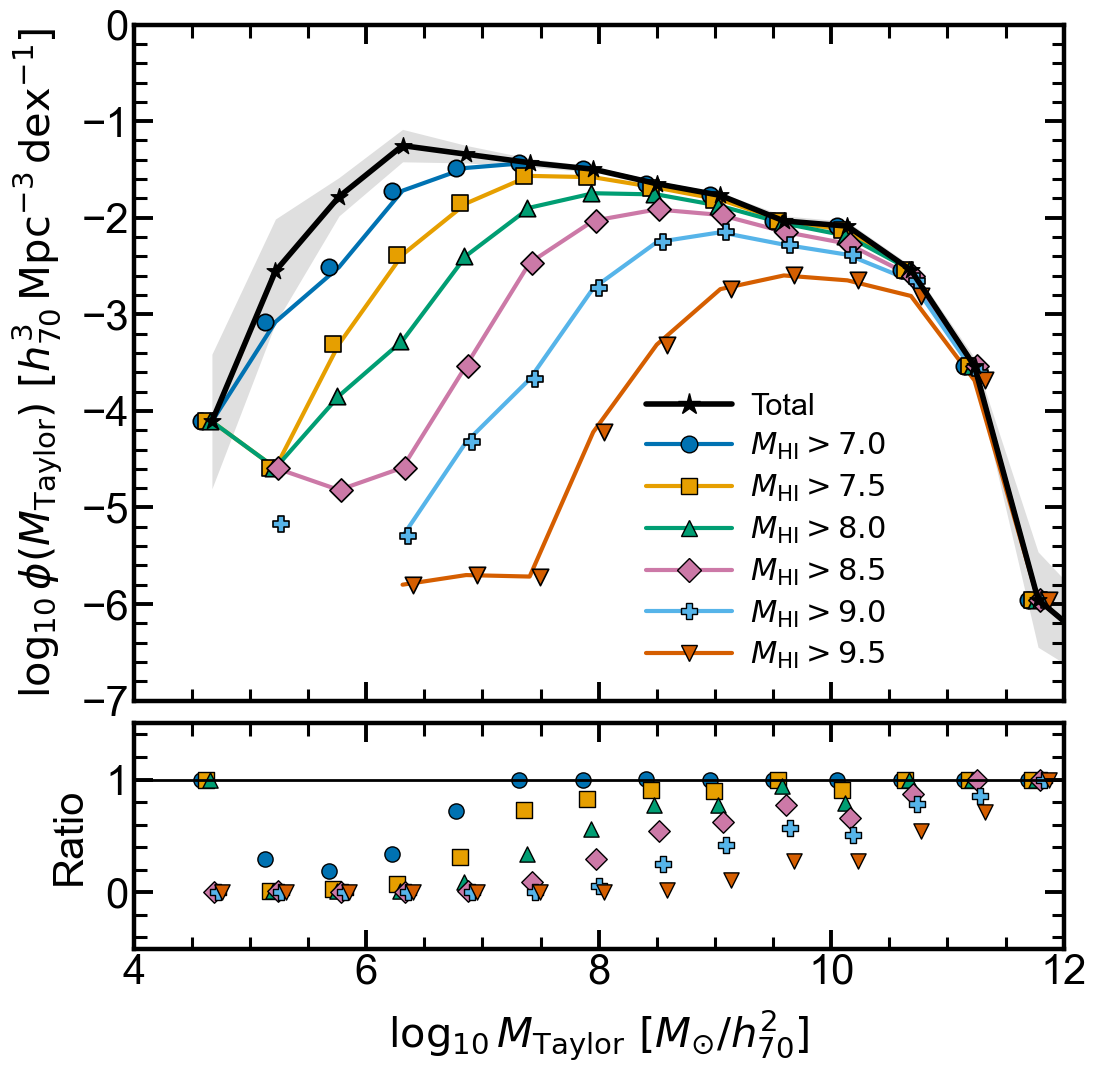} 
    \includegraphics[width=3.4in]{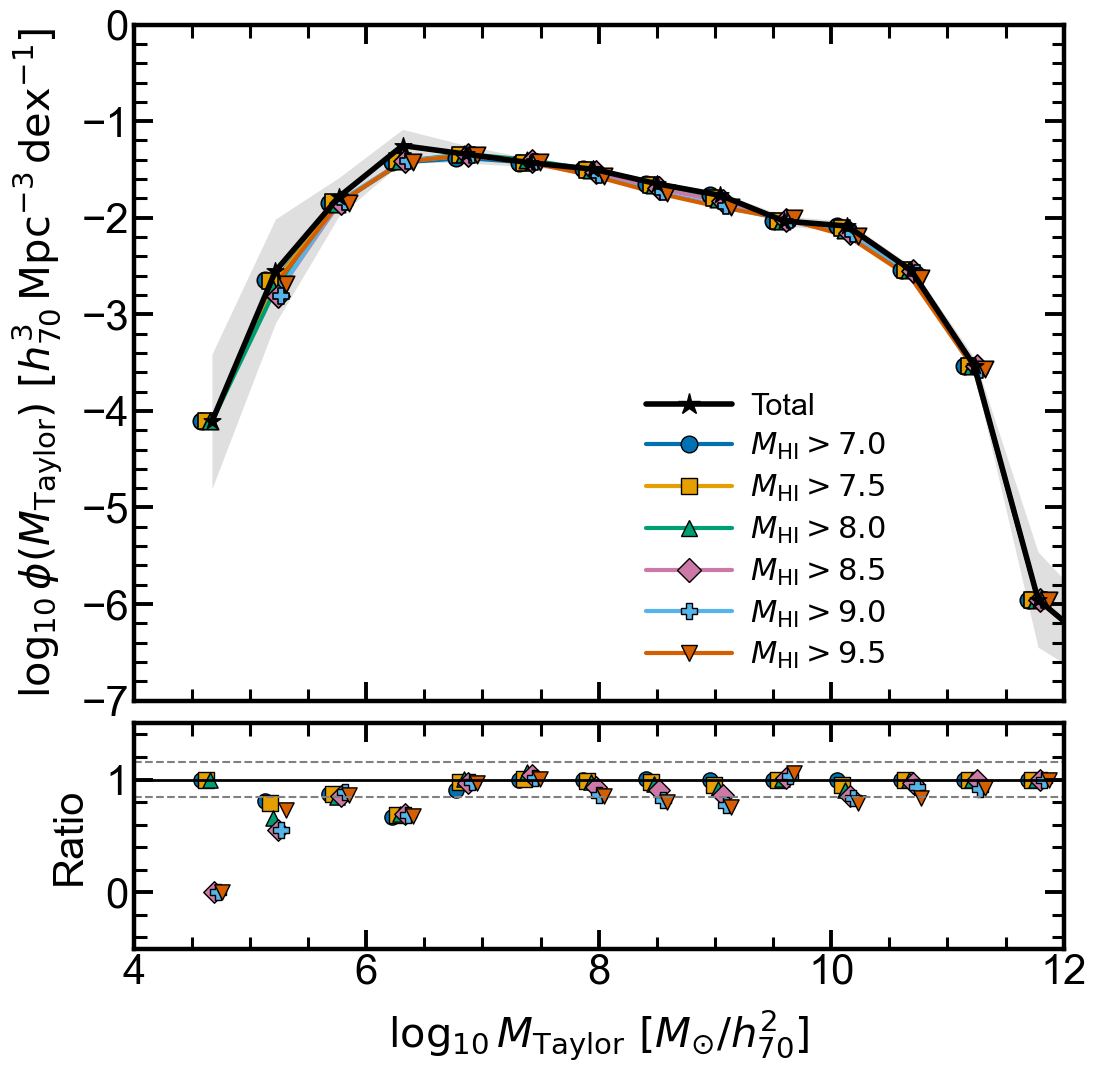} \\
    \end{tabular}
    \caption{\emph{Left}:The stellar mass function (GSMF) derived using \taylor mass estimates for the full sample (black curve and star markers) and for subsamples defined by cumulative \hi mass thresholds $\mhi \geq 7.0,7.5,8.0,8.5,9.0,9.5$
    (colored symbols). The grey shaded region around the total GSMF represents the uncertainty obtained from poisson errors. The lower panel shows the ratio of each subsample GSMF to the total GSMF. The horizontal line indicates unity. The suppression at low masses reflects the imposed \hi mass cuts, while the high-mass end converges to the total GSMF. 
      \emph{Right}: The figure follows the same notation as the \emph{left} panel, but the stellar mass functions for the subsamples are compensated for the $\mhi$ cuts. Colored curves show the reconstructed GSMFs corresponding $\mhi$ thresholds, 
      $\mhi > 7.0,7.5,8.0,8.5,9.0,9.5$ (colored symbols). The black curve represents the total GSMF with its associated uncertainty shown by the grey shaded band. The lower panel shows the ratio of each reconstructed GSMF to the total GSMF.}
    \label{plot_mass_cut_compensation}
\end{figure*}

As shown in figure~\ref{figure_vol_lim} the $\mhi-\mstar$ relation has an intrinsic scatter. Because of this scatter, any cut applied in $\mhi$ will make GSMFs incomplete below a certain stellar mass scale. Our $\alphah$ sample has an intrinsic threshold of $\mhi \geq 6$.  This results in a drastic drop in abundances for $\mstar \leq 6.25$, seen in the left panel of figure~\ref{plot_mass_cut_compensation}. We aim to determine the range of stellar masses for which GSMFs remain unaffected by this cutoff and explore potential correction methods. To achieve this, we first apply various cuts in $\mhi$ on the total sample and look at its effect on the GSMFs, to determine the stellar mass scale which is complete.

We apply $\mhi^{\rm cut}$ cuts ranging from 7.0 to 9.5 on the total $\alphah$ 
galaxy sample, generating six samples in the process. We compute GSMFs for these six samples for $\mtaylor$ stellar mass estimates and compare them with the GSMF of the total sample (black line) in the left panel of figure~\ref{plot_mass_cut_compensation}. The total sample itself 
has an intrinsic cut of $\mhi^{\rm cut} = 6.0$ (see figure~\ref{plot_himf}).

As can be seen from the various GSMFs in the left panel of figure~\ref{plot_mass_cut_compensation}, they show consistent behavior at higher stellar masses but their differences begin to diverge at lower masses. This discrepancy is due to the exclusion of low-mass galaxies due to the imposition of $\mhi \geq \mhi^{\rm cut}$. Assuming a power law behaviour at low masses, we see that GSMF is accurate up to $\mstar \gtrsim 6.2$ for the total population and begins to decline below this mass.
For the $\mhi \geq (\mhi^{\rm cut}=7.0)$ and $\mhi \geq (\mhi^{\rm cut}=7.5)$  
the corresponding GSMFs are complete above $\mstar \geq 7.3$ and $\mstar \geq 8.25$ respectively. Therefore, the GSMF for the 
total population is complete above the mass$ 6.2 > \mhi > 7.3$.

We now look at an alternate approach to obtain the stellar mass completeness 
of the total population. We compensate for these missing galaxies due to the applied $\mhi^{\rm cut}$ and evaluate the success of our corrections. We consider the same 
set of six samples $\mhi^{\rm cut} = \{7.0,7.5,8.0,8.5,9.0,9.5\}$ as before. 
For a sample with a given $\mhi \geq \mhi^{\rm cut}$, 
at fixed $\mstar$ bin, we sample the missing $N$ galaxies with $\mhi < \mhi^{\rm cut}$, 
based on the distribution of \hi masses 
at the same fixed $\mstar$ bin, $P(\mhi|\mstar)$ of the total population. 
The missing $N$ galaxies, each with masses $\mhi^{i} < \mhi^{\rm cut}$
are then assigned a $\veff$ sampled from the distribution of $P(\veff|\mhi)$, 
where $P(\veff|\mhi)$ is determined from the total sample.
Each of these $N$ missing galaxies, below $\mhi < \mhi^{\rm cut}$ at the 
given stellar mass are binned in that particular stellar mass bin with a weight of 
$1/\veff$. We make 1000 realisations of $\veff$ for each missing object  
for $\mhi < \mhi^{\rm cut}$ at every given bin in stellar mass. 
The median corrections over these 1000 realisations are plotted in the right 
panel of figure~\ref{plot_mass_cut_compensation} for each of the six samples.
The correction works well. By looking at the residuals in this figure 
we find that  all samples converge for $\mstar \gtrsim 6.75$.
A conservative estimate of the GSMF completeness is therefore for $\mstar \geq 7.0$, consistent with the previous method, described above.

\section{A Machine Learning approach to predicting Stellar Masses}
\label{sec_ML}
We have used \gswlc mass estimates to calibrate the remaining \kcorrect mass estimates for dust extinction by using a linear model as discussed in section~\ref{sec._HI_selc._sample} and figure~\ref{fig_mtaylor-mkcorrect-mgswlc}. In this mini exercise, we instead adopt a model-independent approach, using Machine Learning (ML) to predict stellar mass estimates.
These are then used to obtain the HI-selected GSMF with the $1/\veff$ method.
We explore three widely used ensemble machine learning algorithms, 
Random Forests \citep{2001MachL..45....5B}, eXtreme Gradient Boosting 
\citep[XGBoost]{2016arXiv160302754C} and Light Gradient Boosting Machine 
\citep[LightGBM]{Ke2017LightGBMAH} and their predictions of stellar mass estimates when trained on $\mgswlc$ mass estimates. We briefly describe the algorithms below and direct the reader to the original references:  
\begin{itemize}
    \item Random Forest works by constructing a large number of decision trees using bootstrap samples of the training data and random subsets of the input features. The final prediction is made by averaging the predictions from all trees.
    \item XGBoost is a gradient boosting algorithm that makes decision trees one after another, where each new tree corrects the errors of the previous ones. 
    \item LightGBM grows decision trees by expanding the leaf that gives the largest improvement in prediction accuracy. This strategy, along with histogram-based feature binning, makes the algorithm run faster and use less memory for large datasets with many features.
\end{itemize}
Our complete parent catalog contains $29,539$ galaxies. Of these, $14,721$ have $\mgswlc$ mass estimates, forming our \textit{Labeled} sample and the remaining $14,818$ galaxies lacking a GSWLC match which is our \textit{Prediction} set. We partition the labeled sample 50:50 (fixed random seed) into a \textit{Training} set ($7,360$ galaxies) and a held-out \textit{Testing} set ($7,361$ galaxies). Missing values are imputed with the per-feature median, and all features are standardized to zero mean and unit variance prior to training. For all three models iterated over both feature sets, the performance is evaluated using predictions on the held-out test set, from which we calculate our performance metrics ($R^2$, RMSE, and MAE). Following validation, each model is refit on the complete labeled sample and then used to predict the stellar masses and individual $1\sigma$ uncertainty intervals for the complete parent catalog.
\newline
We use the following feature sets for training each of the algorithms.
\begin{itemize}
    \item Feature Set I ($\mathcal{F}_{\rm mag}$): 
    $\left\{M_{\rm u}, M_{\rm g}, M_{\rm r}, M_{\rm i}, M_{\rm z}, \wfifty, \mhi, \zcmb  \right\}$
    \item Feature Set II ($\mathcal{F}_{\rm color-mag}$):  
    
    $\left\{M_{\rm r}, C_{\rm ug},C_{\rm gr},C_{\rm ri},C_{iz}, 
    \wfifty, \mhi, \zcmb  \right\}$
\end{itemize}
where, e.g. $C_{\rm ug}$ is the $u-g$ color.
Although feature sets \fmag and \fcolor are equivalent, being linear combinations of each other, tree-based models cannot access these combinations directly because they can only split data along one feature at a time. 
We have therefore included \fcolor to see if providing pre-computed linear combinations, in this case colors, can act as a shortcut for the model   to find physical stellar population boundaries much faster and with fewer data points. Since this is an \hi-selected sample, we have included both $\wfifty$ and $\mhi$. $\wfifty$ (up to an inclination angle) probes the potential of the dark matter halo which in turn is a strong indicator of the stellar mass. $\mhi$ may be a biased and indirect indicator of stellar mass since gas-rich galaxies sample, primarily, the blue cloud.
By providing redshift as an explicit feature, we allow the tree-based model to draw a boundary in the feature space that maps out the survey's completeness limits. It learns where in redshift space it is physically allowed to find a galaxy of a given mass. 

For each model and feature set, we use a fixed default hyperparameter configuration without performing hyperparameter tuning. Random Forest requires only one fitted model per evaluation stage, so the spread of predictions across the constituent trees gives the $1\sigma$ prediction interval. For XGBoost and LightGBM, we trained additional models with quantile regression loss, as detailed in implementations by \citep{friedman2001greedy}, at $16^{\rm th}$, $50^{\rm th}$, and $84^{\rm th}$ to construct the corresponding $1\sigma$ intervals, alongside the primary point estimate model used for performance metrics evaluation. The quantile models share the base learning rate and depth settings of the respective point estimate models. We have verified that the metrics obtained with the squared-error objective and the $50^{\rm th}$-percentile quantile models are empirically consistent and hence we used the former for evaluating the performance metrics for all models for consistency. This serves our purpose of estimating reliable $1\sigma$ prediction intervals without spending significant computational time on hyperparameter tuning.

\begin{figure*}
    \begin{tabular}{cc}
  \includegraphics[width=3.4in]{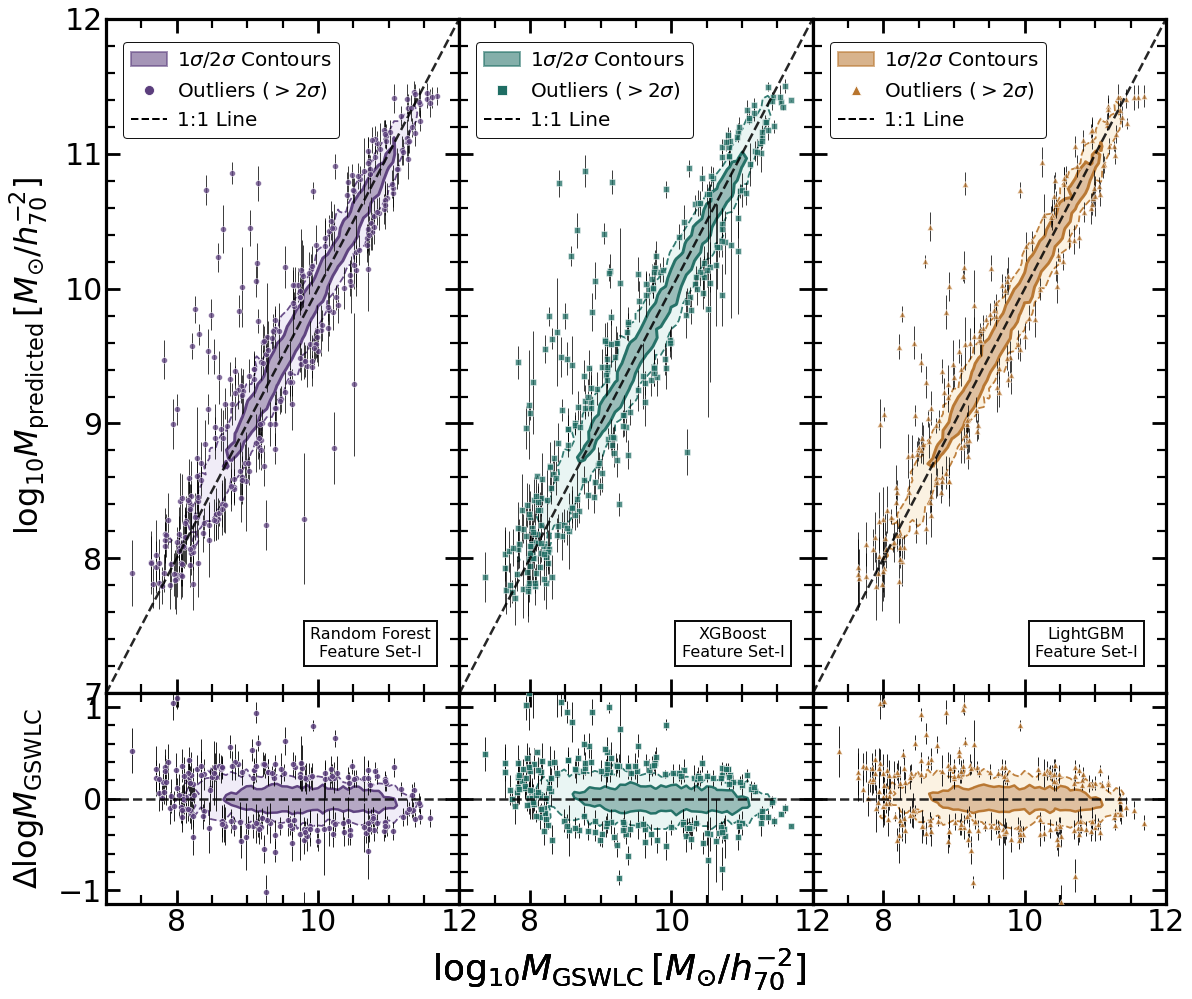} 
  \includegraphics[width=3.4in]{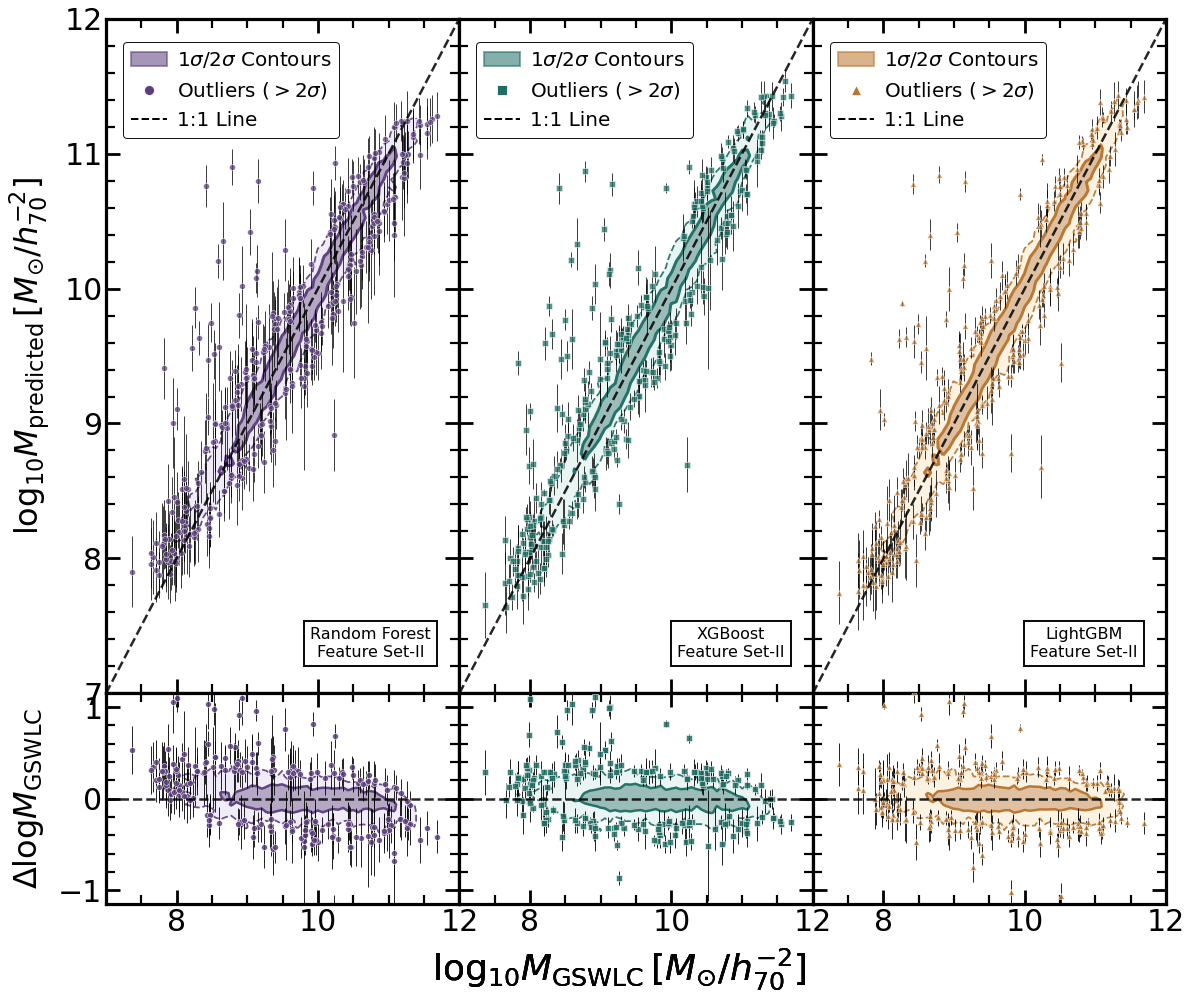}\\
      \end{tabular}
  \caption{\emph{Top}: Stellar mass estimates from three machine learning models - Random Forest(left, purple), XGBoost(middle, teal) and LightGBM(right, orange) trained on two feature sets Feature Set-1 (\fmag)[\emph{Left}]and Feature Set-2 (\fcolor)[\emph{Right}] plotted against GSWLC. Shaded region denotes 1$\sigma$ and 2$\sigma$ density contours and $N=200$ linearly subsampled points from the outlier population lying outside the 2$\sigma$ contour are discretely marked with error bars representing their associated $1\sigma$ uncertainties. The black dashed line is the y=x (1:1) curve. \emph{Bottom}: Corresponding residuals $\Delta \log_{10}\mgswlc=\log_{10}M_{\textrm{predicted}}-\log_{10}\mgswlc$ are plotted as a function of $\log_{10}\mgswlc$ with the dashed horizontal line depicting zero residual.}
\label{plot_scat_comparison}
\end{figure*}

To evaluate the performance of our models on the Test sample, we use the following standard metrics:
\begin{enumerate}
    \item The coefficient of determination ($R^2$): 
    \begin{equation}
    R^2 = 1 - \frac{\displaystyle\sum_{i=1}^{n} (y_i - \hat{y}_i)^2}
                    {\displaystyle\sum_{i=1}^{n} (y_i - \bar{y})^2}
    \label{eq:r2}
\end{equation}
where, $y_i$ is the true stellar mass, $\mgswlc$ of the $i$-th galaxy in the test set, $\hat{y}_i$ is the predicted stellar mass, $\bar{y}$ is the mean of the true stellar masses over the test set and $n$ is the total number of galaxies in the test set. $R^2$ quantifies the fraction of variance in stellar mass explained by the model. $R^2 = 1$ corresponds to a perfect prediction, $R^2 = 0$ corresponds to a model with no more predictive power than simply guessing the sample mean $\bar{y}$ for every galaxy and negative values indicate performance worse than this baseline.
\item Root Mean Square Error (RMSE):
\begin{equation}
    \mathrm{RMSE} = \sqrt{\frac{1}{n} \sum_{i=1}^{n} (\hat{y}_i - y_i)^2}
    \label{eq:rmse}
\end{equation}
RMSE is the square root of mean squared residual and quantifies the prediction scatter of the model which is more sensitive to outliers.
\item Mean Absolute Error (MAE):
\begin{equation}
    \mathrm{MAE} = \frac{1}{n} \sum_{i=1}^{n} \left| \hat{y}_i - y_i \right|
    \label{eq:mae}
\end{equation}
MAE is the average absolute residual. For all three metrics, a higher $R^2$ shows a better model fit, while lower MAE and RMSE scores show higher predictive accuracy. 
\end{enumerate}

\begin{table}
\begin{center}
\begin{tabular}{|l|c|c|c|}
\hline
\textbf{Model} & $R^2$ & RMSE(dex) & MAE(dex) \\
\hline
\multicolumn{4}{|c|}{\textbf{Feature Set-1 (\fmag)}} \\
\hline
 Random Forest & $0.9633$ & $0.1406$ & $0.0913$ \\
 XGBoost        & $0.9626$ & $0.1418$ & $0.0922$ \\
 LightGBM       & $0.9632$ & $0.1407$ & $0.0916$ \\
\hline
\multicolumn{4}{|c|}{\textbf{Feature Set-2 (\fcolor)}} \\
\hline
 Random Forest & $0.9633$ & $0.1406$ & $0.0918$ \\
 XGBoost        & $0.9655$ & $0.1363$ & $0.0877$ \\
 LightGBM       & $0.9657$ & $0.1358$ & $0.0875$ \\
\hline
\end{tabular}
\end{center}
\caption{
Model performance comparison across feature sets, showing the coefficient of determination ($R^2$), root mean squared error (RMSE), and mean absolute error (MAE) for the Random Forest, XGBoost, and LightGBM models.
}
\label{tab:model_performance}
\end{table}

\begin{figure*}
    \begin{tabular}{cc}
    \includegraphics[width=3.4in]{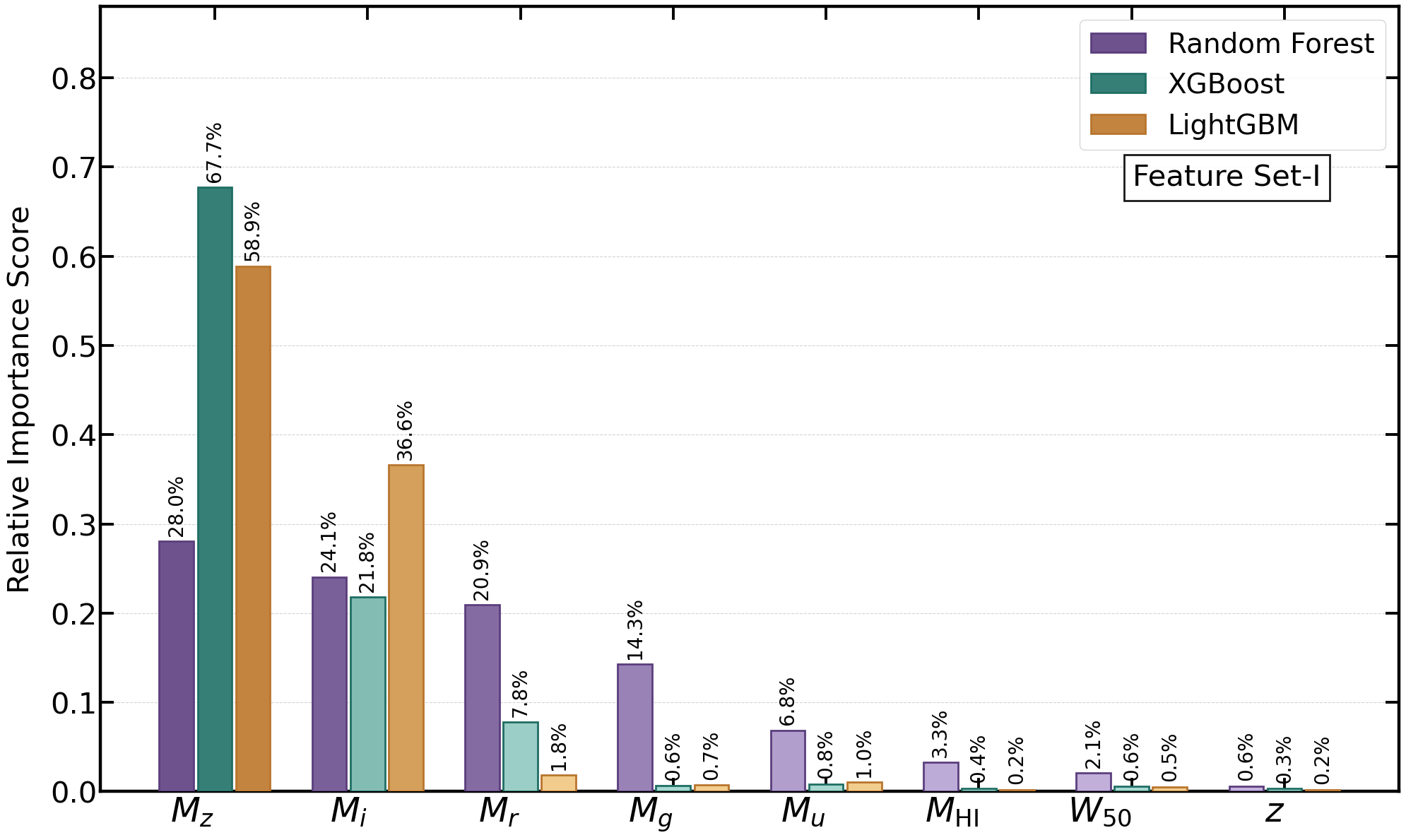} 
    \includegraphics[width=3.4in]{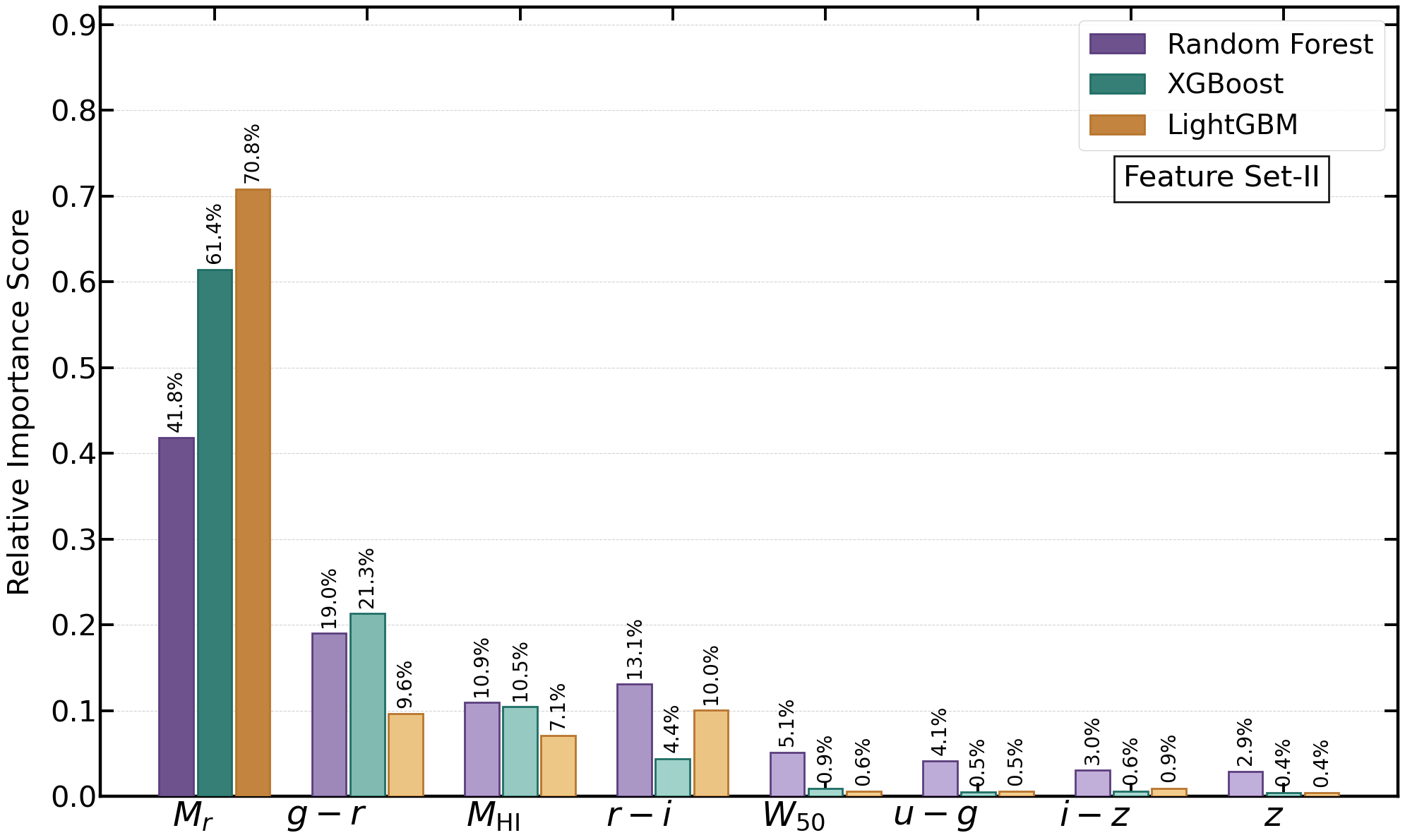} \\
    \end{tabular}
    \caption{Relative feature importance scores for the three machine learning models - Random Forest(purple), XGBoost(teal) and LightGBM(orange) trained on Feature Set-1 \fmag (\emph{Left}) and Feature Set-2 \fcolor (\emph{Right}). Percentage labels above each bar denote a feature's share of the total importance for the corresponding model.}
    \label{plot_rel_imp}
\end{figure*}
In Fig. \ref{plot_scat_comparison}, we have plotted the predicted stellar mass against the true $\mgswlc$ masses across the bin range $(\mstar / M_{\odot}) \in [7, 12]$ evaluated on the Test Dataset ($N = 7,361$). To visualize the distribution density, we plot the 1D-equivalent $1\sigma$ ($68\%$) and $2\sigma$ ($95\%$) contours of the distribution. 
In order to avoid overcrowding we only plot randomly sampled 200 ma outside the 
$2\sigma$ contour. All mass estimates have uncertainties, shown as error bars.
The lower panel shows mass residuals ($\Delta \log \mstar = \log_{10} M_{\rm predicted} - \log_{10} \mgswlc$) vs true stellar mass. 

To understand which features drive the predictions, we also plot the global feature importance from the models on the complete labeled sample ($N = 14,721$). For the Random Forest model, we calculate the feature importance through the Mean Decrease in Impurity (MDI) \citep{2001MachL..45....5B}, which measures how much each feature contributes to reducing prediction error by splitting on a given variable averaged across all constituent trees. We have plotted the ranked MDI scores for both feature sets as percentage contribution. To evaluate feature importance for XGBoost and LightGBM, we use the \textit{Gain} metric \citep{DBLP:conf/nips/LouppeWSG13,friedman2001greedy}, which measures the average reduction in prediction error achieved across all split nodes that use the specific variable during sequential boosting rounds. The results are shown 
in figure~\ref{plot_rel_imp}.

\begin{figure}
  \includegraphics[width=\columnwidth]{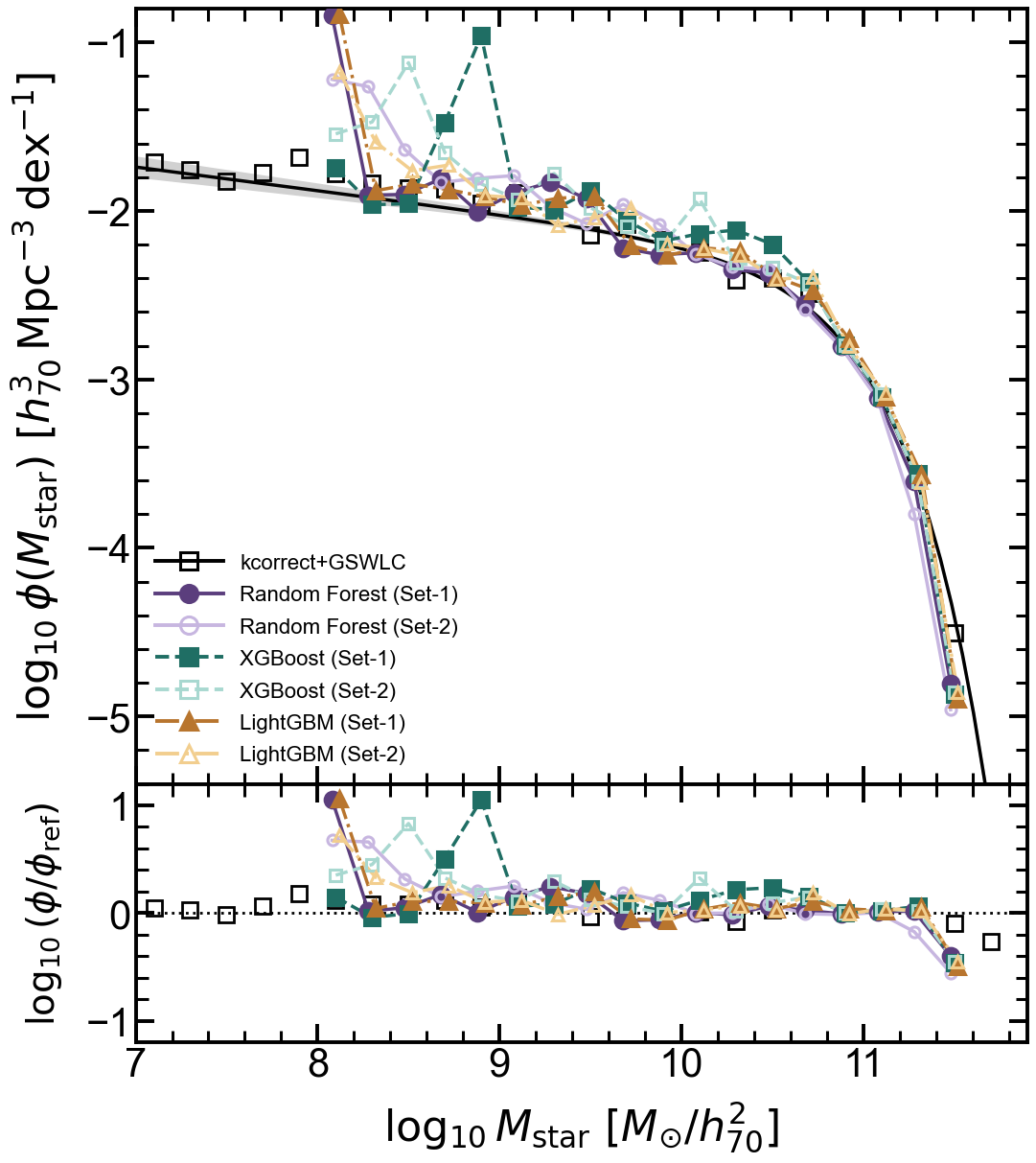} \\
  \caption{Comparison of the stellar mass function (SMF) derived from the $\kgswlc$ mass estimates (black squares with $1\sigma$ uncertainty band) with predictions from three machine learning models - Random Forest, XGBoost, and LightGBM each trained on two feature sets (\fmag  and \fcolor) for each model.
  Line style denotes model (solid: Random Forest, dashed: XGBoost, dash-dotted: LightGBM) and marker style denotes feature set (open/light: Set-1 \fmag, filled/dark: Set-2 \fcolor). The data points corresponding to Random Forest (LightGBM)- estimated masses are shifted to the right (left) by 0.02 dex with respect to the $\mgswlc$ mass estimates for better comparison. The lower panel shows residuals with respect to the SMF derived from $\kgswlc$ mass estimates.}
\label{plot_ml_comparison}
\end{figure}

In figure~\ref{plot_scat_comparison} the predictions for all models 
are tight, with no systematic bias for $2\sigma$ of the population. 
The $M_{\rm predicted}-\mgswlc$ contours follow the 1:1 line closely for mass range $\mstar  \in [8.5, 11.3]$, while below $\mstar \sim 8.5$  the models systematically overestimate the stellar masses and above $\mstar \sim 11.3$ , the predictions underestimate the mass. This behavior depicts the limitation of tree-based algorithms. Decision trees make predictions by averaging groups of similar galaxies at the leaf nodes and at the outer boundaries of the dataset, there are few points to average with, so predictions are intrinsically pulled towards the global sample mean. This edge effect is reinforced by the fact that in our dataset, massive galaxies are rare, giving the models fewer samples to learn from and the faint galaxies at the lower mass end suffer from higher measurement noise and are affected by survey selection, reducing their observed counts.

Random Forest maintains consistent performance across both feature sets for the Testing Set, as shown in table~\ref{tab:model_performance}, while gradient boosting architectures show slightly improved performance metrics when trained using \fcolor, with LightGBM achieving the overall best performance. This difference comes down to how decision trees split data, as depicted by the feature importance plots in figure~\ref{plot_rel_imp}. Because decision trees split data one variable at a time(horizontal or vertical cuts in feature space), they cannot draw diagonal lines in the feature space. When given only independent magnitudes (\fmag) the models must construct staircase like sequence of splits to approximate galaxy colors boundaries(a diagonal feature). Random Forest builds independent trees, which gives it enough depth to construct those approximations using raw magnitudes without overfitting, but XGBoost and LightGBM build much shallower trees sequentially so they run out of splitting depth before mapping the color boundaries. After explicitly providing the color information directly, these algorithms no longer waste splits approximating the diagonal lines and show improved predictive accuracy compared to Random Forest's simple averaging approach.

While all three models qualitatively agree on which features are the most and least important, they distribute this importance very differently. Random Forest spreads its predictive weight across multiple features while XGBoost and LightGBM concentrate their learning onto two noticeably dominant optical features, ($M_r$ and $g-r$). Redshift is the least important feature in both the feature sets since it does not provide any direct information about the galaxy's intrinsic mass, and redshift information is already encoded in the optical features. We have included it for completeness. 

When we apply our individual mass predictions to construct the stellar mass function in  figure~\ref{plot_ml_comparison}, we see that models based on $\fmag$ (solid points, feature set I) agree better with the GSMF based on $\mkcorrectgswlc$ mass estimate as compared to those based on $\fcolor$ (open points, feature set II), in the mass range $8.5 \le \mstar \le 11.3$, although table~\ref{tab:model_performance}
suggests otherwise. We note that the algorithms are trained to predict stellar masses and not the mass functions which depend on a combination of $\{\mhi,\wfifty,\zcmb\}$ to produce $\veff$. We had (wrongly) originally anticipated that by providing selection sensitive features ($\wfifty$,$\mhi$,$\zcmb$), the algorithms would learn the galaxy's effective volume $\veff$ with the stellar mass. But  our models were trained to predict $\mstar$ rather than predicting both $\mstar$ and $\veff$ simultaneously. They treat every galaxy equally and operate without awareness of  $\veff$ volume weights. As a result, even a minor prediction scatter or a small systematic bias at the low-mass end gets heavily amplified by large $1/\veff$ corrections which distort the derived GSMF in the lower mass range. 
At higher masses, the $1/\veff$ corrections are negligible but all models under-predict the
few objects that exist in the sample, resulting in  lower predicted abundances.
However all models are able to, reasonably, capture the overall shape of the GSMF without requiring hyperparameter tuning in the mass range that is well sampled.
Of the six models, Random Forest and LightGBM with $\fmag$ agree with the GSMF estimate 
based on $\mkcorrectgswlc$ masses, the best. In order to better predict the GSMF, one may need to predict a pair of properties $(\mstar,\veff)$. We will explore such possibilities 
and improvements in the future.

\bsp	
\label{lastpage}
\end{document}